\documentclass[fleqn,14pt]{wlscirep}

\usepackage[utf8]{inputenc}
\usepackage[T1]{fontenc}
\usepackage{mathrsfs}
\usepackage{amsmath}
\usepackage{xargs}
\usepackage{bm}
\usepackage{babel,csquotes,xpatch}

\newcounter{mainbib}

\usepackage{CJKutf8}
\newcommand{\CNnames}[1]{{\begin{CJK}{UTF8}{gbsn}~(#1)~\end{CJK}}}

\newcommand{\apss}{Astrophysics and Space Science}
\newcommand{\aap}{Astronomy and Astrophysics}
\newcommand{\apj}{Astrophysical Journal}
\newcommand{\apjs}{Astrophysical Journal Supplement Series}
\newcommand{\nat}{Nature}
\newcommand{\mnras}{Monthly Notices of the Royal Astronomical Society}
\newcommand{\araa}{Annual Review of Astronomy and Astrophysics}
\newcommand{\aapr}{The Astronomy and Astrophysics Review}
\newcommand{\pasj}{Publications of the Astronomical Society of Japan}

\newcommand{\ri}{r_{\rm i}}
\newcommand{\ro}{r_{\rm o}}

\newcommand{\omg}{\langle\Omega\rangle_{\rm g}}
\newcommand{\omp}{\langle\Omega\rangle_{\rm p}}

\newcommand{\dpun}{\Delta\Pi_1}
\newcommand{\dn}{\Delta\nu}
\newcommand{\numax}{\nu_{\rm max}}
\newcommand{\epsg}{\varepsilon_{\rm g}}
\newcommand{\kepler}{\textit{Kepler}}
\newcommand{\mesa}{{\textsc{MESA}}}
\newcommand\ddfrac[2]{\frac{\displaystyle #1}{\displaystyle #2}}

\newcommand\T{\rule{0pt}{2.6ex}}
\newcommand\B{\rule[-1.2ex]{0pt}{0pt}}

\newcommand{\Lop}{\ensuremath{{\cal L}}}
\newcommand{\LopL}{\ensuremath{\Lop_\mathrm{L}}}
\newcommand{\vb}{\bm}
\newcommand{\vxi}{\bm{\xi}}

\newcommand{\er}{\ensuremath{\boldsymbol{e}_r}}
\newcommand{\etheta}{\ensuremath{\boldsymbol{e}_\theta}}
\newcommand{\ephi}{\ensuremath{\boldsymbol{e}_\phi}}
\newcommandx{\Ylm}[2][1=l, 2=m]{\ensuremath{Y_{#1}^{#2}}}
\newcommandx{\Ylmt}[2][1=l, 2=m]{\ensuremath{\hat Y_{#1}^{#2}}}
\newcommand{\Iner}{\ensuremath{{I}}}
\newcommandx{\sint}[1][1={}]{\ensuremath{\sin^{#1}\theta}}
\newcommandx{\cost}[1][1={}]{\ensuremath{\cos^{#1}\theta}}
\newcommandx{\dth}[2][1={}]{\ensuremath{\frac{\partial^{#1}{#2}}{\partial\theta^{#1}}}}

\newcommandx{\dphi}[2][1={}]{\ensuremath{\frac{\partial^{#1}{#2}}{\partial\phi^{#1}}}}
\newcommand{\omegaB}{\ensuremath{\omega_\mathrm{B}}}

\newcommand{\modif}[1]{{\textcolor{blue}{ #1}}}

\title{30 to 100-kG magnetic fields in the cores of red giant stars}
\author[1]{Gang Li \CNnames{李刚}}
\author[1, *]{S\'{e}bastien Deheuvels}
\author[1]{J\'{e}r\^ome Ballot}
\author[1]{Fran\c{c}ois Ligni\`eres}

\affil[1]{IRAP, Université de Toulouse, CNRS, CNES, UPS, Toulouse, France}
\affil[*]{sebastien.deheuvels@irap.omp.eu}
\begin{abstract}
\large

A red giant star is an evolved low- or intermediate-mass star that has exhausted its central hydrogen content, leaving a helium core and a hydrogen-burning shell. Oscillations of stars can be observed as periodic dimmings and brightenings in the optical light curves. In red giant stars, non-radial acoustic waves couple to gravity waves and give rise to mixed modes, which behave as pressure (p) modes in the envelope and gravity (g) modes in the core. These modes were previously used to measure the internal rotation of red giants\cite{deheuvels12,Gehan2018}, leading to the conclusion that purely hydrodynamical processes of angular momentum transport from the core are too inefficient\cite{Marques2013}. Magnetic fields could produce the additional required transport\cite{Gough1998Natur,Fuller2019,Gouhier2022}. However, due to the lack of direct measurements of magnetic fields in stellar interiors, very little is currently known about their properties. Asteroseismology can provide direct detection of magnetic fields because, like rotation, the fields induce shifts in the oscillation mode frequencies\cite{Unno1989, Gough1990,Hasan2005,Gomes2020,Bugnet2021,Loi2021}. Here we report the measurement of magnetic fields in the cores of three red giant stars observed with the \kepler\cite{Borucki2010Sci} satellite. The fields induce shifts that break the symmetry of dipole mode multiplets. We thus measure field strengths ranging from $\sim$\,30 to $\sim$\,100~kG in the vicinity of the hydrogen-burning shell and place constraints on the field tolopolgy.

\end{abstract}
\begin{document}

%\linenumbers%\modulolinenumbers[5]
\large
\flushbottom
\maketitle
\thispagestyle{empty}

Rotation lifts the degeneracy between the angular frequencies $\omega$ of oscillation modes with same degree $l$ and radial order $n$ but different azimuthal order $m$. This produces multiplets with $(2l+1)$ components, which can be used to probe the internal rotation of stars. At first order, rotational multiplets are symmetric with respect to the central ($m=0$) component, as is the case for all red giants studied so far\cite{Gehan2018}. Magnetic fields are known to break this symmetry \cite{Hasan2005, Gomes2020, Bugnet2021, Loi2021, Mathis2021}.

\begin{figure}
    \centering
    \includegraphics[width = 0.8\linewidth]{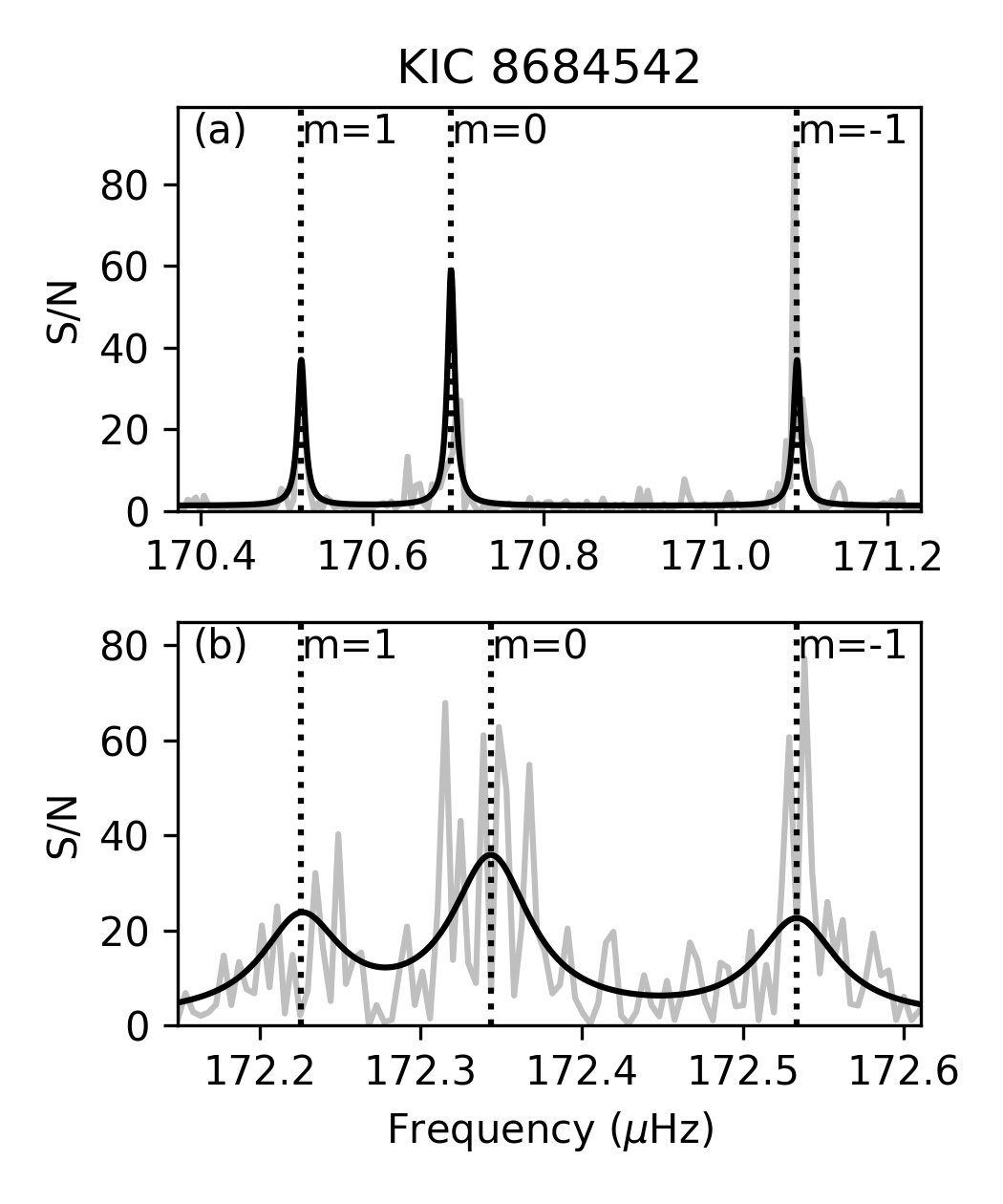} %0.6
    \caption{\textbf{Asymmetric splittings of two mixed modes in KIC\,8684542.} Panel (a) shows a g-dominated splittings, whereas panel (b) is p-dominated. %Two consecutive asymmetric splittings with the same p-mode radial order in KIC\,8684542. 
    In each panel, the grey lines show the observed power spectrum and the black lines show the best-fitting model spectrum. %results.
    The vertical dotted lines mark the locations of the frequencies. Note that the x-axis is re-scaled to align the $m=\pm 1$ modes. }\label{fig:KIC8684542_asymmetry_change}
\end{figure}

\begin{figure}
    \centering
    \includegraphics[width=0.8\linewidth]{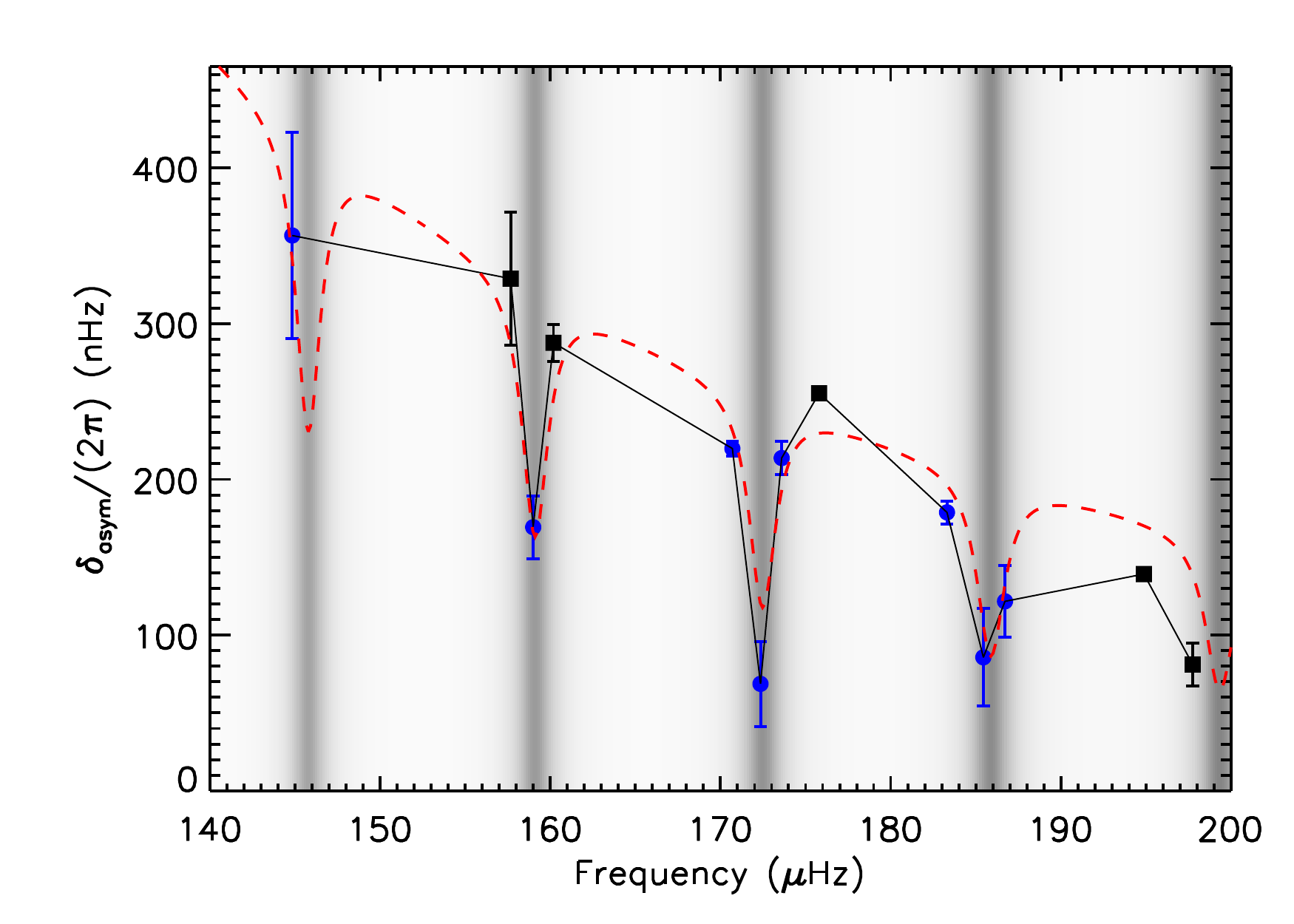} %0.8
    \caption{\textbf{Multiplet asymmetries in KIC\,8684542 as a function of mode frequency.} Blue circles indicate multiplets with three detected components, and black squares, multiplets with only two. Vertical errorbars indicate 1-$\sigma$ standard deviations. The background shows the $\zeta$ value. Light areas correspond to g-dominated modes and dark ridges show the locations of p-dominated modes. The red dashed curve corresponds to a fit of 
    %Eq. \ref{eq_fit_asym} 
    the theoretical expression of magnetic asymmetries (Eq. \ref{eq_dnub} and \ref{eq_asym}) to the observed asymmetries. See also Extended Data Fig.~\ref{fig_asym_freq_115} and \ref{fig_asym_freq_751} for KIC\,11515377 and KIC\,7518143, respectively.} % y-axis title deta_asym / 2pi ?
    \label{fig_asym_freq_868}
\end{figure}

We detected clear asymmetries in the multiplets of three hydrogen-shell burning giants observed with \kepler, namely KIC\,8684542, KIC\,7518143, and KIC\,11515377 
%These stars have masses ranging from 1.43 to 1.60$\,M_\odot$ and radii between 5.16 and 5.65$\,R_\odot$.
%In the oscillation spectra of these stars, we detected between 11 and 13 dipole multiplets showing very significant asymmetries 
(see Fig.~\ref{fig:KIC8684542_asymmetry_change}). 
%We define the asymmetry as the difference between the right side and the left side of the splitting, hence $\delta_\mathrm{asym}=\omega_{m=-1}+\omega_{m=+1}-2\omega_\mathrm{m=0}$\cite{Deheuvels2017}. 
We notice several common characteristics for the three stars. First, the asymmetries of dipole multiplets, defined as $\delta_{\rm asym} = \omega_{m=-1}+\omega_{m=+1}-2\omega_{m=0}$\cite{Deheuvels2017}, consistently have the same sign for each star: they are all positive (or consistent with zero) for KIC\,8684542 (Fig.~\ref{fig_asym_freq_868}) and KIC\,7518143, but negative for KIC\,11515377. %(see Fig.~\ref{fig_asym_freq_868} for KIC\,8684542 as an example). 
Secondly, the absolute values of the asymmetries are systematically lower for p-dominated modes (dark shaded regions in Fig.~\ref{fig_asym_freq_868}) than for g-dominated modes (light shaded regions). This indicates that the cause of the asymmetries is located in the core. Finally, the detected asymmetries sharply decrease with frequency. 
%Thus, the characteristics of the detected asymmetries match those that are expected from magnetic fields.

%Before showing that the detected asymmetries arise from internal magnetic fields, we examine other mechanisms that can also produce multiplet asymmetries. Such features can arise for fast rotators, owing to higher-order terms in the rotational perturbation\cite{Dziembowski1992}. Using the $m=\pm1$ components of dipole multiplets, we measured the core rotation rates of the three stars and found values consistent with typical red giants\cite{SM}. High-order rotational effects are thus expected to be negligible for these stars. Secondly, asymmetries can be produced by near-degeneracy effects, when the frequency separation between consecutive mixed modes is comparable to the rotational splitting\cite{Deheuvels2017}. However, in this case, only p-dominated modes are expected to show significant asymmetries, which is the opposite of what is observed here. Besides, near-degeneracy effects should produce series of alternate positive-negative asymmetries, whereas in our case, all asymmetries have the same sign in each star. We can thus safely rule out these two mechanisms as the source of the observed asymmetries.

In the presence of magnetic fields, we showed that, under very general assumptions, the average frequency shift of the components in a dipole multiplet is given by%\cite{SM}
%\begin{linenomath*}
\begin{equation}
\delta\!\omega_{\rm B} = \frac{\zeta\mathcal{I}}{\mu_0\omega^{3}} \int_{\ri}^{\ro} K(r) \overline{B_r^2} \,\hbox{d}r,
\label{eq_dnub}
\end{equation}
%\end{linenomath*}
where $\overline{B_r^2}$ is a horizontal average of the squared radial field $B_r$, $\ri$ and $\ro$ are the turning points of the g-mode cavity, and $\mathcal{I}$ is a term depending on the core structure% (Eq. \ref{eq_integral})
\cite{SM}. The weight function $K(r)$ %(Eq. \ref{eq_kernel})
sharply peaks near the hydrogen burning shell, so that $\delta\!\omega_{\rm B}$ essentially measures $\overline{B_r^2}$ near this shell (see Extended Data Fig.~\ref{fig_kernel_751}). The dependence of $\delta\!\omega_{\rm B}$ in $\omega^{-3}$ shows that magnetic perturbations are expected to sharply decrease with frequency. The factor $\zeta$ corresponds to the fraction of the mode kinetic energy that is trapped in the g-mode cavity ($\zeta=1$ for pure gravity modes). Eq. \ref{eq_dnub} shows that magnetic shifts are expected to be larger for g-dominated modes than for p-dominated modes. 
The characteristics of magnetic shifts are thus very similar to those of the detected asymmetries.

Asymmetries originate from the dependence of magnetic shifts on $|m|$.
%Magnetic shifts depend on $|m|$, and thus produce asymmetries in splittings.
%Another important property of magnetic shifts is that they depend on $|m|$, and thus produce asymmetric splittings.
%Another important property of magnetic shifts is that they depend on $|m|$, and thus produce asymmetries in rotational multiplets.
The asymmetry of dipole multiplets %, defined as $\delta_{\rm asym} = \omega_{m=-1}+\omega_{m=+1}-2\omega_{m=0}$\cite{Deheuvels2017}, 
can be directly related to the average magnetic shift by the expression
%\begin{linenomath*}
\begin{equation}
    \delta_{\rm asym} = 3a\delta\!\omega_{\rm B}.
    \label{eq_asym}
\end{equation}
%\end{linenomath*}
The coefficient $a$ involves an average of $B_r^2$ weighted by the second degree Legendre polynomial $P_2(\cos\theta) =  (3\cos^2\theta-1)/2$, where $\theta$ is the colatitude, and we have shown that $-1/2\leqslant a \leqslant 1$\cite{SM}.
%We then have $\delta_{\rm asym} = 3a\delta\nu_{\rm B}$, where we have shown that $-1/2\leqslant a \leqslant 1$, so that the measurement of the asymmetry can be used to place lower limits on the intensity of the magnetic field. 
For instance, a dipole magnetic field ($B_r \sim \cos\theta$) yields a positive asymmetry ($a = 2/5$). A field that is entirely concentrated on the poles produces maximal asymmetry ($a=1$). Conversely, a field concentrated near the equator gives minimal asymmetry ($a=-1/2$). %If the average of $B_r^2$ weighted by $P_2(\cos\theta)$ vanishes, then $m=0$ modes have the same magnetic shift as $m=\pm1$ modes and the asymmetry cancels out. %Therefore, the non-detection of multiplet asymmetries does not necessarily mean that magnetic fields have negligible effect on the mode frequencies\cite{Loi2021}.

%Non-axisymmetric 
%Magnetic fields can have more complex effects on the oscillation modes, for instance by increasing the number of possible components in a multiplet up to $(2l+1)^2$\cite{Unno1989,Gough1990}. However, we can take only $2l+1$ components into account for some specific magnetic configurations (such as an axisymmetric field) or when magnetic shifts do not exceed rotational splittings. Equations~\ref{eq_dnub} and \ref{eq_asym} are especially verified in those cases\cite{SM}.
%However, if the effects of non-axisymmetry are small or in special conditions, only three components per dipole multiplet are visible and the expressions for $\delta\!\omega_{\rm B}$ and $\delta_{\rm asym}$ developed for axisymmetric fields still hold\cite{SM}. 

We then compared the measured asymmetries to those that would be produced by internal magnetic fields. We fit an expression of $\delta_{\rm asym}$ based on Eq. \ref{eq_dnub} and \ref{eq_asym} to the observed asymmetries.
%Following Eq. \ref{eq_dnub} and \ref{eq_asym}, we fit an expression of the type
%%\begin{linenomath*}
%\begin{equation}
%\delta_{\rm asym} = \zeta\delta_\mathrm{g}(\omega_\mathrm{max}/\omega)^3
%\label{eq_fit_asym}
%\end{equation}
%%\end{linenomath*}
%to the observed asymmetries, where 
%$\omega_\mathrm{max}\equiv 2\pi\numax$
%corresponds to the angular frequency of maximum power of the oscillations and 
%$\delta_\mathrm{g}$
%is the asymmetry of pure g modes at $\numax$. 
The results are shown in Fig. \ref{fig_asym_freq_868}. % and \ref{fig_asym_freq_other}. 
The agreement with the observed asymmetries is quite good, 
%as Eq.~\ref{eq_fit_asym}  reproduces 
the overall decrease of the asymmetries with frequency being well reproduced, as well as their modulation with the trapping of the modes in the g- and p-cavities. This confirms that the detected asymmetries are indeed produced by magnetic fields in the cores of these red giants. For completeness, we have also shown that other mechanisms that can produce multiplet asymmetries, such as higher-order rotational effects\cite{Dziembowski1992} or near-degeneracy effects\cite{Deheuvels2017}, cannot account for the observed asymmetries\cite{SM}.
%, and opens the exciting opportunity to characterise them using asteroseismology. 

Eq.~\ref{eq_dnub} %and \ref{eq_fit_asym} 
can then be used to estimate the squared radial magnetic field $\langle B_r^2\rangle$ averaged in the horizontal direction, and weighted by the function $K(r)$ in the radial direction, yielding
%\begin{linenomath*}
\begin{equation}
\langle B_r^2\rangle = \frac{\mu_0\delta_{\rm g}\omega_{\rm max}^3}{3a\mathcal{I}},
\end{equation}
%\end{linenomath*}
where $\omega_\mathrm{max}\equiv 2\pi\numax$ corresponds to the angular frequency of maximum power of the oscillations and $\delta_\mathrm{g}$ is the asymmetry of pure g modes at the frequency of maximum power of the oscillation $\numax$. The minimal magnetic field that can produce the observed asymmetries is obtained by assuming $a=1$ for stars with positive asymmetries, and $a=-1/2$ for stars with negative asymmetries. To estimate $\mathcal{I}$, we computed models that reproduce the seismic properties of each of the three stars, using the stellar evolution code \mesa\cite{paxton11, SM}. We thus obtained minimal magnetic field intensities of 65~kG for KIC\,8684542, 26~kG for KIC\,7518143, and 70~kG for KIC\,11515377. 
%These values represent lower limits for the intensity of the (horizontally averaged) radial magnetic field in the vicinity of the hydrogen burning shell. 

These fields are expected to produce magnetic shifts in the mode frequencies. We found that magnetic shifts can be detected by their impact on the pattern of pure gravity modes. In the non-magnetic case, the periods of pure g modes with high radial orders $n_{\rm g}$ are given by $P(n_{\rm g}) = \dpun(n_{\rm g}+\epsg)$, where $\dpun$ is the asymptotic period spacing. The offset term $\epsg$ is strongly constrained for hydrogen-shell burning giants. From \kepler\ observations, it was found that $\epsg = 0.28\pm0.08$\cite{mosser18}, in agreement with theoretical predictions\cite{takata16}. We have shown\cite{SM} that magnetic shifts are expected to produce an increase in the measured value of $\epsg$.

The two red giants KIC\,8684542 and KIC\,11515377 have measured values of $\epsg$ that significantly deviate from the expected range ($\epsg=0.50\pm0.02$ and $0.50\pm0.03$, respectively). This can be used to directly measure the intensity of magnetic shifts in these stars.
%, by assuming that the unperturbed g modes have a value of $\epsg$ typical of red giants. 
For KIC\,7518143, the measured $\epsg = 0.30 \pm 0.03$ is consistent with regular red giants, which can be used to derive an upper limit for the magnetic field intensity.

\begin{figure}
\begin{center}
\includegraphics[width=0.8\linewidth]{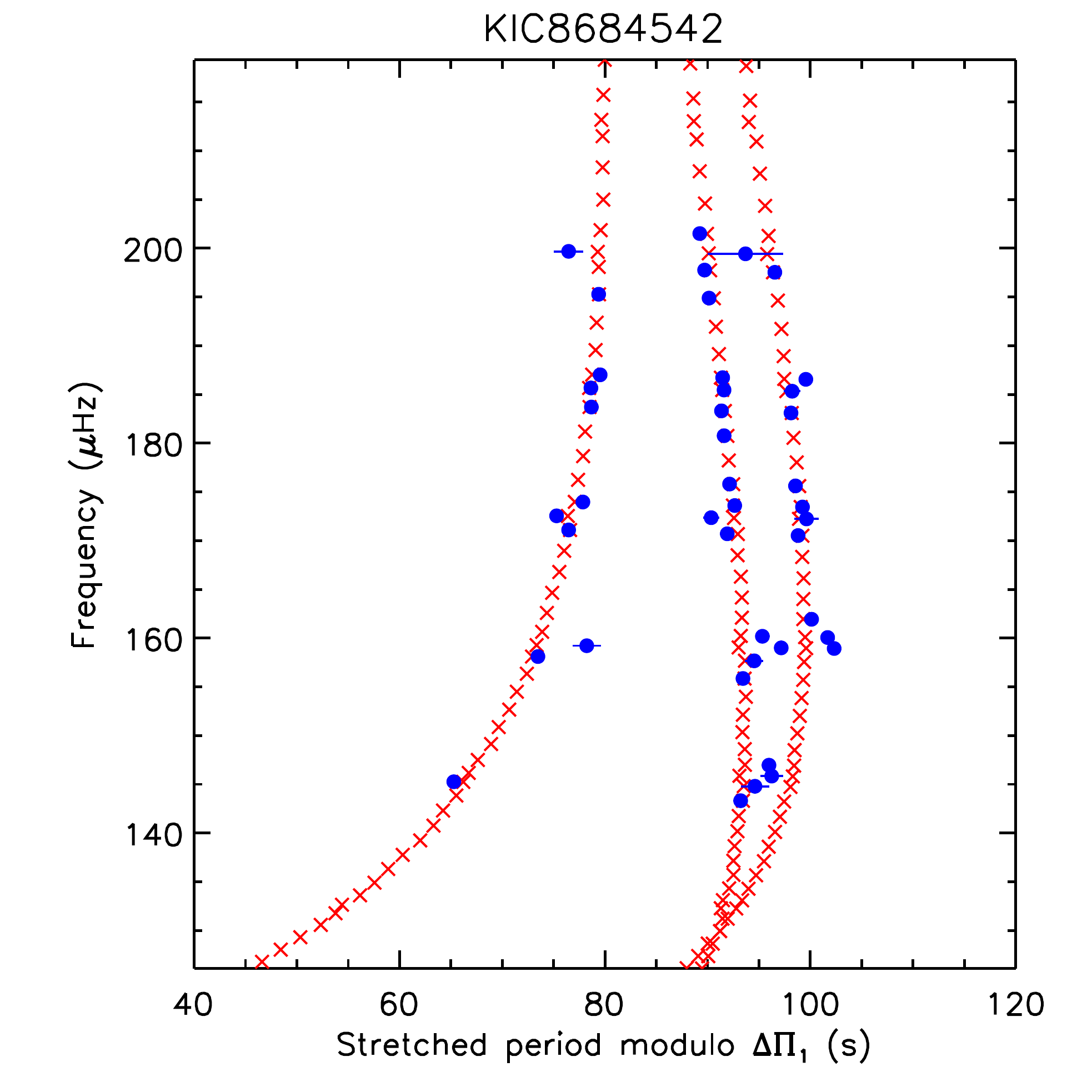}% 7cm
\end{center}
\caption{\textbf{Stretched \'echelle diagram for KIC\,8684542}. The periods of dipole mixed modes have been stretched\cite{mosser15} in order to make them regularly spaced in period by the asymptotic period spacing of g modes, $\dpun$. Thus, modes with the same azimuthal order $m$ appear nearly vertically aligned in an \'echelle diagram. Blue circles show the observed frequencies with horizontal errorbars indicating 1-$\sigma$ standard deviations. Red crosses correspond to the best-fit asymptotic mixed mode frequencies obtained by including a magnetic perturbation. See also Extended Data Fig.~\ref{fig_stretch_115} and \ref{fig_stretch_751} for KIC\,11515377 and KIC\,7518143, respectively.   
\label{fig_stretch_868}}
\end{figure}

To retrieve this information, we computed asymptotic expressions of mixed modes including magnetic and rotational perturbations. 
%Similarly to Eq. \ref{eq_fit_asym}, the average magnetic shift was modelled as $\delta\!\omega_{\rm B} = \zeta\delta\!\omega_\mathrm{g}(\omega_{\rm max}/\omega)^3$, where $\delta\!\omega_\mathrm{g}$ is the magnetic shift of pure g modes at $\numax$. 
We optimised the values of the magnetic shift of pure g modes at $\numax$ ($\delta\!\omega_\mathrm{g}$) and the asymmetry coefficient $a$ to reproduce at best the observed mode frequencies. We found an excellent agreement with the observations for the three stars (see Fig. \ref{fig_stretch_868}). We thus obtained field intensities of $102\pm12$~kG for KIC\,8684542, $98\pm24$~kG for KIC\,11515377, and an upper limit of 41~kG for KIC\,7518143. These measurements are fully compatible with the lower limits that were obtained using multiplet asymmetries, which gives further evidence that the oscillation mode perturbations are indeed caused by magnetic fields.

We also obtained measurements of the coefficient $a$ for the three stars, which can be used to place constraints on the geometry of the internal magnetic field. We found $a=0.47\pm0.12$ for KIC\,8684542 and $a=-0.24_{-0.23}^{+0.08}$ for KIC\,11515377. These measurements would be consistent with a dipolar field aligned with the rotation axis for the first star ($a = 2/5$) and a dipolar field aligned with the equator for the second ($a = -1/5$), although other configurations are naturally possible. %For KIC7518143, whose field intensity is lower, looser constraints on $a$ were obtained ($a > 0.24$).

%As an example, figure~\ref{fig:KIC8684542_asymmetry_change} shows the asymmetry change in KIC\,8684542. The two consecutive splittings with the same p-mode radial order ($n_p = 12$) are displayed in the top and the bottom panels. The best-fitted frequencies and Lorentz profiles are shown (see the method part for more detail). All the three splittings show positive asymmetry, but the splitting of the p-dominated mode (bottom panel) show smaller asymmetry than the g-dominated one, which is consistent with the feature 1 and 2 predicted by eq.~\ref{eq:asymmetry}.

%Figure~\ref{fig:asymmetry_KIC7518143} presents the asymmetry change with frequency in another star KIC\,7518143. We indeed find that the asymmetries of g-dominated modes (light areas) decrease with frequency, and are more significant than those of the p-dominated modes (dark ridges). We hereby conclude that the asymmetric splittings we discovered follow all the features assuming they are caused by the magnetic field. 

%The three red giants in which we detected core magnetic fields are quite similar, having masses ranging from 1.43 to $1.60\,M_\odot$ and radii between 5.16 and $5.65\,R_\odot$ (see Table \ref{tab_wkb}). 

This study provides keys to explore internal magnetic fields along the red giant branch. This type of measurement requires the radial magnetic field to be large enough to produce detectable magnetic splittings, but small enough to remain below a critical value $B_{\rm c}$ above which magnetic tension hampers the propagation of gravity waves\cite{Fuller2015} (the magnetic fields measured here are indeed below $B_{\rm c}$\cite{SM}). The range of field intensities that satisfy both conditions represents about one order of magnitude\cite{SM}. Field strengths above $B_{\rm c}$ have been invoked to account for red giants with suppressed dipole modes\cite{Fuller2015}, implying that about 20\% of red giants are strongly magnetised\cite{Stello2016}. A systematic search for magnetic perturbations in red giants with detected oscillations will give the prevalence of red giants with magnetic fields below $B_{\rm c}$. %Comparing this number with the ratio of stars showing dipole mode suppression will provide a test of the magnetic origin of this phenomenon.

The measurement of magnetic fields in the radiative cores of red giants will allow to progress on the origin and evolution of the stellar magnetic field.\cite{Donati2009,Braithwaite2017}.
%While all stellar convective
%layers are expected to harbour dynamo-generated magnetic fields\cite{Donati2009}, the existence and the origin of magnetic fields in radiative layers is still debated\cite{Braithwaite2017}.} 
%Here we demonstrate that magnetic fields exist in the radiative cores of red giants.
The origin of the detected fields could be a convective core dynamo occurring during the main sequence. Our best-fit models indeed indicate that the hydrogen
burning shell of the three red giants was
convective at the very beginning of the main sequence.
The dynamo fields generated at that time may have survived when the core became radiative as they can relax into
stable configurations\cite{Becerra2022} and ohmic diffusion is negligible\cite{cantiello16}.
Assuming magnetic flux conservation, their strengths
should range from 3 to 5 kG to account for the detected fields\cite{SM}.
This is smaller than the amplitudes predicted by numerical simulations of
core convection\cite{Brun2005ApJ} and dynamo scaling laws\cite{Bugnet2021} by about one order of magnitude, although this could be due to magnetic energy loss during
the relaxation process\cite{Becerra2022} or the post-main sequence evolution\cite{Gouhier2022}. Another possibility is that the detected fields are the remnants of the internal magnetic field of an Ap star. These main sequence intermediate-mass stars possess strong surface oblique dipole\cite{Auriere2007}, whose internal part could have survived the post-main sequence evolution.

Constraints on the field strength and geometry are
key to identifying how the redistribution of angular momentum operates inside stars.
Our discovery of strong magnetic fields in these red giants first suggests that uniform rotation is enforced in their core up to the hydrogen burning
shell. Therefore, the jump in rotation rate between the core and the envelope, which was revealed by
asteroseismology\cite{Deheuvels2014}, must occur at higher radii. Secondly, we show that the Tayler-Spruit
dynamo which has been proposed as a possible
solution for the angular momentum transport in red giants is not at work in the vicinity of their hydrogen burning
shell. Indeed, the predicted amplitude of the radial magnetic
field, of the order of 10$^{-2}$~G\cite{Fuller2019}, is far too low to account for the 10$^5$~G radial fields measured here.

%\printbibliography[title={References1},resetnumbers=true]
%\printbibliography[title={References1},resetnumbers=true]

\section*{Data availability}

\kepler\ data are publicly available from the Mikulski Archive for Space Telescopes (MAST) portal at \url{https://archive.stsci.edu}. Spectra are available at \url{https://doi.org/10.5281/zenodo.6818371}.

\section*{Code availability}

This study makes use of the stellar evolution code MESA, which is available at \url{https://docs.mesastar.org}.

\section*{Acknowledgements}
The authors acknowledge support from from the project BEAMING ANR-18-CE31-0001 of
the French National Research Agency (ANR) and from the Centre
National d’Etudes Spatiales (CNES).

\section*{Author contributions statement}

G.L. discovered the three stars with asymmetric splittings. G.L. and S.D. measured the asymmetries and rotation rates. S.D. measured the absolute magnetic shifts and supervised the whole project. J.B. and F.L. developed the theoretical framework used to interpret the observations. All the authors contributed to writing the manuscript. 

\section*{Competing Interest Declaration}

The authors declare no competing interests.

\section*{Supplementary Information}
Supplementary Information is available for this paper.

\section*{Author Information}
Reprints and permissions information is available at www.nature.com/reprints. Correspondence and requests for materials should be addressed to SD (sebastien.deheuvels@irap.omp.eu). 

\clearpage
\section*{Extended Data Items}

\setcounter{figure}{0}
\renewcommand{\figurename}{Extended Data Figure}
\renewcommand{\tablename}{Extended Data Table}

\begin{figure}[!ht]
\begin{center}
\includegraphics[width=0.8\linewidth]{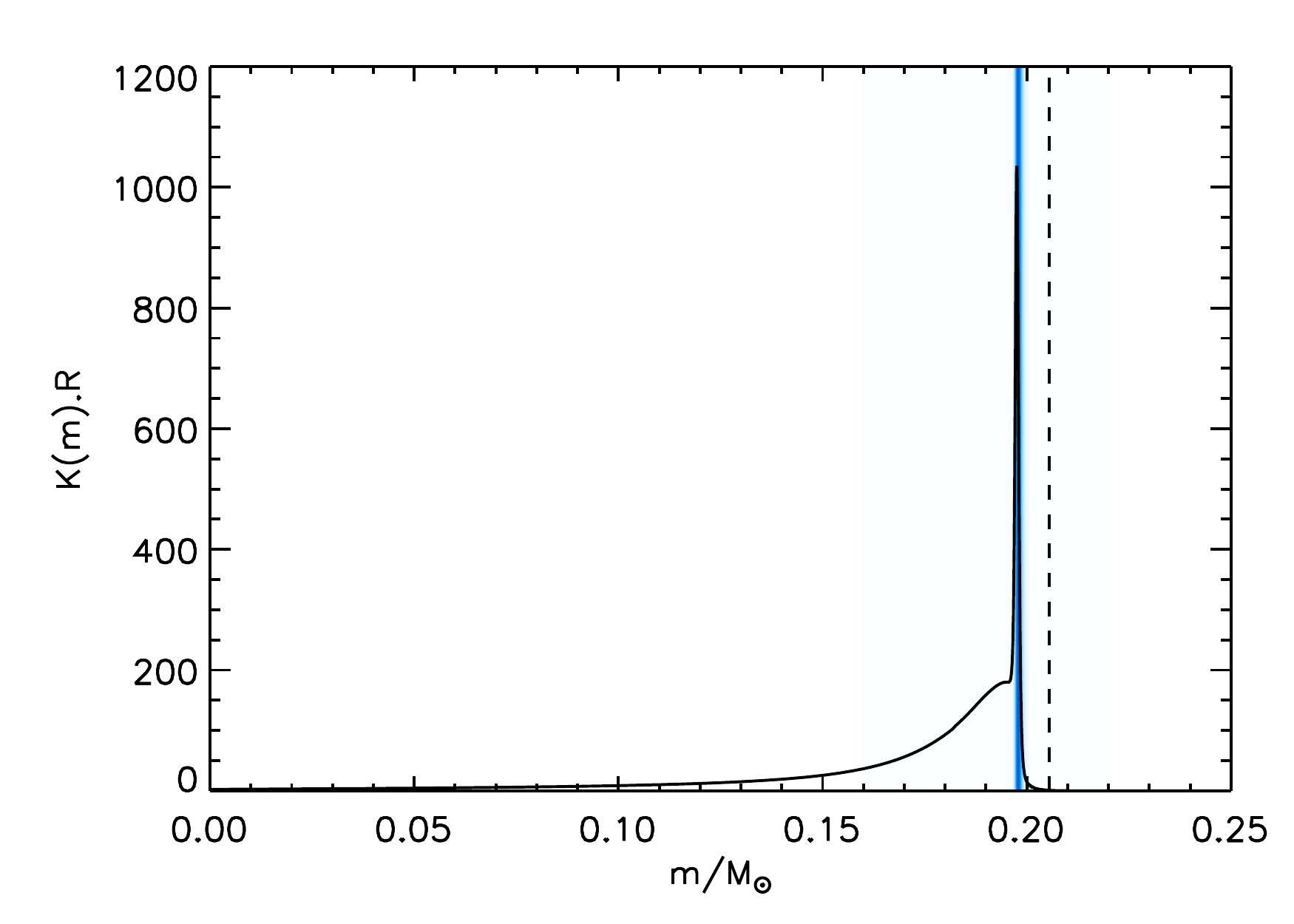}%9cm
\end{center}
\caption{\textbf{Shape of the weight function $K(m)$ as a function of the normalised mass}. The function $K(m)$ is shown for the stellar model representative of KIC\,11515377. The blue shaded region indicates the hydrogen burning shell. The vertical dashed line corresponds to the maximal extent of the initial convective core at the beginning of the main sequence.
\label{fig_kernel_751}}
\end{figure}

\clearpage
\begin{figure}[!htp]
\begin{center}
\includegraphics[width=0.8\linewidth]{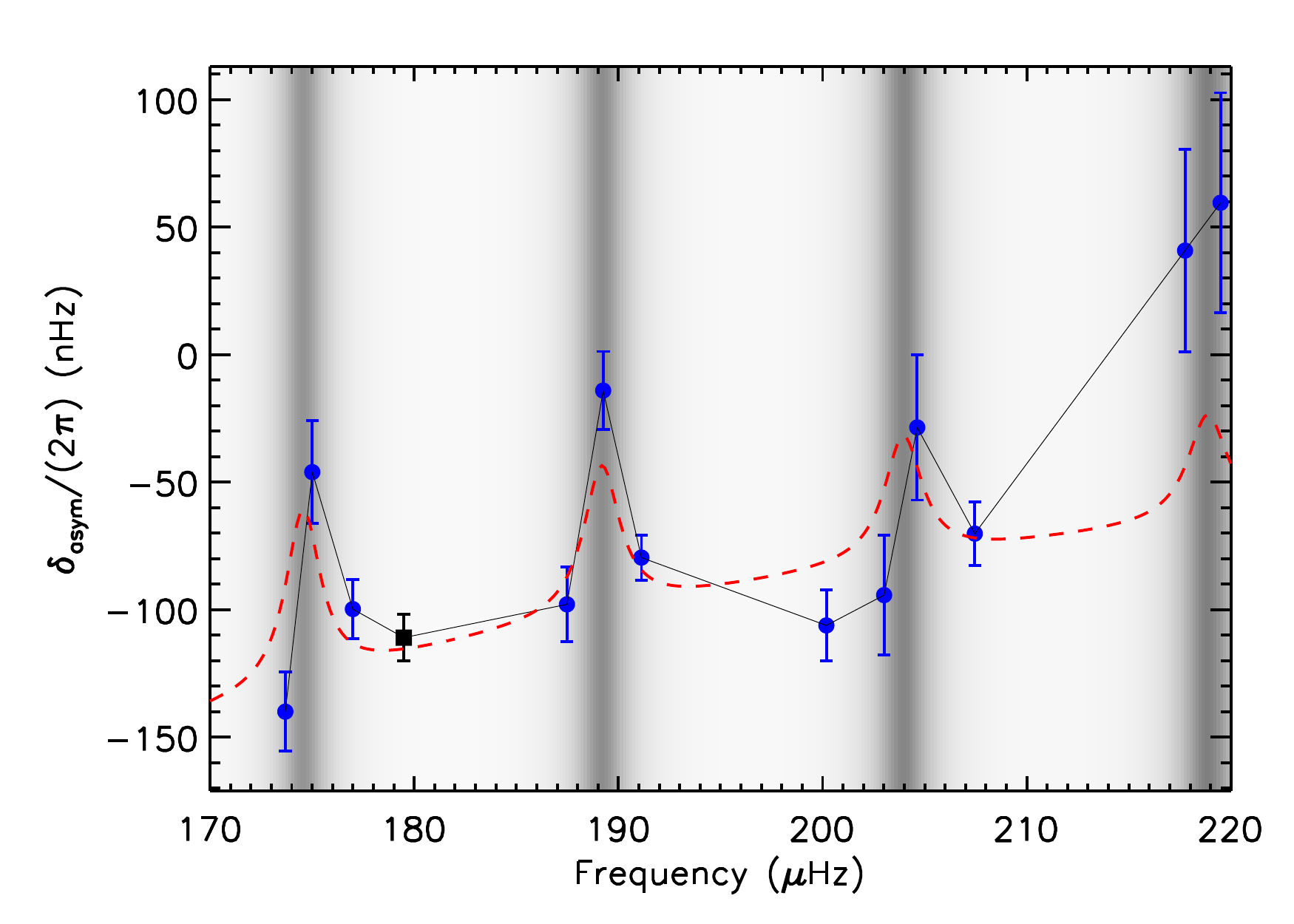}
\end{center}
\caption{\textbf{Multiplet asymmetries in KIC\,11515377 as a function of mode frequency.} Symbols have the same meaning as in Figure \ref{fig_asym_freq_868}.
\label{fig_asym_freq_115}}
\end{figure}

\clearpage
\begin{figure}[!htp]
\begin{center}
\includegraphics[width=0.8\linewidth]{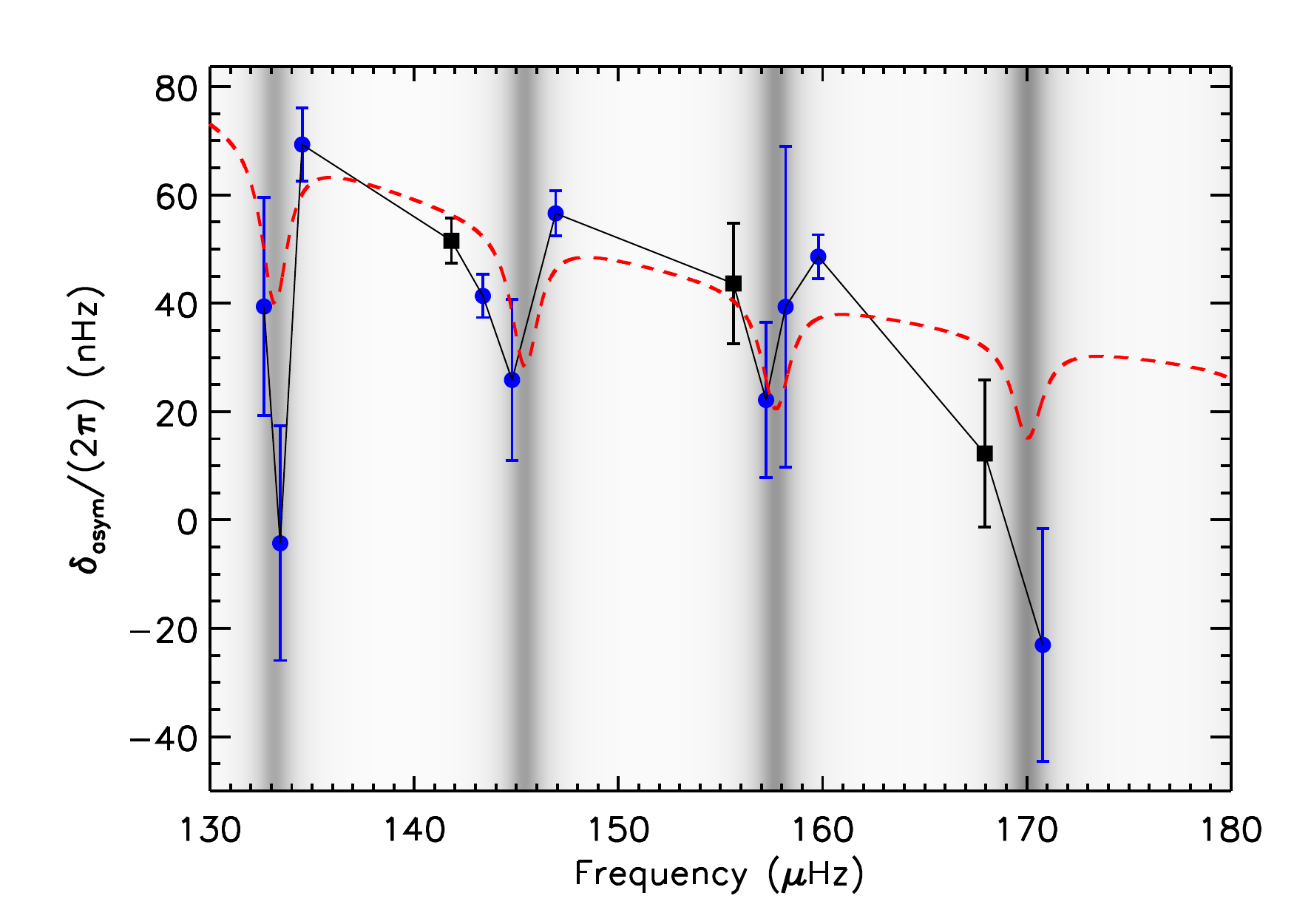}
\end{center}
\caption{\textbf{Multiplet asymmetries in KIC\,7518143 as a function of mode frequency.} Symbols have the same meaning as in Figure \ref{fig_asym_freq_868}.
\label{fig_asym_freq_751}}
\end{figure}

\clearpage
\begin{figure}[!htp]
\begin{center}
\includegraphics[width=0.8\linewidth]{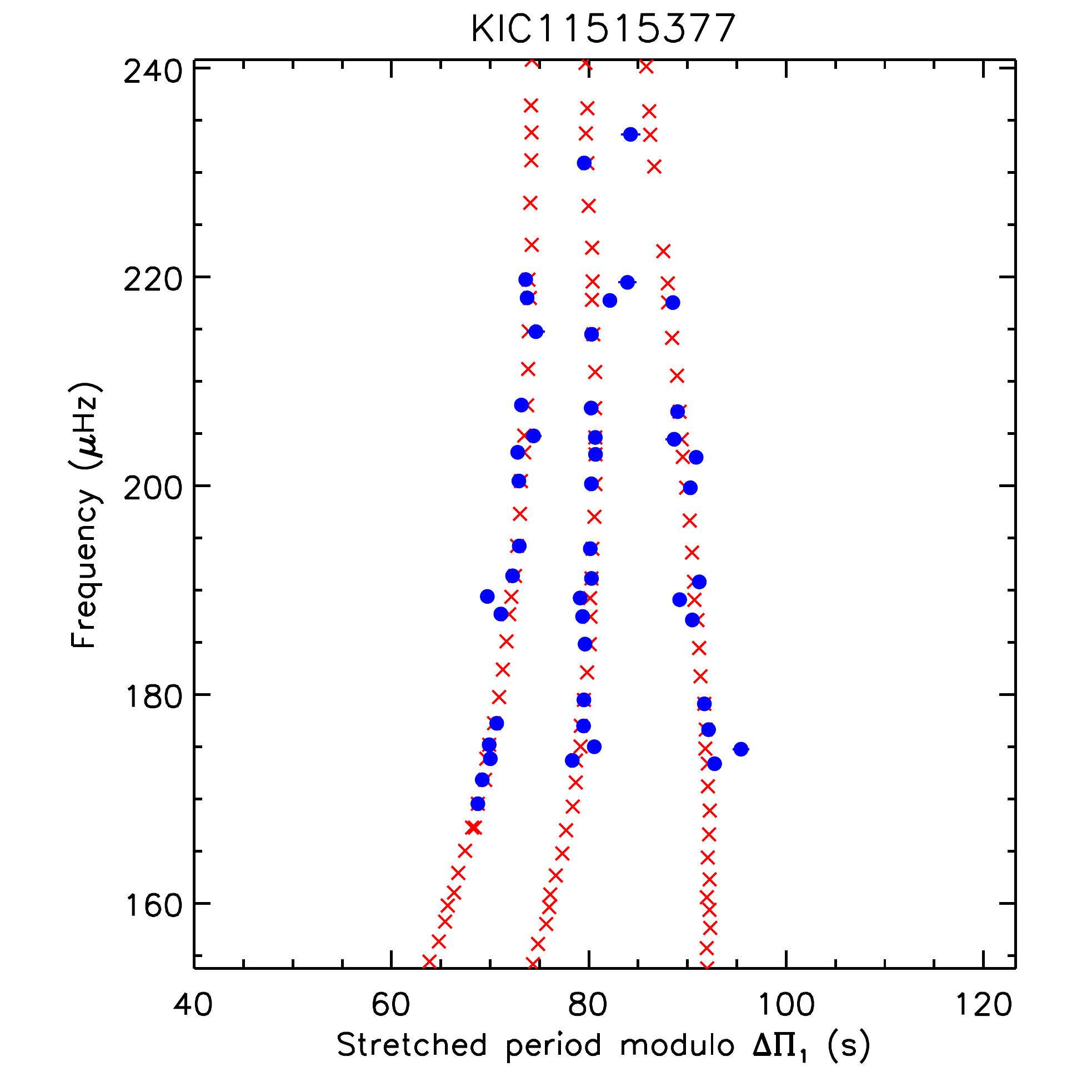}
\end{center}
\caption{\textbf{Stretched \'echelle diagram for KIC\,11515377.}
Symbols have the same meaning as in Figure \ref{fig_stretch_868}.
\label{fig_stretch_115}}
\end{figure}

\clearpage
\begin{figure}[!htp]
\begin{center}
\includegraphics[width=0.8\linewidth]{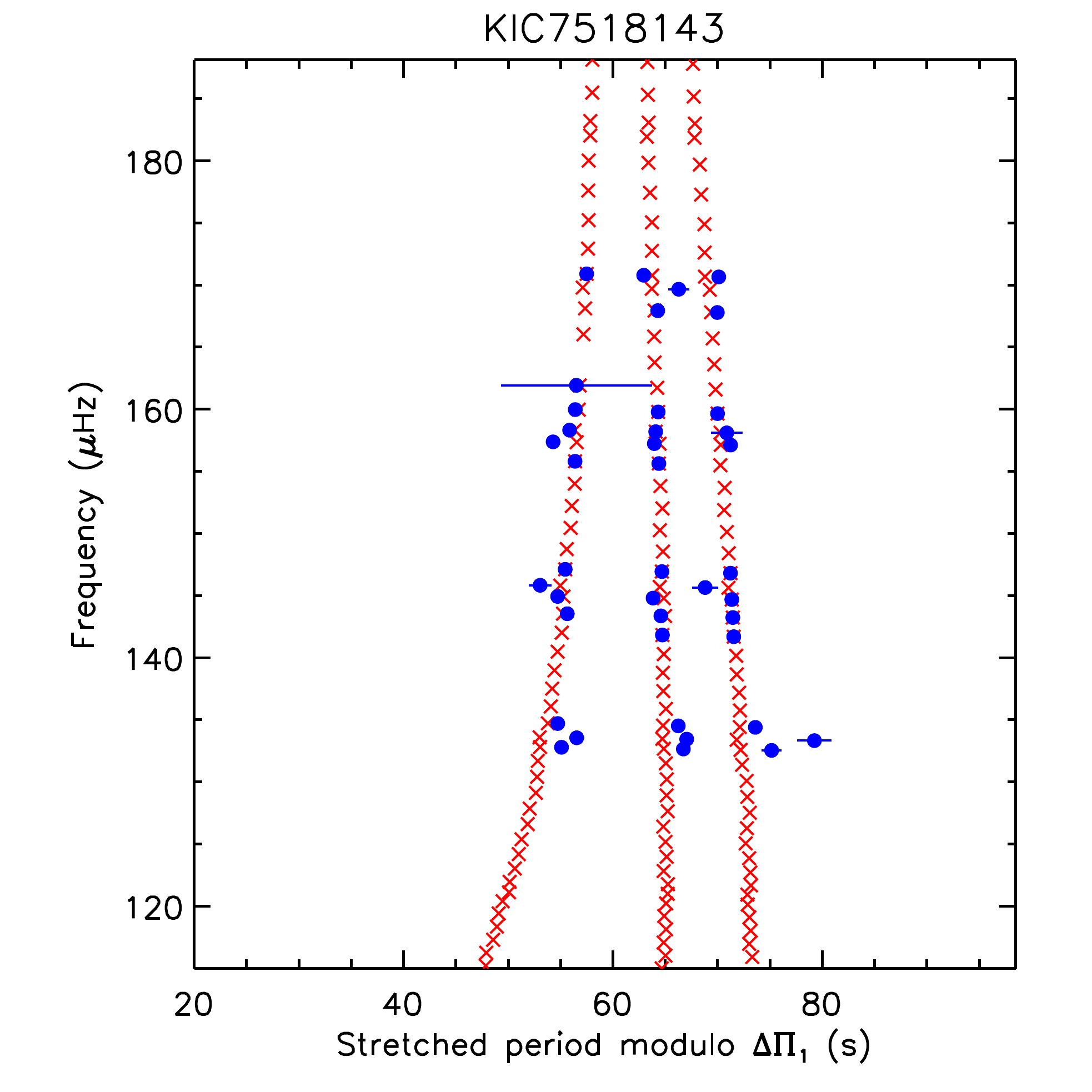}
\end{center}
\caption{\textbf{Stretched \'echelle diagram for KIC\,7518143.}
Symbols have the same meaning as in Figure \ref{fig_stretch_868}.
\label{fig_stretch_751}}
\end{figure}

\clearpage
\newpage
\setcounter{page}{1}

\section*{Supplementary Information}

\renewcommand{\thesubsection}{S\arabic{subsection}}

\subsection{Seismic characterisation and detection of mutliplet asymmetries in three \kepler\ red giants}\label{sect_analysis}

We here describe the seismic analysis of the three \kepler\ red giants KIC\,8684542, KIC\,11515377, and KIC\,7518143, which led us to detect multiplet asymmetries in these stars.
Hereafter, ordinary frequencies are denoted as $\nu$. They are related to angular frequencies through $\omega\equiv 2\pi\nu$. %suggestion...

\subsubsection{Oscillation spectra and global seismic properties \label{sect_PSD}}

The three stars were selected from the asteroseismic red giant catalogue by Yu et al.\cite{Yu2018}, which provides the measurements of seismic parameters (such as large separations $\Delta \nu$ and frequencies of maximum power $\nu_\mathrm{max}$) and the evolutionary stages (hydrogen-shell burning or core-helium burning) of \kepler\ red giants with detected oscillations. We downloaded the \kepler\ long-cadence light curves from the Mikulski Archive for Space Telescopes (MAST, \url{https://archive.stsci.edu}), corrected them\cite{Garcia11}, and calculated the power density spectra (PSD) \cite{Lomb1976,Scargle1982,Kjeldsen1995}. Global asteroseismic parameters ($\nu_\mathrm{max}$, $\Delta\nu$) and background properties %\cite{Karoff2008PhDT} %<< not really needed
were calculated following the processes used in the SYD pipeline \cite{Huber2009,Chontos2021}. 
%The signal-to-noise ratio was used in the following study, which is the ratio between the power density and the background. 

The radial modes ($l=0$, where $l$ is the angular degree) 
%and quadrupole ($l=2$) modes 
were identified using an \'echelle diagram folded with the asymptotic large separation of p modes $\Delta \nu$. The detected peaks were then fit by Lorentzian profiles to estimate the mode parameters (frequencies, linewidths, heights) and their corresponding uncertainties 
through maximum likelihood estimation\cite{Anderson1990ApJ}.
%by maximizing the likelihood function \cite{Anderson1990ApJ},
%\begin{equation}
%    \ln L = -\sum_i\left(\ln M_i + \frac{O_i}{M_i}\right),
%\end{equation}
%where $M_i$ is the model and $O_i$ is the data.
To refine our estimates of p-mode properties, we used a polynomial regression to fit the measured radial mode frequencies to a second-order asymptotic expression of p modes, written as
%\begin{linenomath*}
\begin{equation}
    \nu_{l=0}(n)  = \left[ n + \varepsilon_{\rm p} + \frac{\alpha}{2}(n-n_{\rm max})^2 \right] \dn \label{eq_nup_l0}
\end{equation}
%\end{linenomath*}
where $\alpha$ measures the second-order effects in the asymptotic development around $\nu_\mathrm{max}$, $\varepsilon_{\rm p}$ is a phase offset, and $n_{\rm max} = \nu_{\rm max}/\Delta\nu$. The results are given in Supplementary Table \ref{tab_wkb} for the three stars.

\subsubsection{Identifying dipole mixed modes}

Identifying dipole modes in the oscillation spectra is more complicated because these modes show a mixed behaviour and therefore follow neither the asymptotic pattern of p modes, nor the asymptotic pattern of g modes. Representing the oscillation modes in the so-called ``stretched'' period \'echelle diagram\cite{mosser15} is particularly helpful to identify the modes. We here only briefly describe this method. It has been shown\cite{mosser15} that the period spacing between consecutive dipole mixed modes can be expressed as $\Delta P = \zeta\dpun$, where $\zeta$ is the ratio between the kinetic energy of the mode in the g-mode cavity and the total kinetic energy of the mode ($\zeta$ tends to unity for pure g modes and to zero for pure p modes), and $\dpun$ is the asymptotic period spacing of pure g modes. Thus, the so-called ``stretched'' periods $\tau$, defined by the differential equation $\hbox{d}\tau = \hbox{d}P/\zeta$, are regularly spaced by $\dpun$. It is then convenient to show the stretched periods in an \'echelle diagram folded with $\dpun$ because the modes of same azimuthal order $m$ are expected to align nearly vertically in this diagram.

Estimates of $\zeta$ for the detected modes can be obtained from asymptotic relations\cite{goupil13}. Using estimates of $\dpun$ from previous studies\cite{Gehan2018}, we were then able to build stretched \'echelle diagrams for the three red giants under study (Figure \ref{fig_stretch_868} and Extended Data Figure \ref{fig_stretch_115}-\ref{fig_stretch_751}). For each star, three clear ridges are visible, associated with the $m=-1,0,+1$ components of dipole multiplets. It is already visible that the multiplets show significant asymmetries (especially for KIC\,8684542), as is further addressed in Sect. \ref{sect_asym_obs}.

In the stretched \'echelle diagram, the period spacings of modes with the same $m$ value are given by\cite{Gehan2018}
%\begin{linenomath*}
\begin{equation}
    \Delta \tau_m = \dpun \left(1 + m x_\mathrm{rot} \right),
\end{equation}
%\end{linenomath*}
with
%\begin{linenomath*}
\begin{equation}
    x_\mathrm{rot} = 2\frac{\mathcal{N}_\mathrm{max}}{\mathcal{N}_\mathrm{max}+1}\frac{\delta \nu_\mathrm{R, core}}{\nu_\mathrm{max}},
\end{equation}
%\end{linenomath*}
where  $\mathcal{N}_\mathrm{max}=\Delta \nu/ (\Delta \Pi_1 \nu_\mathrm{max}^2 )$ is the mixed mode density at $\nu_\mathrm{max}$, and $\delta \nu_\mathrm{R, core}$ is the rotational splitting of pure g modes. Therefore, the identification of the azimuthal order $m$ can be done by comparing the period spacing in the stretched \'echelle diagram: the larger $m$, the larger $\Delta \tau_m$. 

\subsubsection{Detecting multiplet asymmetries \label{sect_asym_obs}}

We then extracted the frequencies of the identified dipole modes by fitting Lorentzian profiles to the PSD as described in Sect.~\ref{sect_PSD}. We note that several codes performing the extraction of oscillation mode properties from the PSD are public\cite{corsaro14} and can be used to reproduce our results. We show in Supplementary Figures \ref{fig_8684542_mode_fit}-\ref{fig_7518143_mode_fit} our optimal fits to the PSD for the three stars. The measured mode frequencies are given in Supplementary Table \ref{tab_freq}. Using the identification from the previous section, we could then compute the asymmetries of the multiplets for which all $m$ components were detected, using the expression
%\begin{equation}
%    \frac{\delta_{\rm asym}}{2\pi} = \nu_{m=-1}+\nu_{m=+1}-2\nu_{m=0}
%\end{equation}
%\begin{linenomath*}
\begin{equation}
    \delta_{\rm asym} = \omega_{m=-1}+\omega_{m=+1}-2\omega_{m=0}.
\end{equation} %suggestion since omega is used for splitting in section S1.5
%\end{linenomath*}
The values of the measured asymmetries are given in Supplementary Tables \ref{tab_868}-\ref{tab_751}. We found very significant asymmetries for all three stars. The measured asymmetries are shown as a function of the mode frequency in Figure \ref{fig_asym_freq_868} and Extended Data Figure \ref{fig_asym_freq_115}-\ref{fig_asym_freq_751}.

\subsubsection{Refining estimates of g-mode properties \label{sect_gmodes}}

We then used an asymptotic expression of mixed modes\cite{shibahashi79} to refine our estimates of g-mode properties. For mixed modes, the matching of solutions corresponding to g-modes in the core and to p-modes in the envelope requires that
%\begin{linenomath*}
\begin{equation}
\tan(\theta_{\rm p}) = q \tan(\theta_{\rm g}),
\label{eq_wkb_mixed}
\end{equation}
%\end{linenomath*}
where $q$ corresponds to the coupling strength between the two cavities, and $\theta_{\rm p}$, $\theta_{\rm g}$ are phase terms that can be expressed as a function of the asymptotic expressions of p- and g-modes. We have\cite{mosser12a}
%\begin{linenomath*}
\begin{align}
\theta_{\rm p} & = \frac{\pi}{\Delta\nu} \left( \nu -  \nu_{\rm p} \right) \label{eq_thetap}, \\
\theta_{\rm g} & = \frac{\pi}{\Delta\Pi_1} \left( P - P_{\rm g} \right), \label{eq_thetag}
\end{align}
%\end{linenomath*}
where $\nu_{\rm p}$ correspond to the frequencies of $l=1$ pure p modes and $P_{\rm g}$ are the periods of $l=1$ pure g modes. Their asymptotic expressions are given by
%\begin{linenomath*}
\begin{align}
    \nu_{\rm p}(n_{\rm p})&  = \nu_{l=0}(n_{\rm p}) + \dn/2 - d_{01} \label{eq_nup},\\
    P_{\rm g}(n_{\rm g}) & = (n_{\rm g} + \varepsilon_{\rm g})\Delta\Pi_1, \label{eq_Pg}
\end{align}
%\end{linenomath*}
%where $\alpha$ measures the second-order effects in the asymptotic development, $\varepsilon_{\rm p}$ is a phase offset, $n_{\rm max} = \nu_{\rm max}/\Delta\nu$, and 
where $d_{01}$ corresponds to the mean small separation built with $l=0$ and 1 pressure modes. Pure g modes are characterised by the asymptotic period spacing $\dpun$ and the phase offset $\varepsilon_{\rm g}$. 

The parameters $\dn$, $\alpha$, and $\varepsilon_{\rm p}$ were already measured in Sect. \ref{sect_PSD}. For any given set of parameters $(\dpun,~\epsg,~q,~d_{01})$, asymptotic mixed mode frequencies with $m=0$ can be obtained by solving Eq.~\ref{eq_wkb_mixed} using the Newton-Raphson algorithm. For each star, we optimised these four parameters to reproduce at best the observed frequencies of $m=0$ dipole modes. For this purpose, we computed grids, exploring wide intervals encompassing reasonable values for the four parameters and we calculated the agreement with the measured frequencies for each grid point. We then computed more refined grids to fine-tune the values of the parameters. The results are given in Supplementary Table \ref{tab_wkb}.

\subsubsection{Measurement of internal rotation \label{sect_rotation}}

We then performed a seismic measurement of the internal rotation of the three red giants. Since rotational multiplets are asymmetric in the three stars, the rotational splittings were measured as $\delta\!\omega_{\rm R} = (\omega_{n,m=-1}-\omega_{n,m=+1})/2$ (see Supplementary Tables \ref{tab_868}-\ref{tab_751}). Using asymptotic relations, we have\cite{goupil13}
%\begin{linenomath*}
\begin{equation}
\delta\!\omega_{\rm R} = \frac{\zeta}{2}\omg  +  (1-\zeta)\omp,
\label{eq_rot}
\end{equation}
%\end{linenomath*}
where $\omg$ and $\omp$ are the average rotation rates in the g-mode and p-mode cavities, respectively. 

The parameter $\zeta$ can be conveniently estimated from the mode frequencies by the relation\cite{goupil13,hekker17}
%\begin{linenomath*}
\begin{equation}
\zeta = \left[ 1 + q \frac{\nu^2\dpun}{\dn} \frac{1}{\sin^2\theta_{\rm p} + q^2 \cos^2\theta_{\rm p}} \right]^{-1}.
\label{eq_zeta}
\end{equation}
%\end{linenomath*}
Using the asymptotic properties of pressure and gravity modes measured in Sect.~\ref{sect_PSD} and \ref{sect_gmodes}, we were able to estimate the value of $\zeta$ for each detected mode.

Supplementary Figure \ref{fig_zeta_split} shows the measured rotational splittings as a function of $\zeta$ for the three stars. We recover a linear relation as predicted by Eq. \ref{eq_rot}. The coefficients of a linear regression of this relation directly provide estimates of the average rotation rates in the g-mode cavity ($\omg$) and in the p-mode cavity ($\omp$). The results are given in Supplementary Table \ref{tab_rot}. The measurements of $\omg$ for the three stars are in line with the typical core rotation rates of red giants (an average of $\omg/(2\pi) = 694$~nHz with a standard deviation of 288~nHz was found for 846 \kepler\ red giants\cite{Gehan2018}). The measurements of $\omp$ indicate a slow envelope rotation, in agreement with the expansion of the star as it ascends the red giant branch and in line with the measurements of the average envelope rotation in other red giants\cite{goupil13}.

Our measurements of the core and envelope rotation rates can yield estimates of the rotational splitting for all detected multiplets using Eq. \ref{eq_rot}. Thus, we can derive measurements of the asymmetry for dipole multiplets in which either the $m=-1$ or the $m=+1$ component could not be detected. For instance, if we could measure the frequencies of the $m=0$ and $m=+1$ components, we have
%\begin{linenomath*}
\begin{equation}
    \delta_{\rm asym} = 2(\omega_{m=1} + \delta\!\omega_{\rm R} - \omega_{m=0}).
\end{equation}
%\end{linenomath*}
This enabled us to obtain a few additional asymmetry measurements for the three stars under study. These data points are shown as black squares in Fig. \ref{fig_asym_freq_868} and Extended Data Fig.~\ref{fig_asym_freq_115}-\ref{fig_asym_freq_751}.
%\subsection{Stretched \'echelle diagrams}

\subsection{Effects of magnetic field: the perturbation method}\label{ssec:perturb}

The effects of magnetic fields on stellar oscillations, treated as a perturbation of the adiabatic oscillations have been studied over the last decades\cite{Unno1989,Gough1990}. In recent years, these approaches have been applied to high-order g modes in stars\cite{Hasan2005}, then to mixed modes in red giants considering the simple case of dipolar magnetic fields with different radial profiles, aligned with the rotation axis\cite{Gomes2020,Bugnet2021}
or possibly inclined\cite{Loi2021}. These various works aimed at determining the impact of given magnetic fields on mode frequencies, especially to predict minimal magnetic strengths required to produce detectable asymmetries in red giants for specific field structures\cite{Gomes2020,Bugnet2021,Loi2021}. The problem we want to solve here is to deduce properties of stellar magnetic fields from seismic observable quantities. We therefore want to avoid any a priori assumptions on the magnetic field topology and are thus led to develop the perturbation theory for a general magnetic field. This is the purpose of the present section.
%In the present section we focus on dipole mixed modes, relaxing the assumptions on the magnetic field topology.

%\subsubsection{Variational principle}\label{sssec:variational}
\subsubsection{First-order perturbations}\label{sssec:variational}
Without magnetic field (or rotation), the equations of oscillations in a spherical star reduce to the eigenvalue problem\cite{Unno1989}
%\begin{linenomath*}
\begin{equation}
    \Lop_0(\vxi) = \omega^2 \vxi,
\end{equation}
%\end{linenomath*}
where the operator $\Lop_0(\vxi)\equiv\rho^{-1}\vb{\nabla} p' + \vb{\nabla}\psi' +\rho^{-1}\vb{\nabla}\psi\rho'$ is a linear functional of the displacement $\vxi$. The quantities $\rho$ and $\psi$ are the equilibrium density and gravitational potential, $p'$, $\rho'$ and $\psi'$ the perturbations to pressure, density and gravitational potential. 

In the presence of a magnetic field, the Lorentz force must be taken into account. The problem becomes\cite{Gough1990}
%\begin{linenomath*}
\begin{equation}
    [\Lop_0 + \LopL ](\vxi) = \omega^2 \vxi,
\end{equation}
%\end{linenomath*}
with
%\begin{linenomath*}
\begin{equation}
    \LopL(\vxi)
    =  -\frac{1}{\rho\mu_0} [(\vb{\nabla} \times \vb{B}') \times \vb{B} + (\vb{\nabla} \times \vb{B}) \times \vb{B}' ] - \frac{\vb{\nabla}\cdot(\rho\vxi)}{\rho^2\mu_0} [(\vb{\nabla} \times \vb{B}) \times \vb{B} ], \label{eq:LopL_all}
\end{equation}
%\end{linenomath*}
where $\vb B$ is the magnetic field, $\vb{B}' = \vb{\nabla} \times (\vxi \times \vb{B})$ is the Eulerian perturbation to $\vb{B}$, and $\mu_0$ is the magnetic permeability.

%When one considers g modes, the last term, related to compressibility, may be neglected. Moreover for large scale fields and short wavelengths, %(autant préciser nos approximations dès ce moment en ecrivant les relations $\ksi_r$ vs $\ksi_h$ en fonction de N/omega, une inegalite à respecter sur les echelles de variation du champ et les anisotropies de champs) 
%the first term only is dominant\cite{Hasan2005,Gomes2020,Mathis2021,Bugnet2021,Loi2021}, thus
%\begin{equation}
%    \LopL(\vxi)
%    \approx  -\frac{1}{\rho\mu_0} [(\nabla \times \vb{B'}) \times \vb{B}  ]. \label{eq:LopLsimple}
%\end{equation}

 We assume that the perturbation introduced by the Lorentz force is small ($\LopL(\vxi) \equiv \Lop_1(\vxi)$, $\|\Lop_1(\vxi)\| \ll \|\Lop_0(\vxi)\|$), and we thus perform a first-order perturbation analysis\cite{Unno1989,Gough1990}.% The application of the variational principle is very well documented in the literature\cite{LyndenBell1967,Unno1989,Gough1990}. When the operator $\Lop_1(\vxi)$ is not axisymmetric, the equations for different $m$ are coupled and must be solved simultaneously\cite{Loi2021}.

Let us denote as $\vxi_{m}^{(0)}= \xi_r(r)\Ylm \er + \xi_h(r)r\vb{\nabla}\Ylm$ the eigenfunction of $\Lop_0$ associated with the eigenfrequency $\omega^{(0)}$ for a given $m$. Spherical harmonics are denoted as $\Ylm$ and are normalised such that $\iint \Ylm \sint \,\mathrm{d}\theta \,\mathrm{d}\phi = 1$, and (\er, \etheta, \ephi) is the usual spherical base. The functions $\xi_r$ and $\xi_h$ are independent of $m$.

Due to the degeneracy with $m$, any linear combination
%\begin{linenomath*}
\begin{equation}
    \vxi^{(0)}=\sum_{m'=-l}^{l} a_{m'} \vxi_{m'}^{(0)} \label{eq:xiprime}
\end{equation} 
%\end{linenomath*}
is an eigenfunction of $\Lop_0$ associated with $\omega^{(0)}$.

The eigenvalues and eigenvectors of the perturbed operator $\Lop_0+\Lop_1$ are written as perturbations of the unperturbed solutions: $\omega=\omega^{(0)}+\omega^{(1)}$ and $\vxi=\vxi^{(0)}+\vxi^{(1)}$, with $\omega^{(1)} \ll \omega^{(0)}$ and $\|\vxi^{(1)}\| \ll \|\vxi^{(0)}\|$.

The problem reads 
%\begin{linenomath*}
\begin{equation}
  \left(\omega^{(0)}+\omega^{(1)}\right)^2\left(\vxi^{(0)}+\vxi^{(1)}\right) =
[\Lop_0 + \Lop_1]\left(\vxi^{(0)}+\vxi^{(1)}\right).  \label{eq:Ltot}
\end{equation}
%\end{linenomath*}
We expand this equation and project it on $\vxi_{m}^{(0)}$. By keeping first-order terms only, it reads
%\begin{linenomath*}
\begin{equation}
2\omega^{(0)}\omega^{(1)}\left\langle\vxi_{m}^{(0)},\vxi^{(0)}\right\rangle = \left\langle\vxi_{m}^{(0)},\Lop_1\left(\vxi^{(0)}\right)\right\rangle \quad \forall m
\end{equation}
%\end{linenomath*}
where we have defined the usual inner product
%\begin{linenomath*}
\begin{equation}
    \left\langle \vb{f}, \vb g \right\rangle = \iiint  \vb f^* \cdot \vb g \ \rho r^2\sint \,\mathrm{d}r \,\mathrm{d}\theta \,\mathrm{d}\phi,
\end{equation}
%\end{linenomath*}
and taken advantage of the Hermitian character of the operator $\Lop_0$ \cite{LyndenBell1967}.
By using Eq.~\ref{eq:xiprime} and after some algebra, we obtain 
%\begin{linenomath*}
\begin{equation}
    2\omega^{(0)}\omega^{(1)} a_{m}\left\langle\vxi_{m}^{(0)},\vxi_{m}^{(0)}\right\rangle  =\sum_{m'=-l}^l  a_{m'}\left\langle\vxi_{m}^{(0)},\Lop_1\left(\vxi^{(0)}_{m'}\right)\right\rangle\quad \forall m.
\end{equation}
%\end{linenomath*}
Since the operator $\Lop_1$ is not axisymmetric, the equations for different $m$ are coupled and must be solved simultaneously. We thus face an algebraic eigenvalue problem\cite{Loi2021}:
%\begin{linenomath*}
\begin{equation}
    \omega^{(1)}\mathsf{a}= \mathsf{Ma},
\end{equation}
%\end{linenomath*}
%$\omega_{1,m}$ are the eigenvalues associated to the eigenvectors $\mathsf{a}_m = (\mathsf{a}_{m,k})^\top$ of 
where the components of eigenvectors $\mathsf{a}$ are the coefficients $a_{m'}$, and the elements of the matrix $\mathsf{M}$ are
%\begin{linenomath*}
\begin{equation}
    \mathsf{M}_{m,m'}=
    \frac{\left\langle\vxi_{m}^{(0)},\LopL\left(\vxi^{(0)}_{m'}\right)\right\rangle}{2\omega^{(0)}\left\langle\vxi_{m}^{(0)},\vxi_{m}^{(0)}\right\rangle}. \label{eq:pert_gen}
\end{equation}
%\end{linenomath*}
We recognise in the denominator the mode inertia
%\begin{linenomath*}
\begin{equation}
    \left\langle\vxi_{m}^{(0)},\vxi_{m}^{(0)}\right\rangle= \int (|\xi_r|^2+l(l+1) |\xi_h|^2) \rho r^2 \,\mathrm{d}r \equiv \Iner, \label{eq:Iner_gen}
\end{equation}
%\end{linenomath*}
which does not depend on $m$.

\subsubsection{Application to high-order g modes}
To go further, we make some simplifying assumptions\modif{\cite{Hasan2005}}. We consider from now on high-order pure g modes, that is, short radial wavelengths such that $k_r/k_h\sim N/\omega \gg 1$, where $k_r$ and $k_h$ are the radial and horizontal components of the wave vector and $N$ denotes the Brunt-Väisälä frequency.
As $\xi_h/\xi_r \sim k_r/k_h$, this implies that $\xi_r \ll \xi_h$ and that the radial derivatives of $\xi_h$ and $\xi_r$ are dominant terms (e.g. $r\partial_r\xi_h \sim rk_r\xi_h \gg \xi_h$).
Moreover, the magnetic field is supposed to vary over large scales $L$, such that $Lk_r \gg 1$, and the anisotropy between its components is also limited, that is, $B_\phi/B_r \ll k_r/k_h \sim N/\omega$. In the three studied red giant stars the ratio $N/\omega$ is worth $\sim 10^2$.

Under these assumptions, we simplify $\LopL$ (Eq.~\ref{eq:LopL_all}). 
Since we consider g modes, the last term related to compressibility may be neglected, and among the two others only the first term is dominant\cite{Hasan2005,Gomes2020,Bugnet2021,Loi2021}, thus
%\begin{linenomath*}
\begin{equation}
    \LopL(\vxi)
    \approx  -\frac{1}{\rho\mu_0} [(\vb{\nabla} \times \vb{B}') \times \vb{B}  ]. \label{eq:LopLsimple}
\end{equation}
%\end{linenomath*}
%Using Eq. \ref{eq:LopLsimple},
Using this simplified expression, we show after some algebra that
%\begin{linenomath*}
\begin{equation}
    \left\langle\vxi_{m}^{(0)},\LopL\left(\vxi^{(0)}_{m'}\right)\right\rangle = \frac{1}{\mu_0} \iiint {{\vb{B}'}_{m}^{(0)}}^{\mbox{*}}\cdot{\vb{B}'}_{m'}^{(0)} \,r^2 \sint \,\mathrm{d}r \,\mathrm{d}\theta \,\mathrm{d}\phi, \label{eq:opLproj}
\end{equation}
%\end{linenomath*}
where ${\vb{B}'}_{m}^{(0)} = \vb{\nabla} \times (\vxi_{m}^{(0)} \times \vb{B})$. We recover the expression obtained for axisymmetric fields\cite{Hasan2005,Gomes2020,Bugnet2021} when $m'=m$. To obtain this expression, we neglected a surface term by assuming that it is negligible with respect to the volume integral term. 
This is relevant in particular when the field decays towards the outer part of the g-mode cavity\cite{Loi2021}.
The Eulerian perturbation to the field simplifies to
%\begin{equation}
%    {\vb B}' = \dfrac{1}{r}\dfrac{\partial(r\xi_{\theta} B_r)}{\partial r} \etheta +  \dfrac{1}{r}\dfrac{\partial(r\xi_{\phi} B_r)}{\partial r} \ephi, \label{eq:Bprime}
%\end{equation}
%\begin{linenomath*}
\begin{equation}
    {\vb{B}'}_{m}^{(0)} = \dfrac{\partial(r\xi_{h})}{\partial r} B_r \vb{\nabla}\Ylm. \label{eq:Bprime}
\end{equation}
%\end{linenomath*}
%where $B_r$ is the radial component of the magnetic field, $\xi_{\theta}$ and $\xi_{\phi}$ the latitudinal and azimuthal components of $\vxi$.
%We did not consider a specific magnetic field topology, but we implicitly assumed that $B_{\phi,\theta}\xi_r \ll B_r\xi_h$, that is verified as long as $B_{\phi,\theta}/ B_r \ll \xi_h/\xi_r \sim N/\omega \sim 10^2$ for the red giants we study ($N$ denotes the Brunt-Väisälä frequency). 
Thus we get
%\begin{equation}
%\begin{linenomath*}
\begin{multline}
    \left\langle\vxi_{m}^{(0)},\LopL\left(\vxi^{(0)}_{m'}\right)\right\rangle = \\ \frac{1}{\mu_0}\int_{\ri}^{\ro} \left[\frac{\partial (r\xi_h)}{\partial r}\right]^2 \int_0^{2\pi}\!\!\int_0^{\pi} B_r^2 e^{i(m'-m)\phi}
    \left[ \dth{\Ylmt[l][m]}\dth{\Ylmt[l][m']}+\frac{mm'}{\sint[2]}\Ylmt[l][m]\Ylmt[l][m']
    \right]
    \sint \,\mathrm{d}\theta \,\mathrm{d}\phi \,\mathrm{d}r, \label{eq:xipLxi_g}
\end{multline}
%\end{linenomath*}
%\end{equation}
where $\ri$ and $\ro$ are the inner and outer boundaries of the g-mode cavity, and $\Ylmt (\theta)$ is defined as $\Ylm(\theta,\phi)= \Ylmt (\theta) e^{im\phi}$.
We notice that, 
%as in the axisymmetric case, 
in the limit of short radial wavelengths, g-mode frequencies are only sensitive to the square of the radial component of the magnetic field.
%\modif{The simplifications leading to this expression have been shown to be valid by computing all neglected terms in a specific axisymmetric case\cite{Mathis2021}.}

Finally, the inertia (Eq.~\ref{eq:Iner_gen}) of high-order g modes is
%\begin{linenomath*}
\begin{equation}
    \Iner \approx l(l+1)\int_{\ri}^{\ro} |\xi_h|^2 \rho r^2 \,\mathrm{d}r. \label{eq:Iner_g}
\end{equation}
%\end{linenomath*}

\subsubsection{Effects of magnetic field on high-order dipole g modes} \label{ssec:mag_dipolegmode}
For clarity we drop in the following sections the superscript ``$(0)$'' in $\omega^{(0)}$. Using Eqs.~\ref{eq:xipLxi_g} and \ref{eq:Iner_g}, the matrix elements $\mathsf{M}_{m,m'}$ (Eq.~\ref{eq:pert_gen}) for $l=1$ high-order g modes read
%\begin{linenomath*}
\begin{eqnarray}
     \mathsf{M}_{1,1}=\mathsf{M}_{-1,-1}&=&\frac{1}{2\mu_0\omega \Iner}\frac{3}{4} \int_{\ri}^{\ro} [\partial_r(r\xi_h)]^2 \int_0^{\pi} \left[B_r^2\right]_0 (1+\cost[2]) \sint \,\mathrm{d}\theta \,\mathrm{d}r \label{eq:M11},\\
     \mathsf{M}_{0,0}&=&\frac{1}{2\mu_0\omega \Iner}\frac{3}{2} \int_{\ri}^{\ro} [\partial_r(r\xi_h)]^2 \int_0^{\pi} \left[B_r^2\right]_0 (1-\cost[2]) \sint \,\mathrm{d}\theta \,\mathrm{d}r \label{eq:M00},\\
     \mathsf{M}_{0,1}=-\mathsf{M}_{-1,0}&=&\frac{1}{2\mu_0\omega \Iner}\frac{3}{2\sqrt{2}} \int_{\ri}^{\ro} [\partial_r(r\xi_h)]^2 \int_0^{\pi} \left[B_r^2\right]_1 (\sint\cost) \sint \,\mathrm{d}\theta \,\mathrm{d}r,\\
     \mathsf{M}_{1,0}=-\mathsf{M}_{0,-1}&=& \mathsf{M}_{0,1}^*,\\
     \mathsf{M}_{-1,1} = \mathsf{M}_{1,-1}^* &=&\frac{1}{2\mu_0\omega \Iner}\frac{3}{4} \int_{\ri}^{\ro} [\partial_r(r\xi_h)]^2 \int_0^{\pi} \left[B_r^2\right]_2 (1-\cost[2]) \sint \,\mathrm{d}\theta \,\mathrm{d}r,
\end{eqnarray}
%\end{linenomath*}
where $\left[B_r^2\right]_k$ are the Fourier coefficients along $\phi$ of $B_r^2$ defined as
%\begin{linenomath*}
\begin{equation}
    \left[B_r^2\right]_k (r,\theta) = \frac{1}{2\pi}\int_0^{2\pi} B_r^2 e^{ik\phi} \,\mathrm{d}\phi.
\end{equation}
%\end{linenomath*}
%The operator $\LopL$ (Eq. \ref{eq:LopLsimple}) is Hermitian, thus $\mathsf{M}$ is Hermitian too and its eigenvalues are real.
The matrix $\mathsf{M}$ is Hermitian and thus its eigenvalues are real.
We detail hereafter some properties of matrix $\mathsf{M}$ that are useful to exploit the observed spectra. 

\paragraph{Trace} The trace of $\mathsf{M}$, $\mathrm{Tr}(\mathsf{M})$, is also the sum of its eigenvalues, i.e. $\omega_{m=-1}^{(1)}+\omega_{m=0}^{(1)}+\omega_{m=+1}^{(1)}$. It reads
%\begin{linenomath*}
\begin{equation}
\mathrm{Tr}(\mathsf{M}) = 3\omegaB
\quad\mbox{with}\quad\omegaB=
\frac{1}{2\mu_0\omega I}
\int_{\ri}^{\ro} [\partial_r(r\xi_h)]^2 \int_0^{\pi} \left[B_r^2\right]_0 \sint \,\mathrm{d}\theta \,\mathrm{d}r . \label{eq:TraceM}
\end{equation}
%\end{linenomath*}
Let us define the horizontal average of the squared radial magnetic field as
%\begin{linenomath*}
\begin{equation}
    \overline{B_r^2} = \frac{1}{4\pi}\iint B_r^2 \sint \,\mathrm{d}\theta \,\mathrm{d}\phi,
\end{equation}
%\end{linenomath*}
then
%\begin{linenomath*}
\begin{equation}
   \omegaB= \frac{1}{2\mu_0\omega} \ddfrac{\int_{\ri}^{\ro} [\partial_r(r\xi_h)]^2 \overline{B_r^2} \,\mathrm{d}r }{ \int_{\ri}^{\ro} \xi_h^2 \rho r^2 \,\mathrm{d}r }. \label{eq:omBini}
\end{equation}
%\end{linenomath*}
We use the asymptotic expression of the displacement for high-order g modes\cite{Unno1989} 
%\begin{linenomath*}
\begin{equation}
    \xi_h \sim \rho^{-1/2} r^{-3/2} N^{1/2} \sin\left[\Phi(r)\right] \label{eq:xh_asympt}
\end{equation}
%\end{linenomath*}
with the phase $\displaystyle\Phi(r) =\int_{\ri}^r k_r(r') \,\mathrm{d}r' - \frac{\pi}{4}$. %, where $k_r$ is the radial component of the wave vector. 
Thus Eq.~\ref{eq:omBini} reduces to:
%\begin{linenomath*}
\begin{equation}
    \omegaB= \frac{1}{\mu_0\omega^3} \ddfrac{\int_{\ri}^{\ro} \left(\frac{N}{r}\right)^3 \frac{\overline{B_r^2}}{\rho} \,\mathrm{d}r }{ \int_{\ri}^{\ro} \frac{N}{r}  \,\mathrm{d}r }. \label{eq:omB_gen}
\end{equation}
%\end{linenomath*}
We have used the Stationary Phase Approximation, which is valid for rapidly oscillating functions such as $\Phi(r)$, to take out the phase terms\cite{Mathis2021}. This expression we found here to characterise the trace is similar to the equations describing magnetic shifts for axisymmetric fields\cite{cantiello16,Mathis2021}. We can rewrite $\omegaB$ as
%\begin{linenomath*}
\begin{equation}
\omegaB = \frac{\mathcal{I}}{\mu_0\omega^{3}} \int_{\ri}^{\ro} K(r) \overline{B_r^2} \,\hbox{d}r \equiv  \frac{\mathcal{I}}{\mu_0\omega^{3}} \langle B_r^2 \rangle, \label{eq:omB_Bmean}
\end{equation}
%\end{linenomath*}
where  $K(r)$ is a weight function
%\begin{linenomath*}
\begin{equation}
    K(r) = \ddfrac{\frac{1}{\rho} \left(\frac{N}{r}\right)^3}{ \int_{\ri}^{\ro} \left(\frac{N}{r}\right)^3 \frac{\hbox{d}r}{\rho} },
    \label{eq_kernel}
\end{equation}
%\end{linenomath*}
and $\cal I$ is a factor depending on the core structure
%\begin{linenomath*}
\begin{equation}
    \mathcal{I} = \ddfrac{\int_{\ri}^{\ro} \left(\frac{N}{r}\right)^3 \frac{\hbox{d}r}{\rho}}{\int_{\ri}^{\ro} \left(\frac{N}{r}\right) \,\hbox{d}r}.
    \label{eq_integral}
\end{equation}
%\end{linenomath*}
Magnetic effects thus vary as $1/\omega^3$. This dependency, which was found for dipole fields\cite{Hasan2005,Gomes2020,Bugnet2021,Loi2021}, is generalised here. A typical profile of the function $K$ in red giant stars is plotted in Extended Data Figure~\ref{fig_kernel_751}. Compared to rotation kernels, the function $K$ peaks in a very narrow range located at the hydrogen burning shell due to the cubic dependency of $N/r$.

% FIGURE KERNEL

\paragraph{Range for the matrix elements}
Using inequality relations between integrals, and noticing that $|x\sqrt{1-x^2}| \leqslant (1+x^2)/(2\sqrt{2})$ over $[-1,1]$, we show that
%\begin{linenomath*}
\begin{equation}
    \frac{3}{4}\omegaB \leqslant \mathsf{M}_{11} \leqslant \frac{3}{2}\omegaB, 
    \quad 0 \leqslant \mathsf{M}_{00} \leqslant \frac{3}{2}\omegaB, 
    \quad |\mathsf{M}_{01}| \leqslant \frac{1}{2} \mathsf{M}_{11} \quad{\mbox{and}}
    \quad |\mathsf{M}_{-11}| \leqslant \frac{1}{2} \mathsf{M}_{00}. \label{eq:relMmm}
\end{equation}
%\end{linenomath*}

\paragraph{Asymmetry}
As  shown in Sect.~\ref{ssec:spl_asym}, the asymmetry of multiplets $\delta_\mathrm{asym}$ corresponds in many cases to the quantity $2(\mathsf{M}_{11}-\mathsf{M}_{00})$. %, that is, $\omega_{m=-1}^{(1)}+\omega_{m=+1}^{(1)}-2\omega_{m=0}^{(1)}\approx 2(\mathsf{M}_{11}-\mathsf{M}_{00})$. 
Using Eqs.~\ref{eq:M11} and \ref{eq:M00}, it reads
%\begin{linenomath*}
\begin{equation}
   2(\mathsf{M}_{11}-\mathsf{M}_{00})=
\frac{3}{4\mu_0\omega I}
\int_{\ri}^{\ro} [\partial_r(r\xi_h)]^2 \int_0^{\pi} \left[B_r^2\right]_0 (3\cost[2]-1) \sint \,\hbox{d}\theta \,\hbox{d}r \label{eq:asym_gen}
\end{equation}
%\end{linenomath*}
We recognise an average of $B_r^2$ weighted with the second degree Legendre polynomial $P_2(\cost)=(3\cost[2]-1)/2$.
We can rewrite this equation as
%\begin{linenomath*}
\begin{equation}
   2(\mathsf{M}_{11}-\mathsf{M}_{00}) = 3a\omegaB \label{eq:AsymM}
\end{equation}
%\end{linenomath*}
with the asymmetry parameter $a$ defined as
%\begin{linenomath*}
\begin{equation}
a = \frac{ \displaystyle \int_{\ri}^{\ro} K(r) \iint B_r^2 P_2(\cos\theta) \sin\theta \,\hbox{d}\theta\hbox{d}\phi \,\hbox{d}r} {\displaystyle \int_{\ri}^{\ro} K(r) \iint B_r^2  \sin\theta \,\hbox{d}\theta\hbox{d}\phi \,\hbox{d}r}.
\label{eq_coef_a}
\end{equation}
%\end{linenomath*}
This quantity is a measure of the asymmetry of the azimuthal average of $B_r^2$ between the poles and the equator, averaged over the resonant cavity with the weight function $K$.
%From Eq.~\ref{eq:relMmm}
Since $-1/2\leqslant P_2(\cost) \leqslant 1$, we deduce that the possible range for $a$ is limited to
%\begin{linenomath*}
\begin{equation}
    -\frac{1}{2} \leqslant a \leqslant 1.
\end{equation}
%\end{linenomath*}
Extreme values for $a$ are reached when $B_r^2$ is totally concentrated around the poles ($a=1$) or when it is concentrated along the equator ($a=-1/2$). When $B_r$ has an axisymmetric dipolar structure ($B_r \propto \cost$), we recover $a=2/5$ as already known\cite{Hasan2005}. For an inclined dipole, $a$ decreases as the inclination increases. It vanishes for an inclination of $\sim55^\circ$, then becomes negative and reaches $a=-1/5$ when the dipole is aligned with the equator.

\subsubsection{Effects of magnetic field on mixed modes}\label{ssec:mag_onmixed}
Thus far we have described the perturbations induced on the frequencies of pure g modes. The non-radial modes observed in red giants are mixed modes. The magnetic shift of p modes being proportional to the square of the ratio of the Alfvén velocity to the sound speed in the convective envelope\cite{Mathis2021}, it is expected to be negligible. Therefore, we assume that the magnetic field only affects the g-mode cavity. For mixed modes, the matrix is then simply 
%\begin{linenomath*}
\begin{equation}
   \mathsf{M}^\mathrm{(mixed)} = \zeta \mathsf{M} \label{eq:M_mixed}
\end{equation}
%\end{linenomath*}
where $\zeta$ is the ratio between the kinetic energy of the mode in the g-mode cavity and the total kinetic energy of the mode. This result for magnetic fields is derived with the same approach as for rotation\cite{goupil13}.
%Let us assume that magnetic field only affects the g-mode cavity, then $\left\langle\vxi_m,\LopL(\vxi_{m'})\right\rangle$ is the same as for pure g modes (only the integration on [$r_{i},\ro$] do not vanish). As done for (\ref{eq:rot_mix}), we obtain:
%$$
%\mathsf{M}_{m',m}=
%    \frac{\left\langle\vxi_{0,m'},{\cal L}_1(\vxi_{0,m})\right\rangle}{2\omega (I_p + I_g)}
%    = \frac{I_g}{I}\frac{\left\langle\vxi_{0,m'},{\cal L}_1(\vxi_{0,m})\right\rangle}{2\omega I_g}= \zeta \mathsf{M}^{(g)}_{m',m}
%$$

\subsubsection{Combined effects of rotation and magnetic field on mixed modes} \label{ssec:complete_problem}
Let us consider a star, with a magnetic field, spinning with the rotation profile $\Omega(r)$. We denote as $\omg$ and $\omp$ the average rotation rates in the g-mode and p-mode cavities, respectively. Let us assume that, in the frame rotating with a rotation rate $\omg$, the magnetic field is steady. In this frame, the rotation profile is $\Omega(r)-\omg$.
The first-order perturbations $\omega^{(1)}$ of the frequency $\omega^{(0)}$ of a dipole mixed mode in the rotating frame are calculated by solving
%\begin{linenomath*}
\begin{equation}
    \omega^{(1)} \mathsf{a} = (\zeta\mathsf{M} + \mathsf{R})\mathsf{a}. \label{eq:egpb_corot}
\end{equation}
%\end{linenomath*}
The first term comes from Eq.~\ref{eq:M_mixed}, and
the matrix $\mathsf{R}$ combines the effects of the Coriolis force
and of the residual azimuthal flow in the rotating frame. It is a diagonal matrix with elements
%\begin{linenomath*}
\begin{equation}
    \mathsf{R}_{m,m} =  m \omega_\mathrm{R},
\end{equation}
%\end{linenomath*}
where
%\begin{linenomath*}
\begin{equation}
\omega_\mathrm{R} = \left(1-\frac{\zeta}{2}\right)\omg - (1-\zeta)\omp.
\end{equation}
%\end{linenomath*}
By defining $b=\zeta \omegaB/\omega_\mathrm{R}$, a parameter characterising the effects of the magnetic field relative to those of rotation, we can rewrite the matrix $\zeta\mathsf{M}+\mathsf{R}$ as
%\begin{linenomath*}
\begin{equation}
    \zeta\mathsf{M}+\mathsf{R}=\omega_\mathrm{R}
    \begin{bmatrix}
     b\left(1+\dfrac{a}{2}\right)-1& -bc & bd  \\
     -bc^* & b(1-a) & bc \\
     bd^* & bc^* & b\left(1+\dfrac{a}{2}\right)+1
    \end{bmatrix}, \label{eq:mat_global}
\end{equation}
%\end{linenomath*}
where $a$ is the asymmetry parameter (Eq.~\ref{eq_coef_a}), and $c$ and $d$ parameterise the off-diagonal elements generated by non-axisymmetric components of $B_r^2$. Possible values for $c$ and $d$ are constrained by Eq.~\ref{eq:relMmm}: $|c| \leqslant 1/2+a/4$ and $|d| \leqslant (1-a)/2$.

By solving Eq.~\ref{eq:egpb_corot}, we find three frequencies $\omega_m=\omega^{(0)}+\omega_{m}^{(1)}$ ($m=-1,0,1$) in the rotating frame, associated with three vectors $\mathsf{a}_m$ describing the decomposition of the eigenfunctions on $\vxi_{m'}^{(0)}$ (see Eq.~\ref{eq:xiprime}). In the inertial frame, we obtain up to nine frequencies\cite{Gough1990,Loi2021}
%\begin{linenomath*}
\begin{equation}
\omega_{m,m'} = \omega_m - m'\omg,
\end{equation}
%\end{linenomath*}
each frequency being related to the amplitude $\mathsf{a}_{m,m'}$ ($\equiv$ $m'$-th component of vector $\mathsf{a}_m$).
When $\mathsf{M}$ is diagonal, only $\omega_{m,m} = \omega_m - m\omg$ are associated with non-zero amplitudes.

\subsubsection{Average shift of multiplets} \label{ssec:globalshift}

In the very general case, a $l=1$ multiplet has nine components. 
We introduce the frequency shift of a mode relative to its unperturbed frequency: $\delta\!\omega_{m,m'}=\omega_{m,m'}-\omega^{(0)}$. The average of the frequency shifts of the nine components of a multiplet, denoted as $\delta\!\omegaB$, is
%\begin{linenomath*}
\begin{equation}
    \delta\!\omegaB = \frac{1}{9} \sum_{m,m'} \delta\!\omega_{m,m'} = \frac{1}{9} \sum_{m,m'} \omega_m^{(1)} - m'\omg.
\end{equation}
%\end{linenomath*}
Since the trace of a matrix is the sum of its eigenvalues, 
%\begin{linenomath*}
\begin{equation}
    \omega_{m=+1}^{(1)}+\omega_{m=0}^{(1)}+\omega_{m=-1}^{(1)}=\mathrm{Tr}(\zeta\mathsf{M+R})=\zeta \mathrm{Tr}(\mathsf{M})=3\zeta\omegaB,
\end{equation}
%\end{linenomath*}
by using $\mathrm{Tr}(\mathsf{R})=0$ and Eq.~\ref{eq:TraceM}. We deduce that the average frequency shift is
%\begin{linenomath*}
\begin{equation}
    \delta\!\omegaB = \zeta \omegaB.
\end{equation}
%\end{linenomath*}
In several practical cases, only $m=m'$ components are visible. To simplify the notations, we write $\delta\!\omega_{m}\equiv\delta\!\omega_{m,m}$. Thereby, even when only triplets are visible, we still verify
%\begin{linenomath*}
\begin{equation}
    \delta\!\omegaB = \frac{1}{3}(\delta\!\omega_{m=+1}+\delta\!\omega_{m=0}+\delta\!\omega_{m=-1}) = \zeta \omegaB.
\end{equation}
%\end{linenomath*}
Thus, measuring the average shift $\delta\!\omegaB$ of a multiplet provides a direct measurement of $\zeta \omegaB$. We also show that \omegaB\  represents the average magnetic shift of pure g modes.

%TBC with triplets that does not come from $m=m'$...

\subsubsection{Splitting and asymmetry of triplets} \label{ssec:spl_asym}
Simple analytic expressions of mode frequencies are possible in two cases: (i) when the diagonal elements of the matrix $\zeta\mathsf{M}+\mathsf{R}$ dominate over the other ones; (ii) when $c$ vanishes.

% pb: Diagonally dominated matrix has a math definition that is not strong enough...
\paragraph{Diagonally dominated matrix} Case (i) is achieved either for small values of $b$, that is, when the rotational effects dominate over the magnetic ones, or for small values of $|c|$ and $|d|$, which occurs for example when $B_r^2$ is largely axisymmetric. Nevertheless a field does not need to be axisymmetric to nullify $c$ and $d$, it is sufficient that the Fourier coefficients $\left[B_r^2\right]_1$ and $\left[B_r^2\right]_2$ vanish. The frequency shifts are then:
%\begin{linenomath*}
\begin{eqnarray}
     \delta\!\omega_{m=0} &=& \zeta(1-a)\omegaB,  \label{eq:domB0_diago}\\
     \delta\!\omega_{m=\pm 1} &=& \zeta(1+a/2)\omegaB \pm (\omega_\mathrm{R} - \omg). \label{eq:domB1_diago}
\end{eqnarray}
%\end{linenomath*}
We deduce that the splitting, defined as $\delta\!\omega_\mathrm{R} = \frac{1}{2} (\omega_{m=-1}-\omega_{m=+1})$, is
%\begin{linenomath*}
\begin{equation}
    \delta\!\omega_\mathrm{R} = \omg - \omega_\mathrm{R} = \frac{\zeta}{2} \omg + (1-\zeta) \omp.
\end{equation}
%\end{linenomath*}
In this configuration, the splitting depends only on the rotation, and we recover the same expression as for the non-magnetic case\cite{goupil13}.

The asymmetry of a triplet, defined as $\delta_\mathrm{asym}=\omega_{m=-1}+\omega_{m=+1}-2\omega_{m=0}$, is
%\begin{linenomath*}
\begin{equation}
    \delta_\mathrm{asym} = 3a\zeta \omegaB = 3a \delta\!\omegaB.
\end{equation}
%\end{linenomath*}
The asymmetry is proportional to \omegaB, hence to $\langle B_r^2 \rangle$, but also directly depends on the asymmetry parameter $a$ that we have introduced in Eq.~\ref{eq_coef_a}. As a consequence, when $a = 0$, $\delta_\mathrm{asym}$ also vanishes, even if the magnetic field is not weak: a weak asymmetry does not necessarily mean a weak magnetic field. This relation shows that $\delta_\mathrm{asym}$ can be negative, as we observed in one of our three targets. Without any hypothesis on the magnetic field topology, the asymmetry  $\delta_\mathrm{asym}$ provides a lower limit of $\langle B_r^2 \rangle$. Measuring simultaneously asymmetries $\delta_\mathrm{asym}$ and global shifts $\delta\!\omegaB$ provides a measurement of the rms radial magnetic field as well as a measurement of $a$, that is, information on the latitudinal variations of $B_r^2$.

\paragraph{Non-axisymmetric effects} In case (ii), $c=0$, we find five components with non-null amplitudes. However, we find that in each vector $\mathsf{a}_m$, the component $\mathsf{a}_{m,m}$ are always the largest. %Moreover, the two extra components remain  small (relative heights $< 10\%$) for all values of $d$, as long as $b< 1$. 
As a consequence, among possible components, three of them always have dominant amplitudes: $\omega_{m,m}$ ($m=-1,0,+1$). The frequency  shifts for these three components are
%\begin{linenomath*}
\begin{eqnarray}
     \delta\!\omega_{m=0} &=& \zeta(1-a)\omegaB, \\
     \delta\!\omega_{m=\pm 1} &=& \zeta(1+a/2)\omegaB \pm \left(\alpha\omega_\mathrm{R} - \omg\right).
\end{eqnarray}
%\end{linenomath*}
with $\alpha=\sqrt{1+|bd|^2}$. We recover the previous case for $\alpha=1$. We notice that the asymmetry remains unchanged ($\delta_\mathrm{asym} = 3a\zeta \omegaB$) independently of the value of $\alpha$. However the splitting is affected by the magnetic field in this configuration:
%\begin{linenomath*}
\begin{equation}
    \delta\!\omega_\mathrm{R} = \omg - \alpha\omega_\mathrm{R} %= \omg -\sqrt{\omega_\mathrm{R}^2+\zeta^2\omegaB^2|d|^2}
    = (1-\alpha-\alpha\zeta/2) \omg +\alpha(1-\zeta) \omp.
\end{equation}
%\end{linenomath*}
Since $\alpha$ depends on $\omegaB$, which varies as $1/\omega^3$, $\delta\!\omega_\mathrm{R}$ is not a simple linear function of $\zeta$. The non-axisymmetric terms introduce a spread in this linear relation. As a consequence, when we observe a tight linear relation between $\delta\!\omega_\mathrm{R}$ and $\zeta$, the quasi-axisymmetric approach should be valid.

Having non-null $c$ makes the situation more complex. %A general parametric analysis would go beyond the scope of this paper. 
When $|bc|$ becomes large, the expression for the asymmetry is more complicated. However, a parametric study shows us that, as long as $b < 1$, only the three components $\mathsf{a}_{m,m}$ are non-negligible. For higher values of $b$, the number of high-amplitude components changes and depends on the values of $c$ and $d$, making the spectrum more difficult to interpret. The reward of a successful spectrum interpretation would be to provide information on the non-axisymmetric components of $B_r^2$,
$\left[B_r^2\right]_1$ and $\left[B_r^2\right]_2$, in addition to $a$ and $\omegaB$. 
%Nevertheless, we have verified that the possible presence of off-diagonal elements do not affect the analysis presented in this paper (see Sect.~\ref{ssec:nonaxi_study}).

For test purposes, we applied our expressions to inclined dipolar fields, which have already been studied\cite{Loi2021}. The splittings and asymmetries we derived are in agreement with those already published.

For the three stars presented in this paper, we found that the matrix is diagonally dominated (see Sect.~\ref{ssec:nonaxi_study}). Therefore Eqs.~\ref{eq:domB0_diago} and \ref{eq:domB1_diago} derived within case (i) are valid to interpret the observations.

\subsection{Absolute magnetic shifts \label{sect_shiftmag}}

The magnetic perturbation to the mode frequencies are mainly characterised by a frequency shift, whose intensity depends on $|m|$. Stars that show multiplet asymmetries related to magnetic fields should also exhibit the signature of these shifts. To estimate the intensity of magnetic shifts, we calculated asymptotic expressions of mixed modes that include a magnetic perturbation. The procedure is similar as that followed in Sect. \ref{sect_analysis}, except that the frequencies of pure p and g modes now include frequency shifts $\delta\!\omega_{m}$, which are produced by magnetic and rotational perturbations.

% FIGURE STRETCHED DIAGRAMS

%Assuming an axisymmetric magnetic field, we have
For pure g modes, the expressions of frequency shifts produced by rotation and magnetic fields are given by Eq. \ref{eq:domB0_diago} and \ref{eq:domB1_diago}, in which we set $\zeta = 1$, so that %suggestionJB
%\begin{linenomath*}
\begin{align}
\delta\!\omega_{m=0}^{(\rm g)} & = \left( 1-a \right) \delta\!\omega_\mathrm{g} \left( \frac{\omega_{\rm max}}{\omega} \right)^3  \\
\delta\!\omega_{m=\pm1}^{(\rm g)} & = \left( 1+\frac{a}{2} \right) \delta\!\omega_\mathrm{g} \left( \frac{\omega_{\rm max}}{\omega} \right)^3 \mp\frac{\omg}{2},
\end{align}
%\end{linenomath*}
%For pure g modes, the rotational splitting corresponds to
%and rotational splittings are %suggestionJB
%\begin{equation}
%\delta\!\omega_{{\rm R},m}^{(\rm g)} = -m\frac{\omg}{2}.
%\end{equation}
We thus obtain the perturbed periods of pure gravity modes $P'_{\rm g}$ as
%\begin{linenomath*}
\begin{equation}
%P'_{\rm g}(n_{\rm g},m) = P_{\rm g}(n_{\rm g}) \left[ 1+ \frac{P_{\rm g}(n_{\rm g})}{2\pi} \left(\delta\!\omega_{{\rm B},m}^{(\rm g)} + \delta\!\omega_{{\rm R},m}^{(\rm g)}\right)  \right]^{-1},
P'_{\rm g}(n_{\rm g},m) = P_{\rm g}(n_{\rm g}) \left[ 1+ \frac{P_{\rm g}(n_{\rm g})\delta\!\omega_m^{(\rm g)}}{2\pi} \right]^{-1},
\label{eq_Pg_pert}
\end{equation}
%\end{linenomath*}
which can be substituted to Eq. \ref{eq_Pg} in the asymptotic development.

% suggestion:
%In line with our assumption of a magnetic field buried in the core, w  
Similarly, for pure p modes ($\zeta = 0$), we obtain
%\begin{linenomath*}
\begin{equation}
\nu'_{\rm p}(n_{\rm p},m) = \nu_{\rm p}(n_{\rm p})+ \frac{\delta\!\omega_m^{(\rm p)}}{2\pi}
\label{eq_nup_pert}
\end{equation}
%\end{linenomath*}
where
%\begin{linenomath*}
\begin{equation}
\delta\!\omega_m^{(\rm p)} = -m\omp.
\end{equation}
%\end{linenomath*}
This expression was substituted to Eq. \ref{eq_nup} in the asymptotic development.

Knowing the asymptotic parameters of pure p and g modes, we can obtain expressions for the perturbed mixed mode frequencies for any values of $\delta\!\omega_\mathrm{g}$ (quantifying the intensity of the magnetic field) and $a$ (quantifying the asymmetry of the multiplets) by solving Eq. \ref{eq_wkb_mixed}. For illustration, we computed perturbed mixed mode frequencies assuming an asymmetry coefficient $a = 2/5$ (corresponding to a dipole aligned with the rotation axis) and an intensity of the magnetic shift ranging from moderate ($\delta\!\omega_\mathrm{g}/(2\pi) = 0.2\,\mu$Hz) to strong ($\delta\!\omega_\mathrm{g}/(2\pi) = 2\,\mu$Hz). We assumed $\dpun=80.5$~s, $\epsg = 0.28$, and the rest of the parameters were taken from the inferred values of KIC\,8684542.

The perturbed frequencies are shown in the shape of stretched \'echelle diagrams in Supplementary Figure \ref{fig_stretch_ex}. For strong shifts the stretched periods are no longer regularly spaced, and the ridges are strongly curved. For the three stars in which we have detected multiplet asymmetries, the $m=0$ ridge is approximately vertical in the stretched \'echelle diagram (see Extended Data Figure \ref{fig_stretch_115}-\ref{fig_stretch_751}), and such strong magnetic shifts are thus ruled out.

For moderate shifts, the ridge associated with the $m=0$ component in the stretched \'echelle diagram remains nearly vertical in spite of the perturbations. This means that such moderate magnetic shifts cannot be directly detected in red giants when performing a seismic analysis, if asymmetries are not detected (it can be the case if $a=0$ or if only one or two components per multiplet are visible because of geometric factors). However, if the $m=0$ components are used to measure the asymptotic parameters of p- and g-modes without including a magnetic perturbation (as we have done for the three stars in Sect. \ref{sect_analysis}), we obtain $\dpun^{\rm mes} = 80.1$~s (a value slightly lower than the actual asymptotic period spacing), and $\epsg^{\rm mes} = 0.56$ (significantly larger than the actual value). If the intensity of the magnetic shift increases, the measured $\dpun^{\rm mes}$ decreases, and the measured $\epsg^{\rm mes}$ increases. It is thus clear that a moderate magnetic shift is capable of significantly modifying the measured value of $\epsg$. This is interesting because the actual value of $\epsg$ is strongly constrained for red giants. From \kepler\ data, it was found that $\epsg = 0.28\pm0.08$\cite{mosser18}, in agreement with theoretical predictions that do not include magnetic perturbations\cite{takata16}. Therefore, stars that exhibit magnetic shifts can be identified by their measured value of $\epsg$.

The two red giants that show the strongest multiplet asymmetries (KIC\,8684542 and KIC\,11515377) have measured values of $\epsg$ that significantly deviate from the range of $\epsg$ of typical red giants ($\epsg=0.50\pm0.02$ and $0.50\pm0.03$, respectively). This can be interpreted as the signature of a magnetic shift in the g-mode periods, as shown above. The measured $\epsg$ can thus be used to place constraints on the intensity of the magnetic field. For KIC\,7518143, the measurement of $\epsg$ is consistent with the typical value of $\epsg$ for red giants, which means that vanishing magnetic shifts cannot be excluded using this measurement alone. However, it can be used to derive an upper limit for the magnetic field intensity.

To retrieve this information, we assumed that the actual value of $\epsg$ for these stars corresponds to the value measured for regular red giants, that is, $\epsg = 0.28\pm0.08$. For any set of parameters $(\dpun,\delta\!\omega_\mathrm{g},a)$, we were then able to compute asymptotic frequencies of mixed modes including magnetic and rotational perturbations using Eq. \ref{eq_wkb_mixed}, \ref{eq_Pg_pert} and \ref{eq_nup_pert}. These frequencies could be compared to the observed mode frequencies. We thus optimised these three parameters to reproduce at best all the observed mode frequencies using a grid method (see Supplementary Table \ref{tab_shiftmag}). In Figure \ref{fig_stretch_868} and Extended Data Figure \ref{fig_stretch_115}-\ref{fig_stretch_751}, we overplotted the asymptotic mixed mode frequencies resulting from the best-fit solutions. The agreement with the observations is strikingly good for all three stars. 

We could then use the measured values of $\delta\!\omega_\mathrm{g}$ to derive estimates of $\langle B_r^2\rangle$ as
%\begin{linenomath*}
\begin{equation}
    \langle B_r^2\rangle = \frac{\mu_0 \delta\!\omega_\mathrm{g} \omega_{\rm max}^3}{\mathcal{I}}
\end{equation}
%\end{linenomath*}
The results are given in Supplementary Table \ref{tab_shiftmag}. They are fully consistent with the lower limits of the field intensities that were obtained using multiplet asymmetries.

% TABLE ASYMPT

\subsection{Stellar models \label{sect_models}}

To obtain estimates of the magnetic field intensities using Eq. \ref{eq_dnub} and \ref{eq_asym}, we needed to compute stellar models of the three red giants. For this purpose, we computed a grid of models with varying masses, ages, and metallicities covering the range of \kepler\ red giants using the evolution code \mesa\cite{paxton11}. Among this grid, we searched for models that simultaneously reproduce the asymptotic large separation of p modes $\Delta\nu$ and the asymptotic period spacing of dipole g modes $\dpun$ of each star. This procedure ensures that the selected models reproduce the observed mode frequencies sufficiently well to produce reliable estimates of the weight function $K(r)$ (Eq. \ref{eq_kernel}) and the term $\mathcal{I}$ (Eq. \ref{eq_integral})\cite{deheuvels12}. For illustration, the function $K(r)$ obtained for KIC\,11515377 is shown in Extended Data Figure \ref{fig_kernel_751}.

One potential explanation for the detected magnetic fields is that they could be the remnants of dynamo-generated fields produced in the convective core during the main sequence. Owing to the weak ohmic diffusion, any layer that was convective at some point during the main sequence can retain a strong field until the red giant phase\cite{cantiello16}. The best-fit models of all three stars possess convective cores during the main sequence. We found that the layers corresponding to the current hydrogen burning shell were indeed convective, but only at the very beginning of the main sequence, when the convective core is produced by the burning of $^{3}$He and $^{12}$C outside of equilibrium (regardless of the inclusion or not of core overshooting). Supplementary Figure \ref{fig_evol_conv_core} shows the evolution of the mass of the convective core over the main sequence for KIC\,11515377.

% FIGURE CONV CORE

The amplitude of the dynamo magnetic fields necessary to account for the fields detected on the red giant branch can be estimated assuming magnetic flux conservation, namely $B_{\rm MS} = r_{\rm RGB}^2/r_{\rm MS}^2 B_{\rm RGB}$, where $r_{\rm RGB}$ is the radius of the hydrogen burning shell at current age and $r_{\rm MS}$ the radius of the same shell traced back to the beginning of the main sequence. From our three magnetic field measurements, we find $B_{\rm MS}$ ranging from 3 to 5 kG.

\subsection{Critical magnetic field}\label{ssec:criticalfield}
Using the dispersion relation of magneto-gravity waves\cite{Unno1989}
%\begin{linenomath*}
\begin{equation}
    \omega^2 = \frac{k_h^2}{k^2}N^2+\omega_\mathrm{A}^2\quad\mbox{with }
    \omega_\mathrm{A}^2=\frac{(\vb{B}\cdot\vb{k})^2}{\mu_0\rho}=k^2 v_\mathrm{A}^2\mu^2 \label{eq:rel_dispersion}
\end{equation}
%\end{linenomath*}
where %$k$ is the wave vector, $k_h$ its horizontal component,
$v_\mathrm{A}=B/\sqrt{\mu_0\rho}$ is the Alfvén speed and $\mu$ the cosine of the angle between $\vb{B}$ and $\vb{k}$,
it is straightforward to show that waves can propagate as long as $\omega^4 > 4v_\mathrm{A}^2\mu^2k_h^2N^2$\cite{Fuller2015}.
%Using the dispersion relation of magneto-gravity waves, it is a simple matter\cite{Fuller2015} to show that waves can propagate as long as $\omega^4 > 4v_\mathrm{A}^2\mu^2k_h^2N^2$, where $v_\mathrm{A}=B/\sqrt{\mu_0\rho}$ is the Alfvén speed and $\mu$ the cosinus of the angle between $\vb{B}$ and $\vb{k}$.
Since $k_r \gg k_h$, $B\mu\approx B_r$, so that $\omega^4 > 4 B_r^2 k_h^2N^2 /(\mu_0\rho)$. %Here again (see also \ref{sssec:variational}), it has been implicitly assumed that $B_{\phi,\theta}k_h \ll B_r k_r$, that is verified as long as $B_{\phi,\theta}/ B_r \ll  N/\omega \sim 10^2$.
This inequality defines a critical field\cite{Fuller2015}, which is, for dipole modes,
%\begin{linenomath*}
\begin{equation}
    B_\mathrm{c}^2(r) = \frac{\mu_0 \rho r^2\omega^4}{8 N^2}. \label{eq:Bcdef}
\end{equation}
%\end{linenomath*}
This critical field varies along the radius. It has already been estimated in red giants\cite{Fuller2015,cantiello16,Bugnet2021}. In those stars, it reaches its minimal value $B_\mathrm{c,min}$ where the Brunt-Väisälä frequency $N$ is maximal, that is, in the hydrogen-burning shell located at $r=r_\mathrm{hbs}$. The minimal critical field thus corresponds to
%\begin{linenomath*}
\begin{equation}
    B_\mathrm{c,min}^2 = \frac{\mu_0 \rho_\mathrm{hbs} r_\mathrm{hbs}^2\omega^4}{8 N_\mathrm{hbs}^2}, \label{eq:Bcmin}
\end{equation}
%\end{linenomath*}
where $N_\mathrm{hbs}=N(r_\mathrm{hbs})$ and $\rho_\mathrm{hbs}=\rho(r_\mathrm{hbs})$. 

It is convenient to rewrite the expression of the magnetic shift $\omegaB$ as a function of the ratio between the measured field and the minimum critical field $B_\mathrm{c,min}$. For this purpose, we plug Eqs.~\ref{eq:Bcmin} into Eq.~\ref{eq:omB_Bmean} and we obtain
%\begin{linenomath*}
\begin{equation}
    \omegaB = \frac{\omega}{8}\frac{\langle{B_r^2}\rangle}{B_\mathrm{c,min}^2} \widetilde{\cal I}, \label{eq:omB_Bcmin}
\end{equation}
%\end{linenomath*}
where $\mathcal{\widetilde{I}}$ corresponds to $\mathcal{I}$ normalised by its value in the hydrogen burning shell, that is
%\begin{linenomath*}
\begin{equation}
    \widetilde{\cal I} =  \ddfrac{\int_{\widetilde r_\mathrm{i}}^{\widetilde r_\mathrm{o}} \left(\frac{\widetilde N}{\widetilde r}\right)^3 \,\frac{\mathrm{d}\widetilde{r}}{\widetilde \rho} }{ \int_{\widetilde{r}_\mathrm{i}}^{\widetilde{r}_\mathrm{o}} \frac{\widetilde{N}}{\widetilde{r}}  \,\mathrm{d}\widetilde{r} }. \label{eq_normalised_integral}
\end{equation}
%\end{linenomath*}
The tildes denote dimensionless profiles $\widetilde{X}=X/X(r_\mathrm{hbs})$. Using our best-fit models of the three red giants, we find that $\widetilde{\cal I} \approx 0.12$.
%Taking $\langle{B_r^2}\rangle < B_\mathrm{c,min}^2$, we obtain a maximal value for $\omegaB$:
%\begin{equation}
%    \omegaB < \frac{\omega}{8}{\widetilde{\cal I}}. \label{eq:omBmax}
%\end{equation}

From Eq.~\ref{eq:omB_Bcmin}, we derive the ratio between the measured magnetic field and the minimal critical field:
%\begin{linenomath*}
\begin{equation}
   \frac{\langle B_r^2\rangle}{B_{\rm c, min}^2} = \frac{8}{\widetilde{\mathcal{I}}}\frac{\delta\!\omega_\mathrm{g}}{\omega_{\rm max}},
\end{equation}
%\end{linenomath*}
where $\delta\!\omega_\mathrm{g}$ is the magnetic shift of pure g modes at frequency $\omega_{\max}$ (see Supplementary Table~\ref{tab_shiftmag}).
We eventually find that the measured magnetic fields are below the critical field for the three stars. It amounts to about 28\% of $B_{\rm c, min}$ for KIC\,8684542, 25\% for KIC\,11515377, and less than 12\% for KIC\,7518143.

Using the stellar models computed in Sect. \ref{sect_models}, we can thus estimate the approximate range of field strengths for which magnetic perturbations to oscillation modes are expected to be detectable. Based on the uncertainties reached in our measurements of mode frequencies, we estimate that we should be able to detect magnetic splittings above around 20 nHz. This translates into minimum detectable field strengths ranging from about 25 to 40 kG for the three stars. The upper limit on the detectable field strength is given by the critical field $B_{\rm c,min}$, which ranges from about 300 to 450 kG for the three red giants. The range of detectable fields strength is thus quite narrow, encompassing about one order of magnitude.

\subsection{Validation of the developed model \label{sect_validation}}

\subsubsection{Validity of perturbation methods}
The validity of first-order perturbation treatments for the rotation is well established in red giants, since $2\Omega \ll \omega$ and $\Omega \ll (GM/R^3)^{1/2}$ for these stars. 
The first-order perturbation theory developed in Sect.~\ref{ssec:perturb} requires that $\omega^{(1)} \ll \omega^{(0)}$, that is $\omegaB \ll \omega_{\rm max}$, since $\omegaB$ and $\omega_{\rm max}$ characterise, respectively, the frequency shift induced by magnetic fields and the mode frequencies. For the three stars studied in this paper, we are indeed in this regime since $\omegaB / \omega_{\rm max} \sim 10^{-3}$.
Moreover, the magnetic field needs to be smaller than the critical field, otherwise waves cannot propagate. As discussed in Sect.~\ref{ssec:criticalfield}, this condition is well verified for the three stars.
Finally, our developments rely on the short wavelength approximation. This approximation has already proved to be justified to treat the rotational effects on red giant modes\cite{goupil13}. Here, it also allows us to simplify the magnetic operator $\LopL$ (Eq.~\ref{eq:LopLsimple}). We note that this simplification has been  successfully tested for a particular dipolar field solution involving both poloidal and toroidal components\cite{Mathis2021}.
%Using Eq.~\ref{eq:omBmax}, it means $\omegaB / \omega_{\rm max} < \widetilde{\cal I}/8 \approx 0.015$. This condition is safely verified for the three stars.
%By considering the dispersion relation of magneto-gravity waves (Eq.~\ref{eq:rel_dispersion}), we note that the effects of magnetic field on g modes are weak as long as $\omega_\mathrm{A}^2 \ll \omega^2$. With the same considerations on $B$ and $k$ as the ones mentioned in Sect.~\ref{ssec:criticalfield}, the perturbation method is valid as long as
%\begin{equation}
%    \frac{l(l+1) N^2 B_r^2}{\mu_0\rho r^2 \omega^2} \ll \omega^2.
%\end{equation}
%We introduce the critical magnetic field (Eq.~\ref{eq:Bcdef}) in the relation and we get $B_r^2 \ll 4B_\mathrm{c}^2$. We may drop the factor of 4, since we just compare orders of magnitude. We evaluate this relation in the hydrogen-burning shell, where the effects of the magnetic field on the modes are the strongest, and obtain
%\begin{equation}
%    \frac{\langle B_r^2 \rangle}{B_\mathrm{c,min}^2} \ll 1
%\end{equation}
%For the three stars we have studied in the present paper, this ratio never exceed $0.15$ (see Discussion). This verification \textit{a posteriori} confirms that perturbation treatment of magnetic field is relevant for these stars.

%Il me semble que le plus direct et le plus correct serait de déterminer $\delta \omega_B/\omega$.

\subsubsection{Impact of strong azimuthal fields}
To express matrix $\mathsf{M}$, we have assumed that the horizontal magnetic field does not strongly dominate over its radial component: in practice, we have assumed that $B_\phi/B_r \ll N/\omega \sim 10^2$ for these stars.
Thus our description is not valid if the field is almost purely toroidal. Models based on a Tayler-Spruit dynamo\cite{Fuller2019} predict very strong azimuthal fields $B_\phi/B_r \sim 10^6$ in red giant cores. We studied how such fields would impact matrix $\mathsf{M}$.

%We recalculated $\vb{B}'$ as was done in Sect.~\ref{sssec:variational}, but assuming that $B_\phi$ is strongly dominant. Eq.~\ref{eq:Bprime} then becomes
%\begin{equation}
    %{\vb B}' \approx \frac{1}{r}\xi_h B_\phi \vb{v}(\theta,\phi),
%    {\vb B}' \approx \xi_h \vb{v}(r,\theta,\phi),
%\end{equation}
%where $\vb{v}(\theta,\phi)$ is a vector that does not depend on $r$ and contains geometrical terms. We have used $l(l+1) \xi_h \approx r \partial_{r} \xi_r$ to get this form. 
%where $\vb{v}(\r,\theta,\phi)$ is a vector that contains geometrical terms, $B_\phi$ and its derivatives.
%Keeping only the dominant term, we deduce that 
We determined the dominant term of $\LopL\left(\vxi_{m'}^{(0)}\right)$, assuming that $B_\phi$ is strongly dominant. We deduced that 
%\begin{linenomath*}
\begin{equation}
    \left\langle\vxi_{m}^{(0)},\LopL\left(\vxi_{m'}^{(0)}\right)\right\rangle \approx \frac{1}{\mu_0}\int_{\ri}^{\ro} \xi_h^2 \iint %B_\phi^2
    F_{m,m'}(r,\theta,\phi)
    \,\mathrm{d}\theta \,\mathrm{d}\phi \,\mathrm{d}r  ,
\end{equation}
%\end{linenomath*}
where $F_{m,m'}$ are functions that we do not need to explicit here. %of $\theta$ and $\phi$
They contain geometrical terms, $B_\phi$, and its spatial derivatives.
By using an asymptotic expression for $\xi_h$ (Eq.~\ref{eq:xh_asympt}) and applying the same method as in Sect.~\ref{ssec:mag_dipolegmode}, we showed that the matrix elements have the form
%\begin{linenomath*}
\begin{equation}
    \mathsf{M}_{m,m'}\approx \frac{1}{4\mu_0\omega}\dfrac{\int_{\ri}^{\ro} N/(\rho r^3) \iint %B_\phi^2 
    F_{m,m'}(r,\theta,\phi)
    \,\mathrm{d}\theta \,\mathrm{d}\phi \,\mathrm{d}r}{\int_{\ri}^{\ro} (N/r)\,\mathrm{d}r}.
\end{equation}
%\end{linenomath*}
We notice that the frequency dependency is different: if $B_\phi$ strongly dominates, magnetic shifts vary as $1/\omega$ instead of $1/\omega^3$. We found that this dependency in $1/\omega$ yields very poor fits to the observations, so we can safely exclude such a magnetic topology and validate our initial assumption.

\subsubsection{Impact of non-axisymmetric elements on the analysis}\label{ssec:nonaxi_study}
In this paper, the interpretation of splittings and asymmetries in terms of rotation rates $\langle\Omega\rangle_\mathrm{p,g}$ and magnetic field intensities $\langle B_r^2 \rangle$ has been performed by neglecting the off-diagonal elements of the matrix $\zeta\mathsf{M}+\mathsf{R}$ (Eq.~\ref{eq:mat_global}). We here justify this assumption.

First, the observed splittings $\delta\!\omega_\mathrm{R}$ suggest that the off-diagonal elements are negligible. As discussed in Sect.~\ref{ssec:spl_asym}, these terms should introduce a spread in the linear relation between $\delta\!\omega_\mathrm{R}$ and $\zeta$. As shown in Supplementary Figure~\ref{fig_zeta_split} these relations are very tight, which suggests a weak influence of these terms. Secondly, the global magnetic shifts of g modes (Sect.~\ref{sect_shiftmag}) provide a measurement (or an upper limit) of $\omegaB$, and %$\omega_\mathrm{R}$ can be deduced from the splittings.
the splittings yield a measurement of $\omega_\mathrm{R}$.
By combining both values and taking error bars into account, we find that the parameter $b=\zeta\omegaB/\omega_\mathrm{R}$ is smaller than $1$ for the modes observed in the three stars. In this regime, three components always dominate over the others.

Thanks to these constraints, we can explore the possible values of the unknown complex parameters $c$ and $d$ to quantify their impact on asymmetries and splittings. 
%(Est-ce qu’avant l’etude generale que tu decris ci-dessous, on ne pourrait pas commencer par dire que le faible spread du splitting en fonction de ksi nous indique que les effets non-axisymetrique doivent etre faibles et qu’avec le omegaR et les omegaB obtenus ainsi on a b =. Et que quand on fait une etude des elements propres de la matrice avec ce b et toutes les valeurs autorisées de c et d, la matrice est toujours suffisament dominée par la diagonale pour que les trois composantes du cas diagonal dominent très nettement ? oui je rajoute, mais le deuxième point n'est pas si clair, c'est pourquoi j'ai fait l'analyse)
To do so, we fitted for each star all the observed splittings and asymmetries simultaneously with a model in which $a$, $c$, $d$, $\omegaB$, $\omg$ and $\omp$ are free parameters and $\zeta$ is given by its asymptotic expression (Eq.~\ref{eq_zeta}).

The fitting method is as follows. For a given set of parameters, the eigenvalue problem described in Sect.~\ref{ssec:complete_problem} is solved for each observed mode. For a given mode, among the nine computed components, we select the three ones with the largest amplitudes. From this triplet we compute a modelled splitting 
%$\delta\!\omega_{\mathrm{R},k}^\mathrm{(M)}$
and a modelled asymmetry. 
%$\delta_{\mathrm{asym},k}^\mathrm{(M)}$.
We compute a $\chi^2$ from the modelled and observed values of splittings and asymmetries of all the modes. We explore the parameter space by running a Markov chain Monte Carlo. We set uniform priors for $a$, $c$ and $d$ within the ranges given in Sect.~\ref{ssec:complete_problem}. %Priors for $\omg/(2\pi)$ and $\omp/(2\pi)$ are uniform on loose intervals: $[0,1\:\mathrm{\mu Hz}]$ and $[-0.1,0.1\:\mathrm{\mu Hz}]$, respectively. For $\omegaB$, we used uniform priors on the range $[\omega_-,\omega_+]$ with Gaussian decay (parameterised with $\sigma$) around this interval. Values of $\omega_-$, $\omega_+$, and $\sigma$ are derived from measurements of $\epsg$ that give possible ranges for global shifts (see Sect.~\ref{sect_shiftmag}). In practice, we used $(\omega_-/2\pi,\omega_+/2\pi,\sigma/2\pi)={}$  $(150,250,40\:\mathrm{nHz})$, $(60,220,80\:\mathrm{nHz})$, and $(0,50,50\:\mathrm{nHz})$ for KIC\,8684542, KIC\,11515377, and KIC\,7518143. These priors may look too tight at first glance, but they are quite robust since global shifts are reliable measurements of $\omegaB$, even for non axisymmetric fields (see Sect.~\ref{ssec:globalshift}).
Loose uniform priors are set for $\omg$ and $\omp$. For $\omegaB$, we use priors derived from possible  ranges for global shifts, deduced from $\epsg$ (see Sect.~\ref{sect_shiftmag}). These informative priors are quite robust since global shifts are reliable measurements of $\omegaB$, even for non-axisymmetric fields (see Sect.~\ref{ssec:globalshift}).

%Posterior probability distributions of $b$ remain below or around 1. 
Posterior distributions of $c$ and $d$ are generally loosely constrained. The posterior distributions we recover for $a$, $\omg$, $\omp$, and $\langle B_r^2 \rangle$ (computed from $\omegaB$) are fully compatible with the values quoted in the paper, this confirms that omitting off-diagonal elements does not significantly modify the analysis of the three studied red giant stars.

\subsection{Ruling out other potential sources of multiplet asymmetries \label{sect_other_mechanisms}}

Beside magnetic fields, we also examined other mechanisms that can produce multiplet asymmetries. Such features can arise for fast rotators, owing to higher-order terms in the rotational perturbation\cite{Dziembowski1992}. In Sect. \ref{sect_rotation}, we obtained measurements of the core rotation for the three red giants under study. The values that we found are consistent with typical red giants\cite{Gehan2018}. High-order rotational effects are thus expected to be negligible for these stars.

Secondly, asymmetries can be produced by near-degeneracy effects, when the frequency separation between consecutive mixed modes is comparable to the rotational splitting\cite{Deheuvels2017}. However, in this case, only p-dominated modes are expected to show significant asymmetries, which is the opposite of what is observed here. Besides, near-degeneracy effects should produce series of alternate positive-negative asymmetries, whereas in our case, all asymmetries have the same sign in each star. We can thus safely rule out these two mechanisms as the source of the observed asymmetries.

%\bibliographystyle{naturemag-doi}
%\bibliography{references}

\clearpage
\newpage
\begin{thebibliography}{10}
\urlstyle{rm}
\expandafter\ifx\csname url\endcsname\relax
  \def\url#1{\texttt{#1}}\fi
\expandafter\ifx\csname urlprefix\endcsname\relax\def\urlprefix{URL }\fi
\expandafter\ifx\csname doiprefix\endcsname\relax\def\doiprefix{DOI: }\fi
\providecommand{\bibinfo}[2]{#2}
\providecommand{\eprint}[2][]{\url{#2}}


\bibitem{deheuvels12}
\bibinfo{author}{{Deheuvels}, S.} \emph{et~al.}
\newblock \bibinfo{journal}{\bibinfo{title}{{Seismic Evidence for a Rapidly
  Rotating Core in a Lower-giant-branch Star Observed with Kepler}}}.
\newblock {\emph{\JournalTitle{\apj}}} \textbf{\bibinfo{volume}{756}},
  \bibinfo{pages}{19} (\bibinfo{year}{2012}).

\bibitem{Gehan2018}
\bibinfo{author}{{Gehan}, C.}, \bibinfo{author}{{Mosser}, B.},
  \bibinfo{author}{{Michel}, E.}, \bibinfo{author}{{Samadi}, R.} \&
  \bibinfo{author}{{Kallinger}, T.}
\newblock \bibinfo{journal}{\bibinfo{title}{{Core rotation braking on the red
  giant branch for various mass ranges}}}.
\newblock {\emph{\JournalTitle{\aap}}} \textbf{\bibinfo{volume}{616}},
  \bibinfo{pages}{A24} (\bibinfo{year}{2018}).

\bibitem{Marques2013}
\bibinfo{author}{{Marques}, J.~P.} \emph{et~al.}
\newblock \bibinfo{journal}{\bibinfo{title}{{Seismic diagnostics for transport
  of angular momentum in stars. I. Rotational splittings from the pre-main
  sequence to the red-giant branch}}}.
\newblock {\emph{\JournalTitle{\aap}}} \textbf{\bibinfo{volume}{549}},
  \bibinfo{pages}{A74} (\bibinfo{year}{2013}).

\bibitem{Gough1998Natur}
\bibinfo{author}{{Gough}, D.~O.} \& \bibinfo{author}{{McIntyre}, M.~E.}
\newblock \bibinfo{journal}{\bibinfo{title}{{Inevitability of a magnetic field
  in the Sun's radiative interior}}}.
\newblock {\emph{\JournalTitle{\nat}}} \textbf{\bibinfo{volume}{394}},
  \bibinfo{pages}{755--757} (\bibinfo{year}{1998}).

\bibitem{Fuller2019}
\bibinfo{author}{{Fuller}, J.}, \bibinfo{author}{{Piro}, A.~L.} \&
  \bibinfo{author}{{Jermyn}, A.~S.}
\newblock \bibinfo{journal}{\bibinfo{title}{{Slowing the spins of stellar
  cores}}}.
\newblock {\emph{\JournalTitle{\mnras}}} \textbf{\bibinfo{volume}{485}},
  \bibinfo{pages}{3661--3680} (\bibinfo{year}{2019}).

\bibitem{Gouhier2022}
\bibinfo{author}{{Gouhier}, B.}, \bibinfo{author}{{Jouve}, L.} \&
  \bibinfo{author}{{Ligni{\`e}res}, F.}
\newblock \bibinfo{journal}{\bibinfo{title}{{Angular momentum transport in a
  contracting stellar radiative zone embedded in a large scale magnetic
  field}}}.
\newblock {\emph{\JournalTitle{arXiv e-prints}}}
  \bibinfo{pages}{arXiv:2201.02645} (\bibinfo{year}{2022}).

\bibitem{Unno1989}
\bibinfo{author}{{Unno}, W.}, \bibinfo{author}{{Osaki}, Y.},
  \bibinfo{author}{{Ando}, H.}, \bibinfo{author}{{Saio}, H.} \&
  \bibinfo{author}{{Shibahashi}, H.}
\newblock \emph{\bibinfo{title}{{Nonradial oscillations of stars}}}
  (\bibinfo{publisher}{University of Tokyo Press, Tokyo},
  \bibinfo{year}{1989}).

\bibitem{Gough1990}
\bibinfo{author}{{Gough}, D.~O.} \& \bibinfo{author}{{Thompson}, M.~J.}
\newblock \bibinfo{journal}{\bibinfo{title}{{The effect of rotation and a
  buried magnetic field on stellar oscillations}}}.
\newblock {\emph{\JournalTitle{\mnras}}} \textbf{\bibinfo{volume}{242}},
  \bibinfo{pages}{25--55} (\bibinfo{year}{1990}).

\bibitem{Hasan2005}
\bibinfo{author}{{Hasan}, S.~S.}, \bibinfo{author}{{Zahn}, J.~P.} \&
  \bibinfo{author}{{Christensen-Dalsgaard}, J.}
\newblock \bibinfo{journal}{\bibinfo{title}{{Probing the internal magnetic
  field of slowly pulsating B-stars through g modes}}}.
\newblock {\emph{\JournalTitle{\aap}}} \textbf{\bibinfo{volume}{444}},
  \bibinfo{pages}{L29--L32} (\bibinfo{year}{2005}).

\bibitem{Gomes2020}
\bibinfo{author}{{Gomes}, P.} \& \bibinfo{author}{{Lopes}, I.}
\newblock \bibinfo{journal}{\bibinfo{title}{{Core magnetic field imprint in the
  non-radial oscillations of red giant stars}}}.
\newblock {\emph{\JournalTitle{\mnras}}} \textbf{\bibinfo{volume}{496}},
  \bibinfo{pages}{620--628} (\bibinfo{year}{2020}).

\bibitem{Bugnet2021}
\bibinfo{author}{{Bugnet}, L.} \emph{et~al.}
\newblock \bibinfo{journal}{\bibinfo{title}{{Magnetic signatures on mixed-mode
  frequencies. I. An axisymmetric fossil field inside the core of red
  giants}}}.
\newblock {\emph{\JournalTitle{\aap}}} \textbf{\bibinfo{volume}{650}},
  \bibinfo{pages}{A53} (\bibinfo{year}{2021}).

\bibitem{Loi2021}
\bibinfo{author}{{Loi}, S.~T.}
\newblock \bibinfo{journal}{\bibinfo{title}{{Topology and obliquity of core
  magnetic fields in shaping seismic properties of slowly rotating evolved
  stars}}}.
\newblock {\emph{\JournalTitle{\mnras}}} \textbf{\bibinfo{volume}{504}},
  \bibinfo{pages}{3711--3729} (\bibinfo{year}{2021}).

\bibitem{Borucki2010Sci}
\bibinfo{author}{{Borucki}, W.~J.} \emph{et~al.}
\newblock \bibinfo{journal}{\bibinfo{title}{{Kepler Planet-Detection Mission:
  Introduction and First Results}}}.
\newblock {\emph{\JournalTitle{Science}}} \textbf{\bibinfo{volume}{327}},
  \bibinfo{pages}{977} (\bibinfo{year}{2010}).

\bibitem{Mathis2021}
\bibinfo{author}{{Mathis}, S.} \emph{et~al.}
\newblock \bibinfo{journal}{\bibinfo{title}{{Probing the internal magnetism of
  stars using asymptotic magneto-asteroseismology}}}.
\newblock {\emph{\JournalTitle{\aap}}} \textbf{\bibinfo{volume}{647}},
  \bibinfo{pages}{A122} (\bibinfo{year}{2021}).

\bibitem{Deheuvels2017}
\bibinfo{author}{{Deheuvels}, S.}, \bibinfo{author}{{Ouazzani}, R.~M.} \&
  \bibinfo{author}{{Basu}, S.}
\newblock \bibinfo{journal}{\bibinfo{title}{{Near-degeneracy effects on the
  frequencies of rotationally-split mixed modes in red giants}}}.
\newblock {\emph{\JournalTitle{\aap}}} \textbf{\bibinfo{volume}{605}},
  \bibinfo{pages}{A75} (\bibinfo{year}{2017}).

\bibitem{SM}
\bibinfo{title}{{Supplementary Materials}}.

\bibitem{Dziembowski1992}
\bibinfo{author}{{Dziembowski}, W.~A.} \& \bibinfo{author}{{Goode}, P.~R.}
\newblock \bibinfo{journal}{\bibinfo{title}{{Effects of Differential Rotation
  on Stellar Oscillations: A Second-Order Theory}}}.
\newblock {\emph{\JournalTitle{\apj}}} \textbf{\bibinfo{volume}{394}},
  \bibinfo{pages}{670} (\bibinfo{year}{1992}).

\bibitem{paxton11}
\bibinfo{author}{{Paxton}, B.} \emph{et~al.}
\newblock \bibinfo{journal}{\bibinfo{title}{{Modules for Experiments in Stellar
  Astrophysics (MESA)}}}.
\newblock {\emph{\JournalTitle{\apjs}}} \textbf{\bibinfo{volume}{192}},
  \bibinfo{pages}{3} (\bibinfo{year}{2011}).

\bibitem{mosser18}
\bibinfo{author}{{Mosser}, B.} \emph{et~al.}
\newblock \bibinfo{journal}{\bibinfo{title}{{Period spacings in red giants. IV.
  Toward a complete description of the mixed-mode pattern}}}.
\newblock {\emph{\JournalTitle{\aap}}} \textbf{\bibinfo{volume}{618}},
  \bibinfo{pages}{A109} (\bibinfo{year}{2018}).

\bibitem{takata16}
\bibinfo{author}{{Takata}, M.}
\newblock \bibinfo{journal}{\bibinfo{title}{{Asymptotic analysis of dipolar
  mixed modes of oscillations in red giant stars}}}.
\newblock {\emph{\JournalTitle{\pasj}}} \textbf{\bibinfo{volume}{68}},
  \bibinfo{pages}{109} (\bibinfo{year}{2016}).

\bibitem{Fuller2015}
\bibinfo{author}{{Fuller}, J.}, \bibinfo{author}{{Cantiello}, M.},
  \bibinfo{author}{{Stello}, D.}, \bibinfo{author}{{Garcia}, R.~A.} \&
  \bibinfo{author}{{Bildsten}, L.}
\newblock \bibinfo{journal}{\bibinfo{title}{{Asteroseismology can reveal strong
  internal magnetic fields in red giant stars}}}.
\newblock {\emph{\JournalTitle{Science}}} \textbf{\bibinfo{volume}{350}},
  \bibinfo{pages}{423--426} (\bibinfo{year}{2015}).

\bibitem{Stello2016}
\bibinfo{author}{{Stello}, D.} \emph{et~al.}
\newblock \bibinfo{journal}{\bibinfo{title}{{A prevalence of dynamo-generated
  magnetic fields in the cores of intermediate-mass stars}}}.
\newblock {\emph{\JournalTitle{\nat}}} \textbf{\bibinfo{volume}{529}},
  \bibinfo{pages}{364--367} (\bibinfo{year}{2016}).

\bibitem{Donati2009}
\bibinfo{author}{{Donati}, J.~F.} \& \bibinfo{author}{{Landstreet}, J.~D.}
\newblock \bibinfo{journal}{\bibinfo{title}{{Magnetic Fields of Nondegenerate
  Stars}}}.
\newblock {\emph{\JournalTitle{\araa}}} \textbf{\bibinfo{volume}{47}},
  \bibinfo{pages}{333--370} (\bibinfo{year}{2009}).

\bibitem{Braithwaite2017}
\bibinfo{author}{{Braithwaite}, J.} \& \bibinfo{author}{{Spruit}, H.~C.}
\newblock \bibinfo{journal}{\bibinfo{title}{{Magnetic fields in non-convective
  regions of stars}}}.
\newblock {\emph{\JournalTitle{Royal Society Open Science}}}
  \textbf{\bibinfo{volume}{4}}, \bibinfo{pages}{160271} (\bibinfo{year}{2017}).

\bibitem{Becerra2022}
\bibinfo{author}{{Becerra}, L.}, \bibinfo{author}{{Reisenegger}, A.},
  \bibinfo{author}{{Valdivia}, J.~A.} \& \bibinfo{author}{{Gusakov}, M.~E.}
\newblock \bibinfo{journal}{\bibinfo{title}{{Evolution of random initial
  magnetic fields in stably stratified and barotropic stars}}}.
\newblock {\emph{\JournalTitle{\mnras}}} \textbf{\bibinfo{volume}{511}},
  \bibinfo{pages}{732--745} (\bibinfo{year}{2022}).

\bibitem{cantiello16}
\bibinfo{author}{{Cantiello}, M.}, \bibinfo{author}{{Fuller}, J.} \&
  \bibinfo{author}{{Bildsten}, L.}
\newblock \bibinfo{journal}{\bibinfo{title}{{Asteroseismic Signatures of
  Evolving Internal Stellar Magnetic Fields}}}.
\newblock {\emph{\JournalTitle{\apj}}} \textbf{\bibinfo{volume}{824}},
  \bibinfo{pages}{14} (\bibinfo{year}{2016}).

\bibitem{Brun2005ApJ}
\bibinfo{author}{{Brun}, A.~S.}, \bibinfo{author}{{Browning}, M.~K.} \&
  \bibinfo{author}{{Toomre}, J.}
\newblock \bibinfo{journal}{\bibinfo{title}{{Simulations of Core Convection in
  Rotating A-Type Stars: Magnetic Dynamo Action}}}.
\newblock {\emph{\JournalTitle{\apj}}} \textbf{\bibinfo{volume}{629}},
  \bibinfo{pages}{461--481} (\bibinfo{year}{2005}).

\bibitem{Auriere2007}
\bibinfo{author}{{Auri{\`e}re}, M.} \emph{et~al.}
\newblock \bibinfo{journal}{\bibinfo{title}{{Weak magnetic fields in Ap/Bp
  stars. Evidence for a dipole field lower limit and a tentative interpretation
  of the magnetic dichotomy}}}.
\newblock {\emph{\JournalTitle{\aap}}} \textbf{\bibinfo{volume}{475}},
  \bibinfo{pages}{1053--1065} (\bibinfo{year}{2007}).

\bibitem{Deheuvels2014}
\bibinfo{author}{{Deheuvels}, S.} \emph{et~al.}
\newblock \bibinfo{journal}{\bibinfo{title}{{Seismic constraints on the radial
  dependence of the internal rotation profiles of six Kepler subgiants and
  young red giants}}}.
\newblock {\emph{\JournalTitle{\aap}}} \textbf{\bibinfo{volume}{564}},
  \bibinfo{pages}{A27} (\bibinfo{year}{2014}).

\bibitem{mosser15}
\bibinfo{author}{{Mosser}, B.}, \bibinfo{author}{{Vrard}, M.},
  \bibinfo{author}{{Belkacem}, K.}, \bibinfo{author}{{Deheuvels}, S.} \&
  \bibinfo{author}{{Goupil}, M.~J.}
\newblock \bibinfo{journal}{\bibinfo{title}{{Period spacings in red giants. I.
  Disentangling rotation and revealing core structure discontinuities}}}.
\newblock {\emph{\JournalTitle{\aap}}} \textbf{\bibinfo{volume}{584}},
  \bibinfo{pages}{A50} (\bibinfo{year}{2015}).

\setcounter{mainbib}{\value{enumiv}}
\end{thebibliography}

\begin{thebibliography}{10}
\urlstyle{rm}
\expandafter\ifx\csname url\endcsname\relax
  \def\url#1{\texttt{#1}}\fi
\expandafter\ifx\csname urlprefix\endcsname\relax\def\urlprefix{URL }\fi
\expandafter\ifx\csname doiprefix\endcsname\relax\def\doiprefix{DOI: }\fi
\providecommand{\bibinfo}[2]{#2}
\providecommand{\eprint}[2][]{\url{#2}}

\setcounter{enumiv}{\value{mainbib}}

\bibitem{Yu2018}
\bibinfo{author}{{Yu}, J.} \emph{et~al.}
\newblock \bibinfo{journal}{\bibinfo{title}{{Asteroseismology of 16,000 Kepler
  Red Giants: Global Oscillation Parameters, Masses, and Radii}}}.
\newblock {\emph{\JournalTitle{\apjs}}} \textbf{\bibinfo{volume}{236}},
  \bibinfo{pages}{42} (\bibinfo{year}{2018}).

\bibitem{Garcia11}
\bibinfo{author}{{Garc{\'\i}a}, R.~A.} \emph{et~al.}
\newblock \bibinfo{journal}{\bibinfo{title}{{Preparation of Kepler light curves
  for asteroseismic analyses}}}.
\newblock {\emph{\JournalTitle{\mnras}}} \textbf{\bibinfo{volume}{414}},
  \bibinfo{pages}{L6--L10} (\bibinfo{year}{2011}).

\bibitem{Lomb1976}
\bibinfo{author}{{Lomb}, N.~R.}
\newblock \bibinfo{journal}{\bibinfo{title}{{Least-Squares Frequency Analysis
  of Unequally Spaced Data}}}.
\newblock {\emph{\JournalTitle{\apss}}} \textbf{\bibinfo{volume}{39}},
  \bibinfo{pages}{447--462} (\bibinfo{year}{1976}).

\bibitem{Scargle1982}
\bibinfo{author}{{Scargle}, J.~D.}
\newblock \bibinfo{journal}{\bibinfo{title}{{Studies in astronomical time
  series analysis. II. Statistical aspects of spectral analysis of unevenly
  spaced data.}}}
\newblock {\emph{\JournalTitle{\apj}}} \textbf{\bibinfo{volume}{263}},
  \bibinfo{pages}{835--853} (\bibinfo{year}{1982}).

\bibitem{Kjeldsen1995}
\bibinfo{author}{{Kjeldsen}, H.} \& \bibinfo{author}{{Bedding}, T.~R.}
\newblock \bibinfo{journal}{\bibinfo{title}{{Amplitudes of stellar
  oscillations: the implications for asteroseismology.}}}
\newblock {\emph{\JournalTitle{\aap}}} \textbf{\bibinfo{volume}{293}},
  \bibinfo{pages}{87--106} (\bibinfo{year}{1995}).

\bibitem{Huber2009}
\bibinfo{author}{{Huber}, D.} \emph{et~al.}
\newblock \bibinfo{journal}{\bibinfo{title}{{Automated extraction of
  oscillation parameters for Kepler observations of solar-type stars}}}.
\newblock {\emph{\JournalTitle{Communications in Asteroseismology}}}
  \textbf{\bibinfo{volume}{160}}, \bibinfo{pages}{74} (\bibinfo{year}{2009}).

\bibitem{Chontos2021}
\bibinfo{author}{{Chontos}, A.}, \bibinfo{author}{{Sayeed}, M.} \&
  \bibinfo{author}{{Huber}, D.}
\newblock \bibinfo{title}{{pySYD: Automated Measurements of Global
  Asteroseismic Parameters}}.
\newblock In \emph{\bibinfo{booktitle}{Posters from the TESS Science Conference
  II (TSC2)}}, \bibinfo{pages}{189} (\bibinfo{year}{2021}).

\bibitem{Anderson1990ApJ}
\bibinfo{author}{{Anderson}, E.~R.}, \bibinfo{author}{{Duvall}, J., Thomas~L.}
  \& \bibinfo{author}{{Jefferies}, S.~M.}
\newblock \bibinfo{journal}{\bibinfo{title}{{Modeling of Solar Oscillation
  Power Spectra}}}.
\newblock {\emph{\JournalTitle{\apj}}} \textbf{\bibinfo{volume}{364}},
  \bibinfo{pages}{699} (\bibinfo{year}{1990}).

\bibitem{goupil13}
\bibinfo{author}{{Goupil}, M.~J.} \emph{et~al.}
\newblock \bibinfo{journal}{\bibinfo{title}{{Seismic diagnostics for transport
  of angular momentum in stars. II. Interpreting observed rotational splittings
  of slowly rotating red giant stars}}}.
\newblock {\emph{\JournalTitle{\aap}}} \textbf{\bibinfo{volume}{549}},
  \bibinfo{pages}{A75} (\bibinfo{year}{2013}).

\bibitem{corsaro14}
\bibinfo{author}{{Corsaro}, E.} \& \bibinfo{author}{{De Ridder}, J.}
\newblock \bibinfo{journal}{\bibinfo{title}{{DIAMONDS: A new Bayesian nested
  sampling tool. Application to peak bagging of solar-like oscillations}}}.
\newblock {\emph{\JournalTitle{\aap}}} \textbf{\bibinfo{volume}{571}},
  \bibinfo{pages}{A71} (\bibinfo{year}{2014}).

\bibitem{shibahashi79}
\bibinfo{author}{{Shibahashi}, H.}
\newblock \bibinfo{journal}{\bibinfo{title}{{Modal Analysis of Stellar
  Nonradial Oscillations by an Asymptotic Method}}}.
\newblock {\emph{\JournalTitle{\pasj}}} \textbf{\bibinfo{volume}{31}},
  \bibinfo{pages}{87--104} (\bibinfo{year}{1979}).

\bibitem{mosser12a}
\bibinfo{author}{{Mosser}, B.} \emph{et~al.}
\newblock \bibinfo{journal}{\bibinfo{title}{{Probing the core structure and
  evolution of red giants using gravity-dominated mixed modes observed with
  Kepler}}}.
\newblock {\emph{\JournalTitle{\aap}}} \textbf{\bibinfo{volume}{540}},
  \bibinfo{pages}{A143} (\bibinfo{year}{2012}).

\bibitem{hekker17}
\bibinfo{author}{{Hekker}, S.} \& \bibinfo{author}{{Christensen-Dalsgaard}, J.}
\newblock \bibinfo{journal}{\bibinfo{title}{{Giant star seismology}}}.
\newblock {\emph{\JournalTitle{\aapr}}} \textbf{\bibinfo{volume}{25}},
  \bibinfo{pages}{1} (\bibinfo{year}{2017}).

\bibitem{LyndenBell1967}
\bibinfo{author}{{Lynden-Bell}, D.} \& \bibinfo{author}{{Ostriker}, J.~P.}
\newblock \bibinfo{journal}{\bibinfo{title}{{On the stability of differentially
  rotating bodies}}}.
\newblock {\emph{\JournalTitle{\mnras}}} \textbf{\bibinfo{volume}{136}},
  \bibinfo{pages}{293} (\bibinfo{year}{1967}).

\end{thebibliography}
\renewcommand{\refname}{Supplementary References}

\clearpage
\setcounter{figure}{0}
\renewcommand{\figurename}{Supplementary Figure}
\renewcommand{\tablename}{Supplementary Table}
\section*{Supplementary Figures}
\vfill
\begin{figure}[!h]
    \centering
    \includegraphics[width = \linewidth]{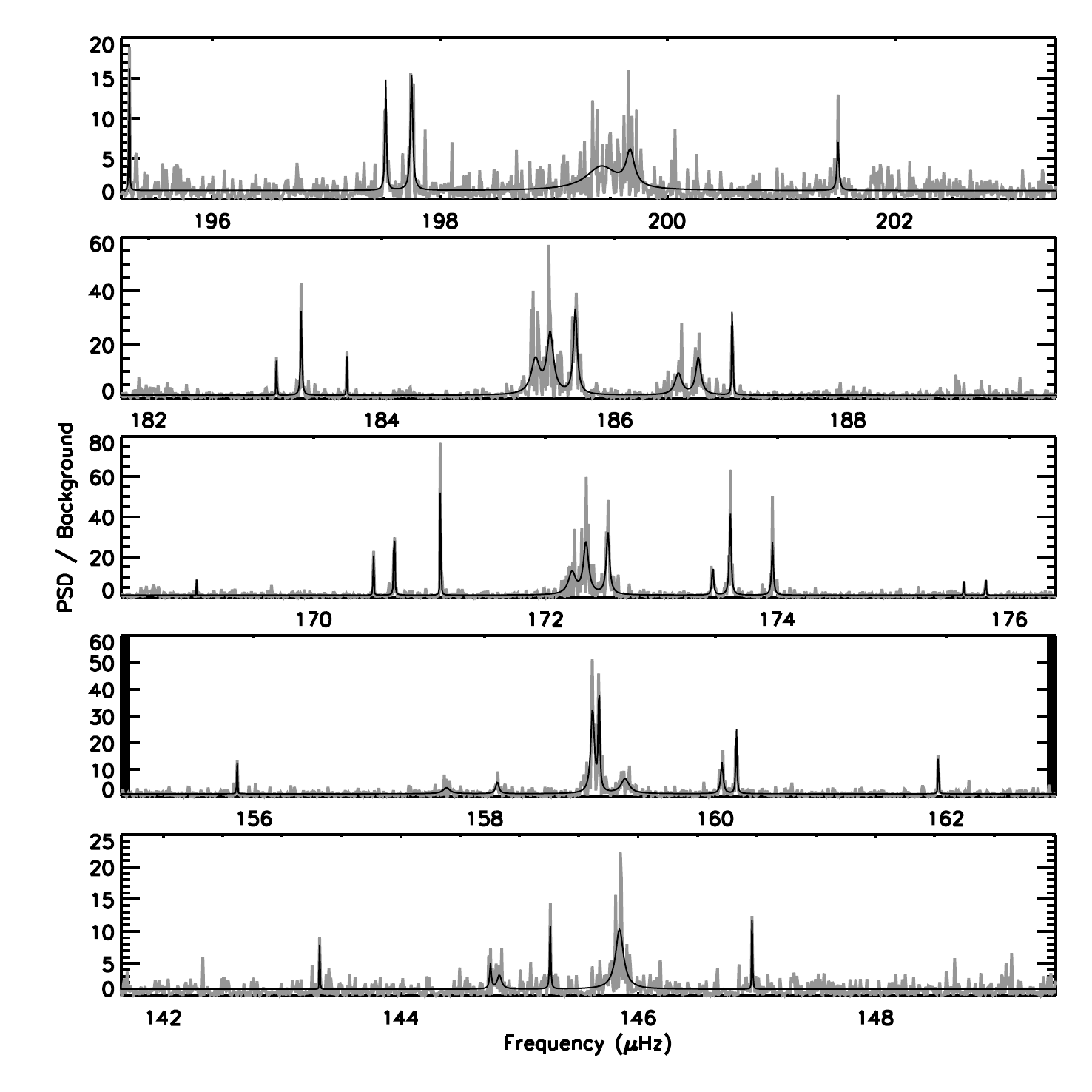}
    \caption{\textbf{Power spectrum density (PSD) of KIC\,8684542.} The grey curve corresponds to the PSD divided by the background noise. The black curve indicates the optimal fit to the PSD obtained in Sect. \ref{sect_analysis}.}\label{fig_8684542_mode_fit}
\end{figure}
\vfill
\clearpage
\begin{figure}
    \centering
    \includegraphics[width = \linewidth]{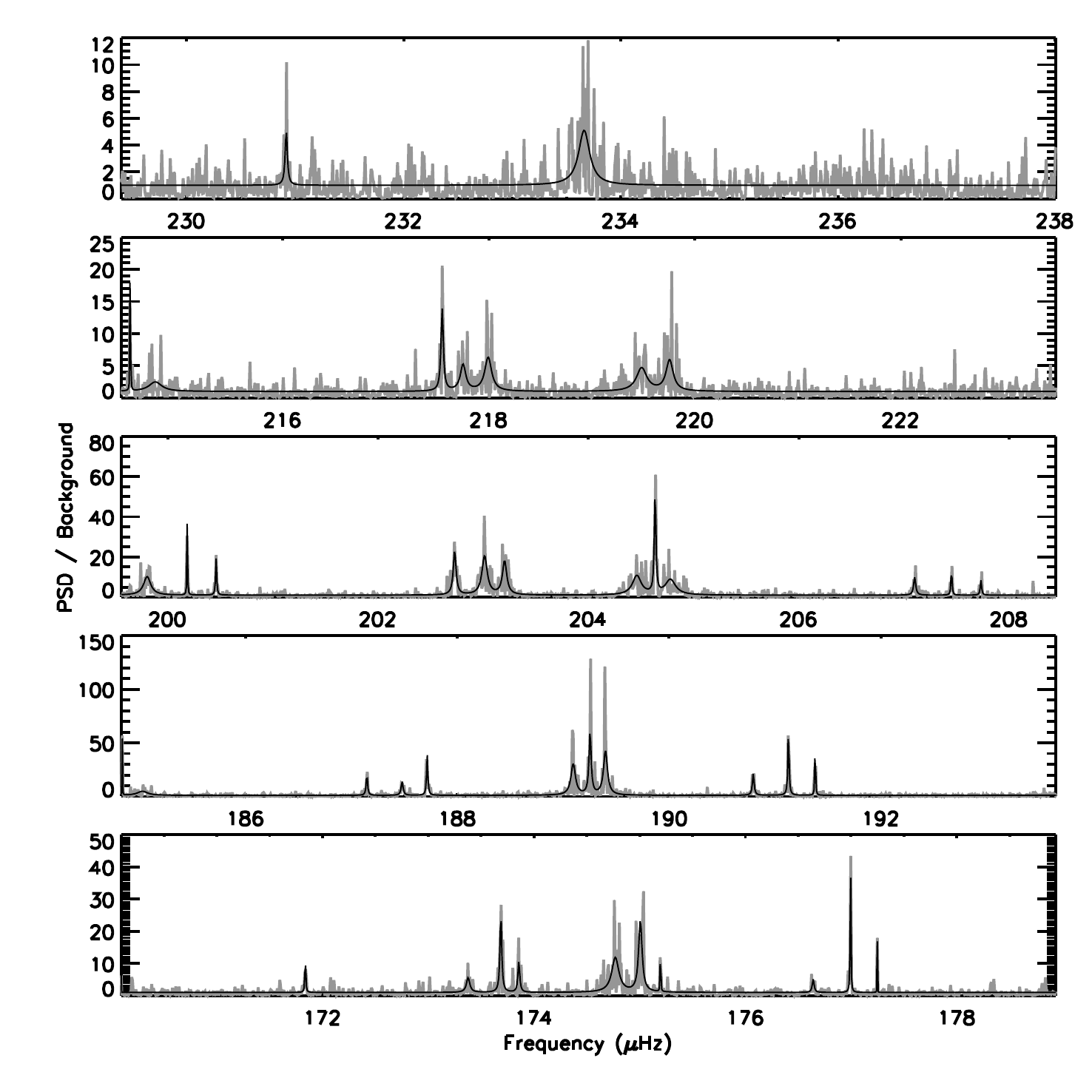}
    \caption{\textbf{Power spectrum density of KIC\,11515377.} Symbols have the same meaning as in Supplementary Figure \ref{fig_8684542_mode_fit}.}\label{fig_11515377_mode_fit}
\end{figure}

\clearpage
\begin{figure}
    \centering
    \includegraphics[width = \linewidth]{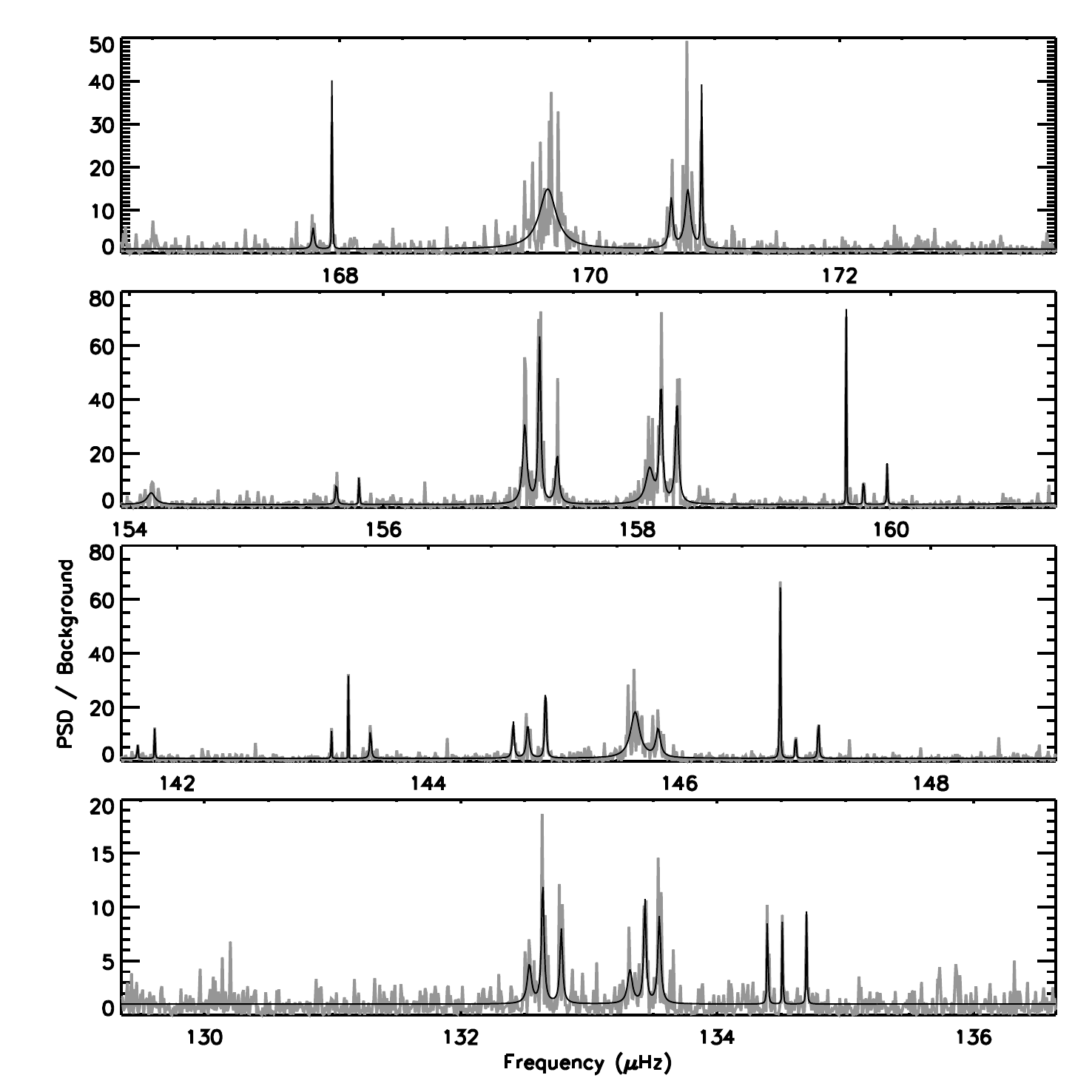}
    \caption{\textbf{Power spectrum density of KIC\,7518143.} Symbols have the same meaning as in Supplementary Figure \ref{fig_8684542_mode_fit}.}\label{fig_7518143_mode_fit}
\end{figure}

\clearpage
\begin{figure}
\begin{center}
\includegraphics[width=0.5\linewidth]{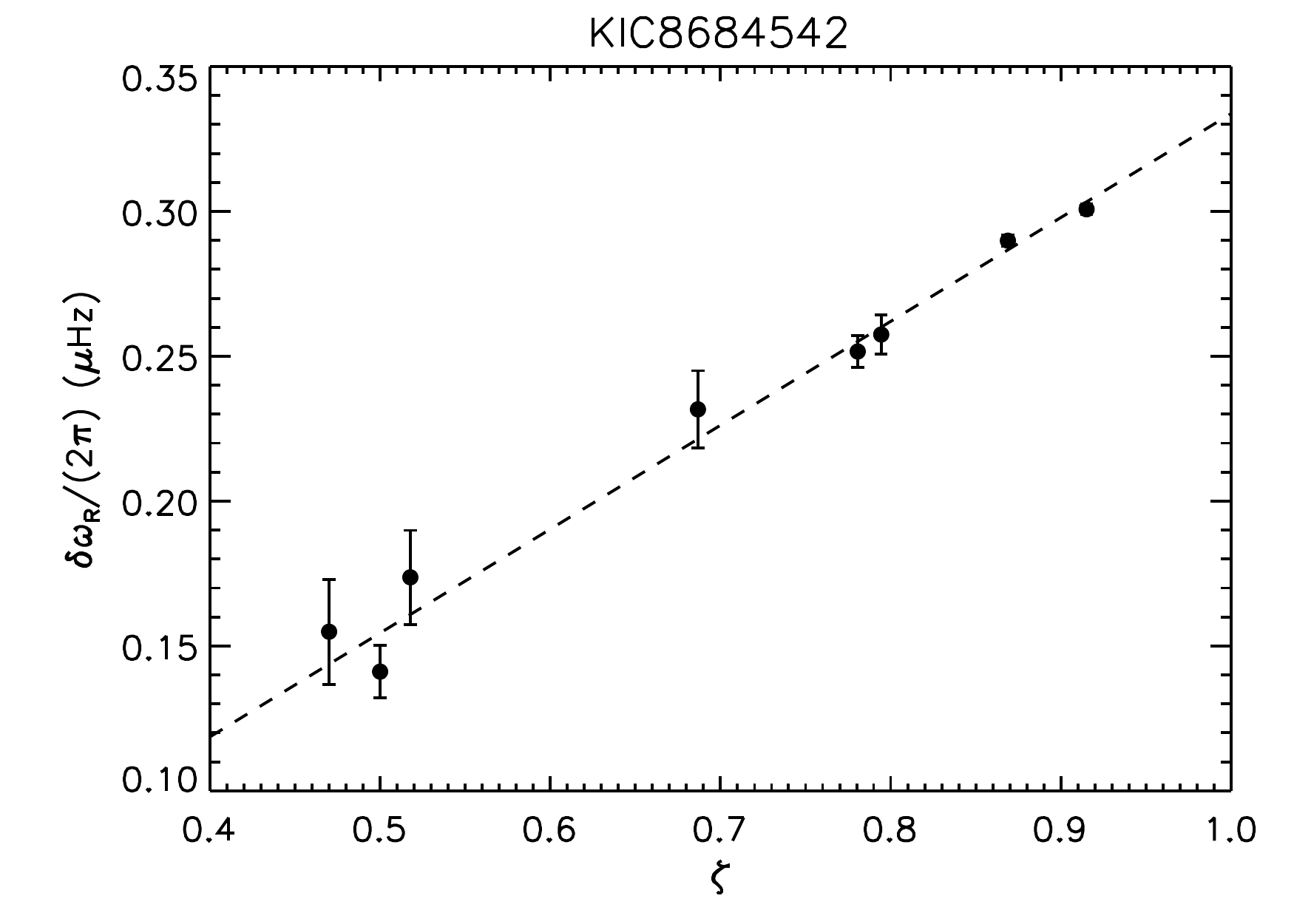} % %6cm
\includegraphics[width=0.5\linewidth]{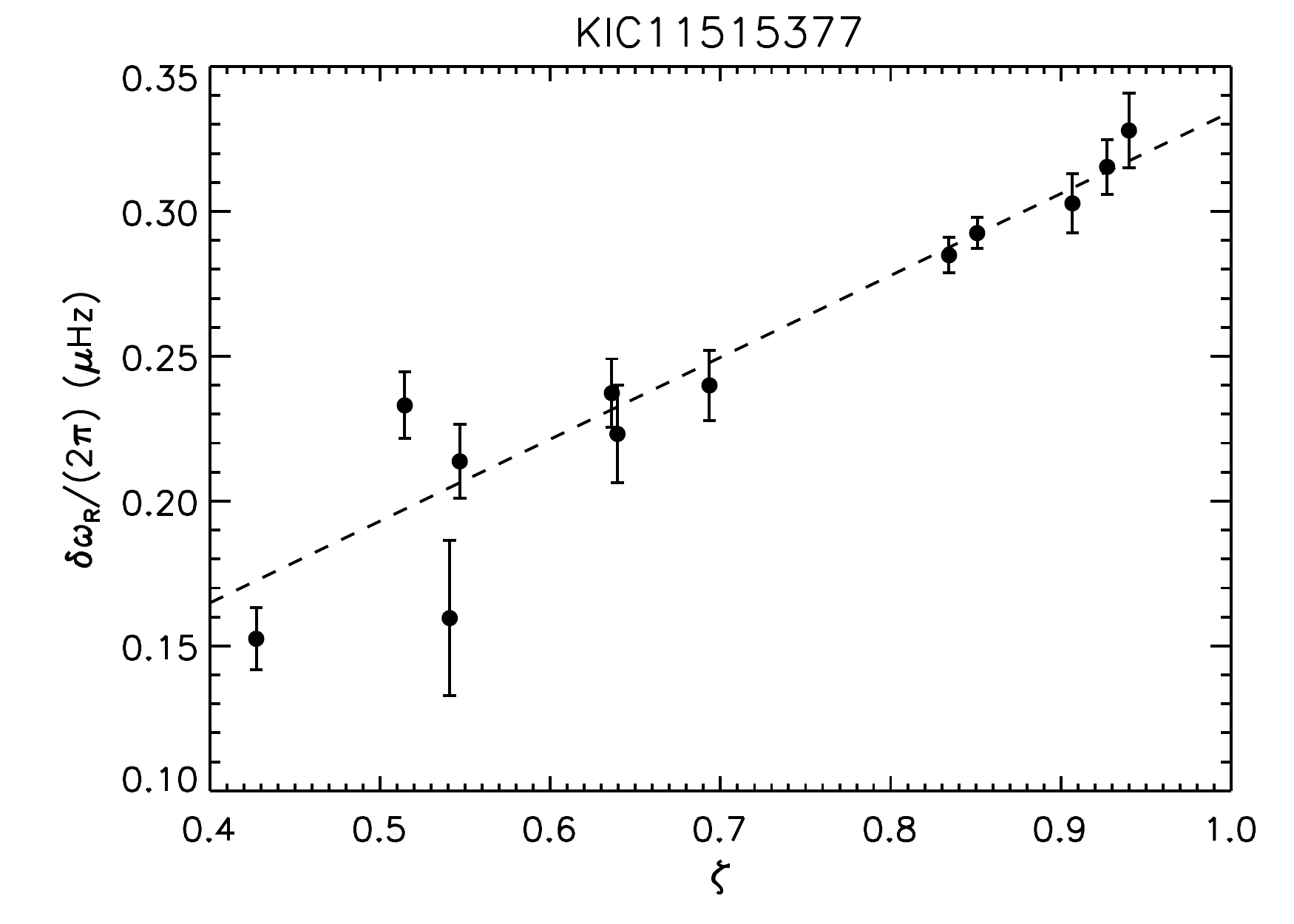}%
\includegraphics[width=0.5\linewidth]{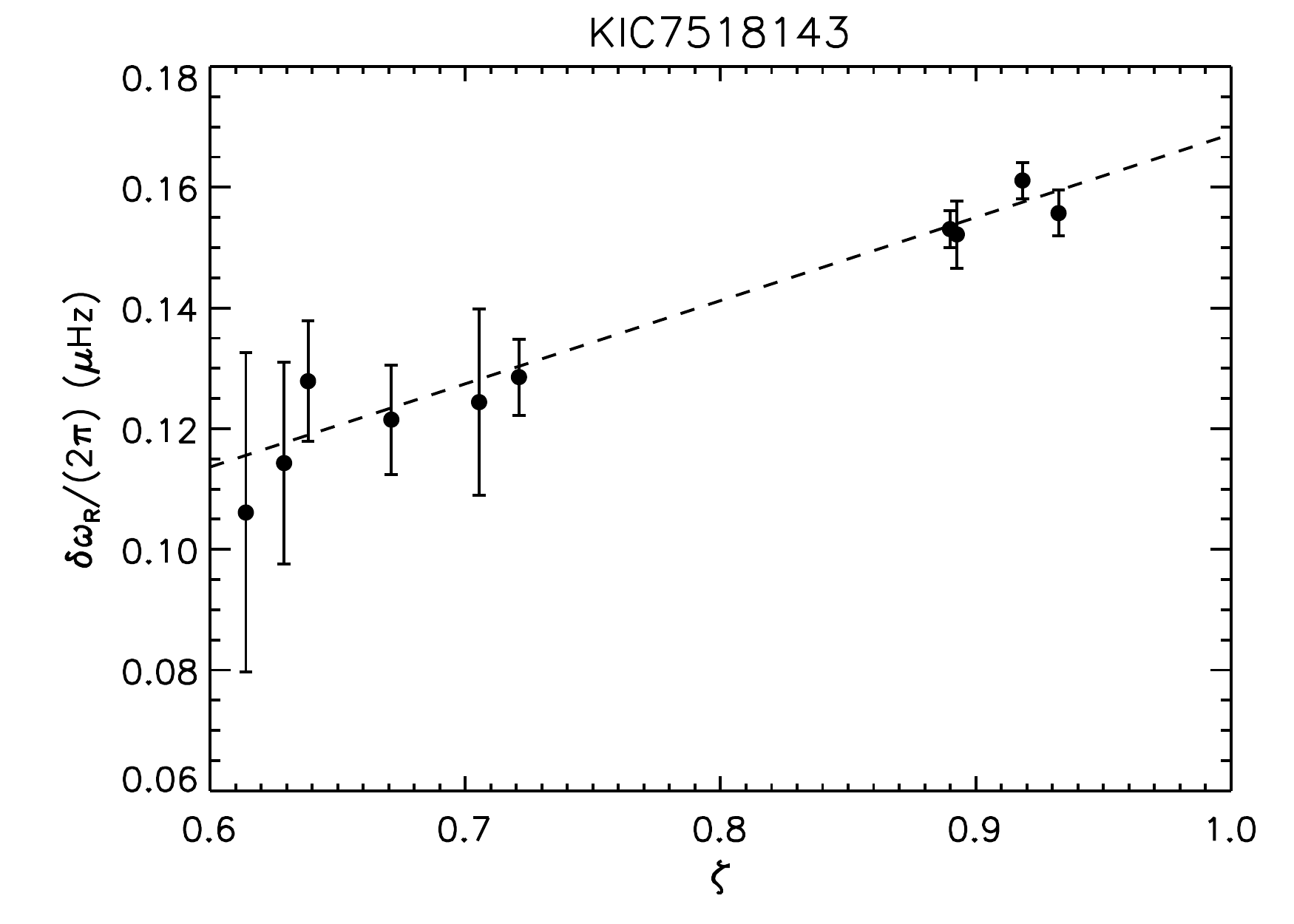}
\end{center}
\caption{\textbf{Observed rotational splittings as a function of the mode trapping.} The mode trapping is measured by the parameter $\zeta$  ($\zeta\rightarrow 0$ for pure p modes, and $\zeta\rightarrow 1$ for pure g modes). The dashed lines correspond to the linear fits.  %of $\delta\nu_{\rm R}(\zeta)$.
\label{fig_zeta_split}}
\end{figure}

\clearpage
\begin{figure}
\begin{center}
\includegraphics[width=0.5\linewidth]{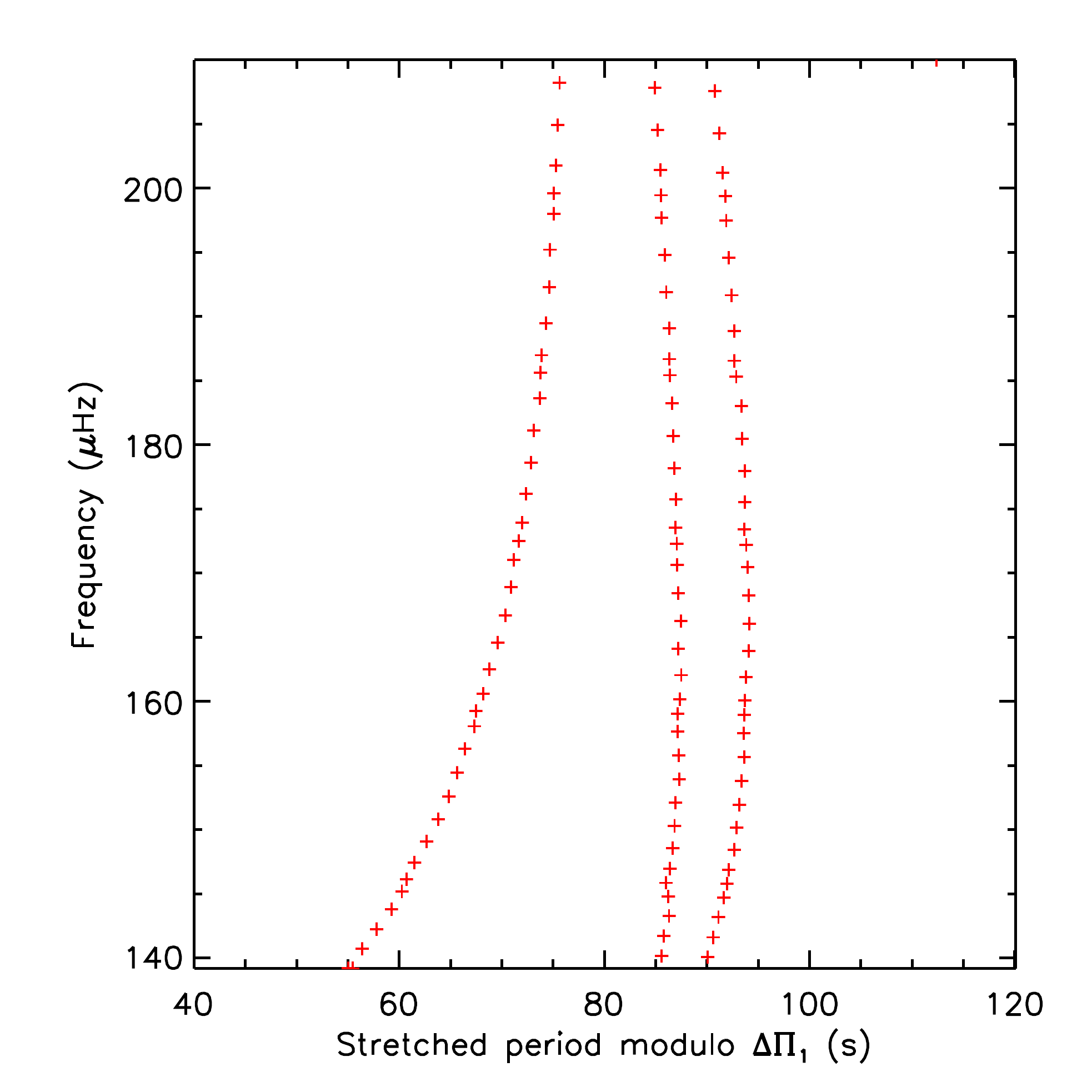}%
\includegraphics[width=0.5\linewidth]{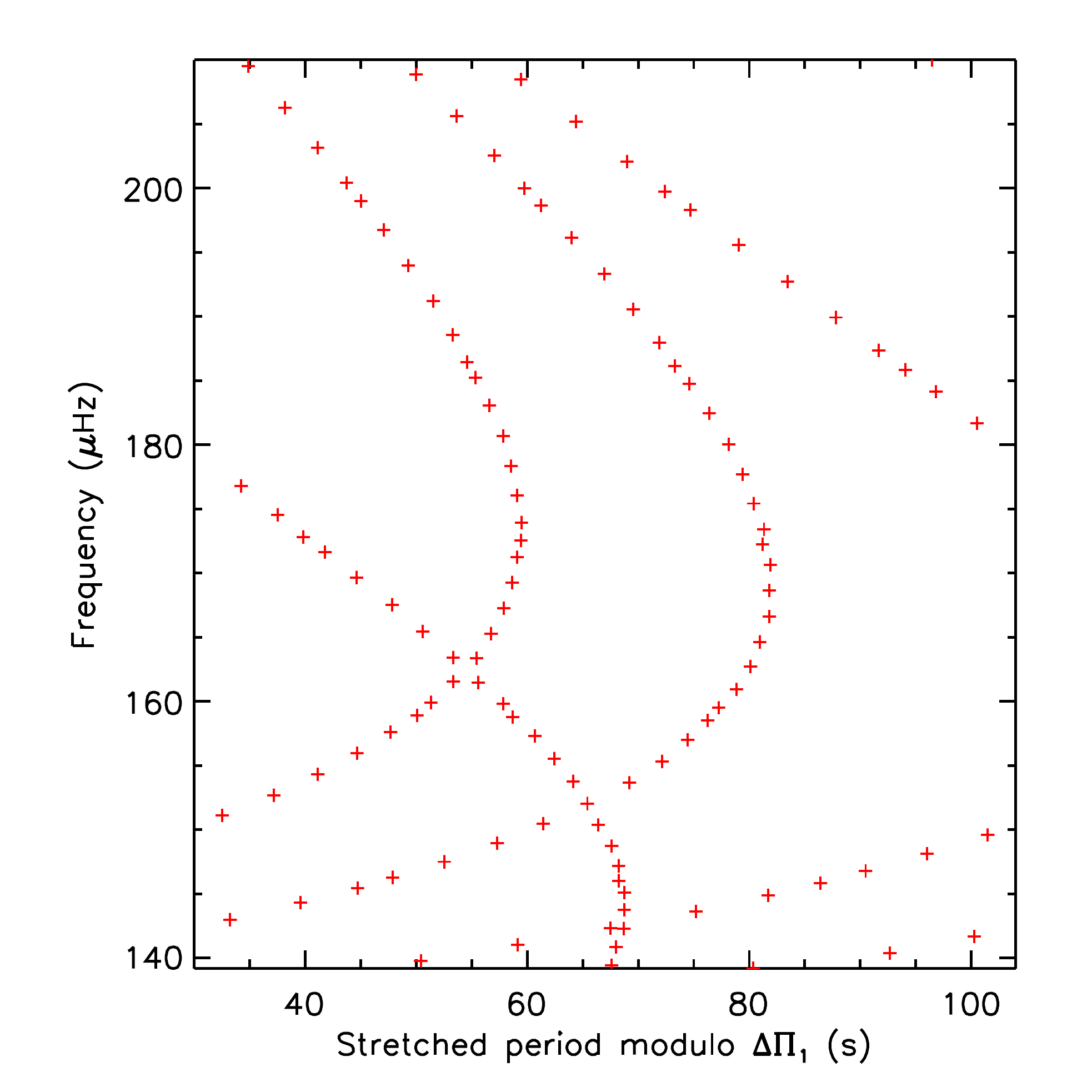}
\end{center}
\caption{\textbf{Stretched period \'echelle diagrams of asymptotic mixed mode frequencies for two different field strengths.} The left panel corresponds to a moderate magnetic shift ($\delta\!\omega_\mathrm{g}/(2\pi) = 0.2\,\mathrm{\mu Hz}$) and the right panel to a strong magnetic shift ($\delta\!\omega_\mathrm{g}/(2\pi) = 2\,\mathrm{\mu Hz}$).
\label{fig_stretch_ex}}
\end{figure}

\clearpage\begin{figure}
\begin{center}
\includegraphics[width=0.7\linewidth]{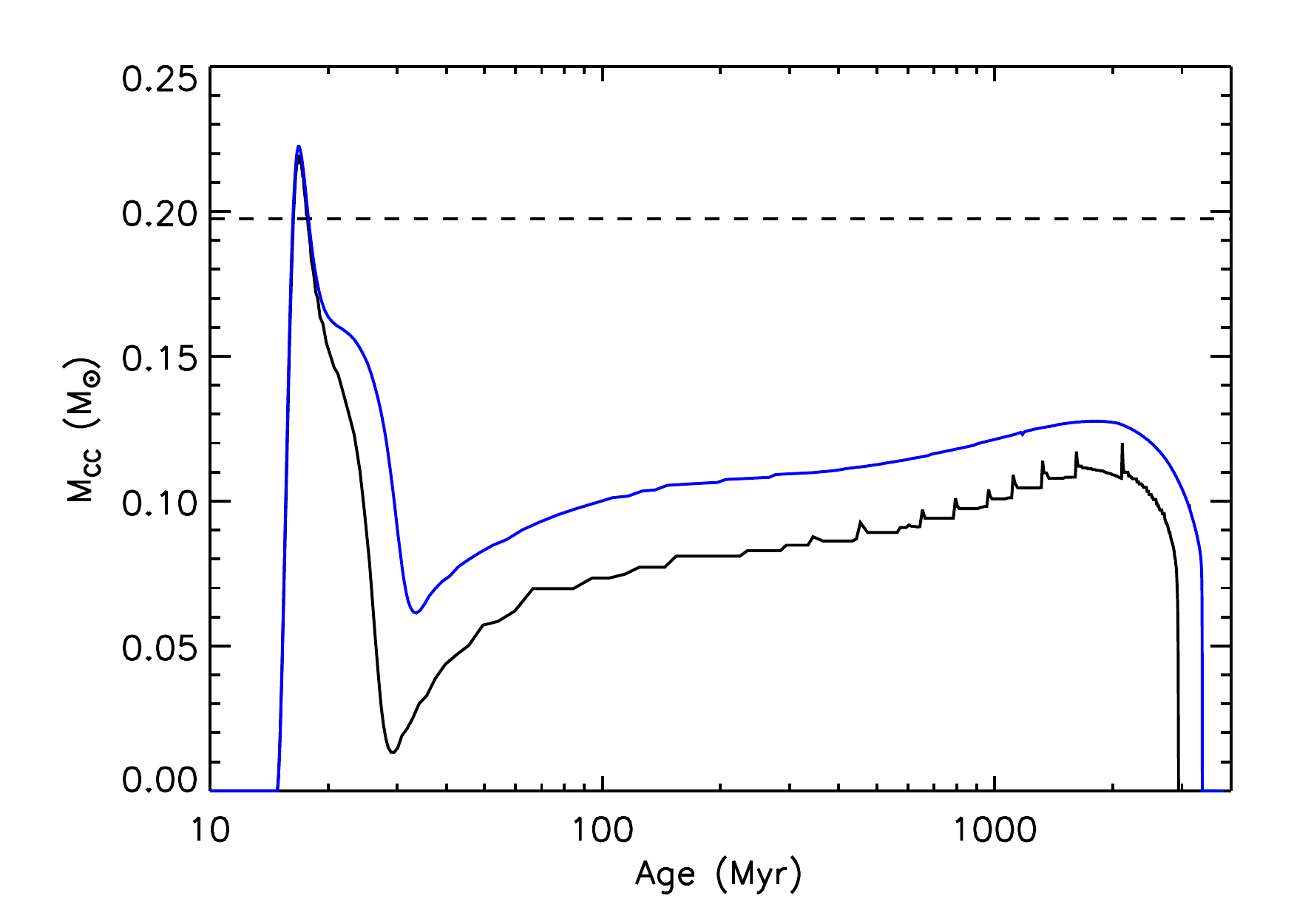}
\end{center}
\caption{\textbf{Evolution of convective core mass during the main sequence for the best-fit model of KIC\,11515377.} Models were computed without core overshooting (black curve), or with overshooting extending over $0.2\,H_p$, where $H_p$ is the pressure scale height (blue curve). The dashed line indicates the layer of the hydrogen burning shell at current age.
\label{fig_evol_conv_core}}
\end{figure}

\clearpage
\section*{Supplementary Tables}
\vfill
\begin{table}[!h]
\begin{center}
\caption{\textbf{Properties of p and g modes of KIC\,8684542, KIC\,11515377, and KIC\,7518143.} These values were obtained by fitting an asymptotic expression of mixed modes to the $m=0$ observed frequencies (magnetic perturbations are not included at this stage). The last two lines give masses and radii derived from seismic scaling relations\cite{Yu2018}. \label{tab_wkb}}
\begin{tabular}{l c c c}
\hline \hline
\T\B parameter & KIC\,8684542 & KIC\,11515377 & KIC\,7518143 \\
\hline
%\T $\dn$ ($\mu$Hz) & $13.46\pm0.17$ & $14.77\pm0.17$ & $12.33\pm0.17$ \\
%$\dpun$ (s) & 80.0 & 83.3 & 78.5 \\
%$\varepsilon_{\rm g}$ & 0.59 & 0.46 & 0.31 \\
%$q$ & 0.126 & 0.155 & 0.120 \\
%\B $d_{01}$ & 0.66 & 0.61 & 0.65 \\
%\T $\dn$ ($\mu$Hz) & $13.46\pm0.17$ & $14.77\pm0.17$ & $12.33\pm0.17$ \\
\T $\numax$ ($\mu$Hz) & $176.8\pm1.1$ & $191.6\pm1.0$ & $156.3\pm0.9$ \\
$\dn$ ($\mu$Hz) & $13.484\pm0.006$ & $14.946\pm0.008$ & $12.374\pm0.008$ \\
$\alpha$ & $(3.2\pm0.4)\times10^{-3}$ & $(9.1\pm0.3)\times10^{-3}$ & $(3.1\pm0.4)\times10^{-3}$ \\
$\varepsilon_{\rm p}$ & $1.275\pm0.006$ & $1.149\pm0.008$ & $1.234\pm0.009$ \\
$\dpun$ (s) & $80.16\pm0.02$ & $83.19\pm0.04$ & $78.49\pm0.03$ \\
$\varepsilon_{\rm g}$ & $0.50\pm0.02$ & $0.50\pm0.03$ & $0.30\pm0.03$ \\
$q$ & $0.122\pm0.04$ & $0.148\pm0.004$ & $0.117\pm0.004$ \\
\B $d_{01}$ ($\mu$Hz) & $-0.11\pm0.02$ & $0.09\pm0.02$ & $-0.08\pm0.02$ \\
\hline
\T $M/M_\odot$ & $1.43\pm0.07$ & $1.60\pm0.09$ & $1.50\pm0.08$ \\ 
\B $R/R_\odot$ & $5.23\pm0.09$ & $5.16\pm0.09$ & $5.65\pm0.11$ \\ 
\hline
\end{tabular}
\end{center}
\end{table}
\vfill

\clearpage
\begin{table}
\begin{center}
\caption{\textbf{Extracted mode frequencies for KIC\,8684542, KIC\,11515377 and KIC\,7518143.}
%The first column lists the observed frequencies of $m=0$ components. $\zeta$ is g-mode kinetic energy over the total mode kinetic energy. $\delta\!\omega_\mathrm{R}/2\pi$ is the rotational splitting in unit of ordinary frequency, while $\delta_\mathrm{asym}/2\pi$ is the asymmetry in unit of ordinary frequency.
\label{tab_freq}}
\begin{tabular}[t]{c}
\hline \hline
KIC\,8684542 \\
\hline
$ 143.318 \pm 0.003 $ \\
$ 144.756 \pm 0.009 $ \\
$ 144.829 \pm 0.033 $ \\
$ 145.259 \pm 0.006 $ \\
$ 145.844 \pm 0.012 $ \\
$ 146.963 \pm 0.005 $ \\
$ 155.852 \pm 0.001 $ \\
$ 157.669 \pm 0.020 $ \\
$ 158.110 \pm 0.011 $ \\
$ 158.935 \pm 0.008 $ \\
$ 158.992 \pm 0.005 $ \\
$ 159.217 \pm 0.016 $ \\
$ 160.057 \pm 0.006 $ \\
$ 160.182 \pm 0.004 $ \\
$ 161.934 \pm 0.002 $ \\
$ 170.516 \pm 0.001 $ \\
$ 170.696 \pm 0.002 $ \\
$ 171.095 \pm 0.001 $ \\
$ 172.229 \pm 0.017 $ \\
$ 172.350 \pm 0.010 $ \\
$ 172.539 \pm 0.007 $ \\
$ 173.443 \pm 0.005 $ \\
$ 173.594 \pm 0.004 $ \\
$ 173.958 \pm 0.004 $ \\
$ 175.608 \pm 0.002 $ \\
$ 175.797 \pm 0.001 $ \\
$ 180.760 \pm 0.001 $ \\
$ 183.099 \pm 0.002 $ \\
$ 183.310 \pm 0.004 $ \\
$ 183.700 \pm 0.001 $ \\
$ 185.315 \pm 0.014 $ \\
$ 185.446 \pm 0.013 $ \\
$ 185.663 \pm 0.007 $ \\
$ 186.547 \pm 0.013 $ \\
$ 186.718 \pm 0.009 $ \\
$ 187.011 \pm 0.002 $ \\
$ 194.882 \pm 0.002 $ \\
$ 195.270 \pm 0.001 $ \\
$ 197.524 \pm 0.005 $ \\
$ 197.751 \pm 0.006 $ \\
$ 199.420 \pm 0.051 $ \\
$ 199.672 \pm 0.023 $ \\
$ 201.499 \pm 0.007 $ \\
\hline
\end{tabular}
\begin{tabular}[t]{c}
\hline \hline
KIC\,11515377\\
\hline
$ 169.540 \pm 0.001 $ \\
$ 171.837 \pm 0.003 $ \\
$ 173.377 \pm 0.010 $ \\
$ 173.687 \pm 0.005 $ \\
$ 173.857 \pm 0.006 $ \\
$ 174.770 \pm 0.013 $ \\
$ 175.006 \pm 0.008 $ \\
$ 175.197 \pm 0.002 $ \\
$ 176.641 \pm 0.010 $ \\
$ 176.994 \pm 0.003 $ \\
$ 177.247 \pm 0.001 $ \\
$ 179.111 \pm 0.002 $ \\
$ 179.491 \pm 0.004 $ \\
$ 184.837 \pm 0.001 $ \\
$ 187.146 \pm 0.004 $ \\
$ 187.480 \pm 0.007 $ \\
$ 187.716 \pm 0.004 $ \\
$ 189.094 \pm 0.008 $ \\
$ 189.254 \pm 0.005 $ \\
$ 189.399 \pm 0.007 $ \\
$ 190.795 \pm 0.005 $ \\
$ 191.127 \pm 0.004 $ \\
$ 191.380 \pm 0.002 $ \\
$ 193.974 \pm 0.001 $ \\
$ 194.231 \pm 0.005 $ \\
$ 199.804 \pm 0.012 $ \\
$ 200.185 \pm 0.002 $ \\
$ 200.460 \pm 0.004 $ \\
$ 202.728 \pm 0.007 $ \\
$ 203.013 \pm 0.010 $ \\
$ 203.203 \pm 0.010 $ \\
$ 204.461 \pm 0.018 $ \\
$ 204.634 \pm 0.005 $ \\
$ 204.780 \pm 0.020 $ \\
$ 207.103 \pm 0.007 $ \\
$ 207.453 \pm 0.004 $ \\
$ 207.733 \pm 0.007 $ \\
$ 214.520 \pm 0.002 $ \\
$ 214.763 \pm 0.040 $ \\
$ 217.547 \pm 0.007 $ \\
$ 217.750 \pm 0.018 $ \\
$ 217.994 \pm 0.015 $ \\
$ 219.287 \pm 0.015 $ \\
$ 219.491 \pm 0.018 $ \\
$ 219.753 \pm 0.018 $ \\
$ 230.918 \pm 0.009 $ \\
$ 233.662 \pm 0.019 $ \\
\hline
\end{tabular}
\begin{tabular}[t]{c}
\hline \hline
KIC\,7518143\\
\hline
$ 132.535 \pm 0.013 $ \\
$ 132.640 \pm 0.006 $ \\
$ 132.784 \pm 0.008 $ \\
$ 133.320 \pm 0.015 $ \\
$ 133.437 \pm 0.007 $ \\
$ 133.549 \pm 0.008 $ \\
$ 134.393 \pm 0.003 $ \\
$ 134.511 \pm 0.002 $ \\
$ 134.698 \pm 0.005 $ \\
$ 141.684 \pm 0.002 $ \\
$ 141.823 \pm 0.002 $ \\
$ 143.228 \pm 0.001 $ \\
$ 143.363 \pm 0.001 $ \\
$ 143.539 \pm 0.004 $ \\
$ 144.676 \pm 0.005 $ \\
$ 144.792 \pm 0.007 $ \\
$ 144.933 \pm 0.004 $ \\
$ 145.646 \pm 0.013 $ \\
$ 145.830 \pm 0.014 $ \\
$ 146.800 \pm 0.001 $ \\
$ 146.925 \pm 0.001 $ \\
$ 147.106 \pm 0.003 $ \\
$ 155.630 \pm 0.005 $ \\
$ 155.810 \pm 0.002 $ \\
$ 157.115 \pm 0.007 $ \\
$ 157.232 \pm 0.005 $ \\
$ 157.371 \pm 0.007 $ \\
$ 158.102 \pm 0.026 $ \\
$ 158.189 \pm 0.007 $ \\
$ 158.315 \pm 0.006 $ \\
$ 159.649 \pm 0.002 $ \\
$ 159.786 \pm 0.001 $ \\
$ 159.971 \pm 0.002 $ \\
$ 161.924 \pm 0.184 $ \\
$ 167.786 \pm 0.008 $ \\
$ 167.937 \pm 0.004 $ \\
$ 169.664 \pm 0.016 $ \\
$ 170.653 \pm 0.008 $ \\
$ 170.786 \pm 0.010 $ \\
$ 170.896 \pm 0.004 $ \\
\hline
\end{tabular}
\end{center}
\end{table}

\clearpage
\begin{table}
\begin{center}
\caption{\textbf{Extracted seismic parameters for KIC\,8684542.}
%The first column lists the observed frequencies of $m=0$ components. $\zeta$ is g-mode kinetic energy over the total mode kinetic energy. $\delta\!\omega_\mathrm{R}/2\pi$ is the rotational splitting in unit of ordinary frequency, while $\delta_\mathrm{asym}/2\pi$ is the asymmetry in unit of ordinary frequency.
\label{tab_868}}
\begin{tabular}{c c c c}
\hline \hline
\rule[-2ex]{0pt}{5.4ex} %$\nu_{m=0} \, (\mu$Hz) 
$\dfrac{\omega_{m=0}}{2\pi}$ ($\mu$Hz)
& $\zeta$ & $\dfrac{\delta\!\omega_{\rm R}}{2\pi}$ (nHz) & $\dfrac{\delta_{\rm asym}}{2\pi}$ (nHz) \\
\hline
\T $ 144.829 \pm 0.033$ & $0.781$ & $ 251.6 \pm   5.4$ & $356.6 \pm  66.2$ \\
$ 157.669 \pm 0.020$ & $0.839$ & - & $328.9 \pm  42.7$ \\
$ 158.992 \pm 0.005$ & $0.500$ & $ 141.1 \pm   9.0$ & $169.3 \pm  20.2$ \\
$ 160.182 \pm 0.004$ & $0.819$ & - & $287.8 \pm  11.8$ \\
$ 170.696 \pm 0.002$ & $0.869$ & $ 289.9 \pm   1.8$ & $219.8 \pm   4.7$ \\
$ 172.350 \pm 0.010$ & $0.470$ & $ 154.9 \pm  18.1$ & $ 68.7 \pm  27.3$ \\
$ 173.594 \pm 0.004$ & $0.794$ & $ 257.5 \pm   6.8$ & $213.7 \pm  10.8$ \\
$ 175.797 \pm 0.001$ & $0.951$ & - & $255.3 \pm   3.3$ \\
$ 183.310 \pm 0.004$ & $0.915$ & $ 300.7 \pm   1.9$ & $178.8 \pm   7.3$ \\
$ 185.446 \pm 0.013$ & $0.518$ & $ 173.7 \pm  16.2$ & $ 85.7 \pm  31.4$ \\
$ 186.718 \pm 0.009$ & $0.687$ & $ 231.7 \pm  13.4$ & $121.7 \pm  23.2$ \\
$ 194.882 \pm 0.002$ & $0.956$ & - & $139.2 \pm   3.7$ \\
\B $ 197.751 \pm 0.006$ & $0.815$ & - & $ 81.1 \pm  14.0$ \\
\hline
\end{tabular}
\end{center}
\end{table}

\clearpage
\begin{table}
\begin{center}
\caption{\textbf{Extracted seismic parameters for KIC\,11515377.} 
%The columns are the same as Table~\ref{tab_868}. 
\label{tab_115}}
\begin{tabular}{c c c c}
\hline \hline
%\T \B Frequency ($\mu$Hz) & $\zeta$ & $\delta\nu_{\rm R}$ (nHz) & $\delta_{\rm asym}$ (nHz) \\
\rule[-2ex]{0pt}{5.4ex} $\dfrac{\omega_{m=0}}{2\pi}$ ($\mu$Hz) & $\zeta$ & $\dfrac{\delta\!\omega_{\rm R}}{2\pi}$ (nHz) & $\dfrac{\delta_{\rm asym}}{2\pi}$ (nHz) \\
\hline
\T $ 173.687 \pm 0.005$ & $0.693$ & $ 240.0 \pm  12.1$ & $-140.0 \pm  15.5$ \\
$ 175.006 \pm 0.008$ & $0.547$ & $ 213.7 \pm  12.7$ & $ -46.1 \pm  20.2$ \\
$ 176.994 \pm 0.003$ & $0.907$ & $ 302.8 \pm  10.2$ & $ -99.8 \pm  11.5$ \\
$ 179.491 \pm 0.004$ & $0.965$ & - & $-111.0 \pm   9.2$ \\
$ 187.480 \pm 0.007$ & $0.834$ & $ 284.9 \pm   6.1$ & $ -97.9 \pm  14.6$ \\
$ 189.254 \pm 0.005$ & $0.427$ & $ 152.5 \pm  10.7$ & $ -14.1 \pm  15.3$ \\
$ 191.127 \pm 0.004$ & $0.851$ & $ 292.5 \pm   5.4$ & $ -79.6 \pm   8.9$ \\
$ 200.185 \pm 0.002$ & $0.940$ & $ 327.9 \pm  12.9$ & $-106.2 \pm  13.8$ \\
$ 203.013 \pm 0.010$ & $0.636$ & $ 237.3 \pm  11.8$ & $ -94.3 \pm  23.5$ \\
$ 204.634 \pm 0.005$ & $0.541$ & $ 159.6 \pm  26.8$ & $ -28.5 \pm  28.5$ \\
$ 207.453 \pm 0.004$ & $0.927$ & $ 315.4 \pm   9.5$ & $ -70.2 \pm  12.4$ \\
$ 217.750 \pm 0.018$ & $0.639$ & $ 223.2 \pm  16.8$ & $  40.8 \pm  39.8$ \\
\B $ 219.491 \pm 0.018$ & $0.514$ & $ 233.0 \pm  11.5$ & $  59.5 \pm  43.1$ \\
\hline
\end{tabular}
\end{center}
\end{table}

\clearpage
\begin{table}
\begin{center}
\caption{\textbf{Extracted seismic parameters for KIC\,7518143.} 
%The columns are the same as Table~\ref{tab_868}. 
\label{tab_751}}
\begin{tabular}{c c c c}
\hline \hline
%\T \B Frequency ($\mu$Hz) & $\zeta$ & $\delta\nu_{\rm R}$ (nHz) & $\delta_{\rm asym}$ (nHz) \\
\rule[-2ex]{0pt}{5.4ex} $\dfrac{\omega_{m=0}}{2\pi}$ ($\mu$Hz) & $\zeta$ & $\dfrac{\delta\!\omega_{\rm R}}{2\pi}$ (nHz) & $\dfrac{\delta_{\rm asym}}{2\pi}$ (nHz) \\
\hline
\T $ 132.640 \pm 0.006$ & $0.705$ & $ 124.4 \pm  15.4$ & $ 39.4 \pm  20.1$ \\
$ 133.437 \pm 0.007$ & $0.629$ & $ 114.3 \pm  16.8$ & $ -4.3 \pm  21.6$ \\
$ 134.511 \pm 0.002$ & $0.893$ & $ 152.2 \pm   5.6$ & $ 69.3 \pm   6.8$ \\
%$ 135.866 \pm 0.004$ & $0.961$ & - & $ 60.2 \pm   8.3$ \\
$ 141.823 \pm 0.002$ & $0.971$ & - & $ 51.5 \pm   4.2$ \\
$ 143.363 \pm 0.001$ & $0.932$ & $ 155.7 \pm   3.8$ & $ 41.4 \pm   4.0$ \\
$ 144.792 \pm 0.007$ & $0.721$ & $ 128.6 \pm   6.3$ & $ 25.9 \pm  14.9$ \\
$ 146.925 \pm 0.001$ & $0.890$ & $ 153.1 \pm   3.0$ & $ 56.6 \pm   4.2$ \\
$ 155.630 \pm 0.005$ & $0.924$ & - & $ 43.7 \pm  11.1$ \\
$ 157.232 \pm 0.005$ & $0.638$ & $ 127.9 \pm  10.0$ & $ 22.1 \pm  14.3$ \\
$ 158.189 \pm 0.007$ & $0.614$ & $ 106.1 \pm  26.5$ & $ 39.3 \pm  29.6$ \\
$ 159.786 \pm 0.001$ & $0.918$ & $ 161.1 \pm   3.0$ & $ 48.6 \pm   4.0$ \\
$ 167.937 \pm 0.004$ & $0.915$ & - & $ 12.3 \pm  13.6$ \\
\B $ 170.786 \pm 0.010$ & $0.671$ & $ 121.5 \pm   9.1$ & $-23.1 \pm  21.5$ \\
\hline
\end{tabular}
\end{center}
\end{table}

\clearpage
\begin{table}
\begin{center}
\caption{\textbf{Average rotation rates in the core $\omg$ and in the envelope $\omp$.} \label{tab_rot}}
\begin{tabular}{l c c}
\hline \hline
\T KIC ID & $\omg/(2\pi)$ & $\omp/(2\pi)$  \\
\B & (nHz) & (nHz) \\
\hline
\T 8684542 & $667\pm43$ & $-24\pm15$   \\
\B 11515377 & $669\pm46$ & $52\pm14$ \\
7518143 & $338\pm53$ & $31\pm17$ \\ %relative order compared to other tables
\hline
\end{tabular}
\end{center}
\end{table}

\clearpage\begin{table}
\begin{center}
\caption{\textbf{Magnetic field properties for the three red giants.} These values were obtained by fitting an asymptotic expression of mixed modes that includes a magnetic perturbation to the observed mode frequencies of KIC\,8684542, KIC\,11515377, and KIC\,7518143. \label{tab_shiftmag}}
\begin{tabular}{l c c c}
\hline \hline
\T\B parameter & KIC\,8684542 & KIC\,11515377 & KIC\,7518143 \\
\hline
\T $\dpun$ (s) & $80.47\pm0.11$  & $83.56\pm0.14$ & $78.52\pm0.08$ \\
$\delta\!\omega_\mathrm{g}/(2\pi)$ (nHz) & $195\pm45$ & $126\pm60$ & $<35$ \\
\B $a$ & $0.47\pm0.12$ & $-0.24_{-0.23}^{+0.08}$ & $>0.24$ \\
\hline
\T \B $\langle B_r^2\rangle^{1/2}$ (kG) & $102\pm12$ & $98\pm24$ & $<41$  \\
\hline
\end{tabular}
\end{center}
\end{table}

%\clearpage
%
%\renewcommand{\contentsname}{List of Supplementary Methods}
%\renewcommand{\listfigurename}{List of Supplementary Figures}
%\renewcommand{\listtablename}{List of Supplementary Tables}
%
%\small
%\tableofcontents
%\listoffigures
%\listoftables

%\noindent LaTeX formats citations and references automatically using the bibliography records in your .bib file, which you can edit via the project menu. Use the cite command for an inline citation, e.g.  \cite{Haogidmaps2014}.

%For data citations of datasets uploaded to e.g. \emph{figshare}, please use the \verb|howpublished| option in the bib entry to specify the platform and the link, as in the \verb|Haogidmaps2014| example in the sample bibliography file.

%\section*{Additional information}

%To include, in this order: \textbf{Accession codes} (where applicable); \textbf{Competing interests} (mandatory statement). 

%The corresponding author is responsible for submitting a \href{http://www.nature.com/srep/policies/index.html#competing}{competing interests statement} on behalf of all authors of the paper. This statement must be included in the submitted article file.

%\begin{table}[ht]
%\centering
%\begin{tabular}{|l|l|l|}
%\hline
%Condition & n & p \\
%\hline
%A & 5 & 0.1 \\
%\hline
%B & 10 & 0.01 \\
%\hline
%\end{tabular}
%\caption{\label{tab:example}Legend (350 words max). Example legend text.}
%\end{table}

\end{document}